\pdfoutput=1

\documentclass[11pt]{article}

\usepackage{ACL2024}

\usepackage{hyperref}

\usepackage{times}
\usepackage{latexsym}
\usepackage{arydshln}
\usepackage{graphicx}
\usepackage{subcaption}
\usepackage{booktabs,arydshln}
\usepackage{amsmath}
\usepackage{enumitem}
\usepackage{multirow}
\usepackage{cleveref}
\usepackage[export]{adjustbox}
\usepackage[utf8]{inputenc}
\usepackage{xurl}
\usepackage{tabularray}
\usepackage{siunitx}
\usepackage{longtable}

\usepackage{amssymb}
\usepackage{pifont}

\newcommand{\circa}{{\raise.17ex\hbox{$\scriptstyle\sim$}}}

\usepackage{array}
\newcolumntype{L}[1]{>{\raggedright\let\newline\\\arraybackslash\hspace{0pt}}m{#1}}
\newcolumntype{C}[1]{>{\centering\let\newline\\\arraybackslash\hspace{0pt}}m{#1}}
\newcolumntype{R}[1]{>{\raggedleft\let\newline\\\arraybackslash\hspace{0pt}}m{#1}}

\usepackage[T1]{fontenc}

\usepackage[utf8]{inputenc}

\usepackage{microtype}

\usepackage{silence}
\usepackage[colorinlistoftodos]{todonotes}
\usepackage{cleveref}
\usepackage{listings}
\usepackage[most]{tcolorbox}

\newtcolorbox{highlightbox}[1][]{%
    colback=yellow!15,
    colframe=yellow!15,
    notitle,
    sharp corners,
    enhanced,
    breakable,
    }

\makeatletter
\newcommand*\iftodonotes{\if@todonotes@disabled\expandafter\@secondoftwo\else\expandafter\@firstoftwo\fi} 
\makeatother

\crefname{section}{\S}{\S\S} 
\Crefname{section}{\S}{\S\S} 
\crefname{table}{Tab.}{Tables}
\crefname{figure}{Fig.}{Figures}
\crefname{algorithm}{Algorithm}{}
\crefname{equation}{eq.}{}
\crefname{appendix}{App.}{}
\crefname{lstlisting}{listing}{listings}
\Crefname{lstlisting}{Listing}{Listings}
\crefformat{section}{\S#2#1#3} 

\definecolor{KUPetrol}{RGB}{0,120,148} 

\definecolor{KUBlue}{RGB}{33,92,175} 
\definecolor{KUGreen}{RGB}{98,115,19} 
\definecolor{KUPurpleDark}{RGB}{140,10,89} 
\definecolor{KUPurple}{RGB}{163,7,116} 
\definecolor{KUGray}{RGB}{111,111,111} 
\definecolor{KURed}{RGB}{183,53,45} 
\definecolor{KUPetrol}{RGB}{0,120,148} 
\definecolor{KUBronze}{RGB}{142,103,19} 

\newtoggle{color-macro}
\settoggle{color-macro}{true} 

\iftoggle{color-macro}{\colorlet{MacroColor}{KUPetrol}
}{
\colorlet{MacroColor}{black}
}

\iftoggle{color-macro}{
\colorlet{TokenColor}{KUBronze}
}{
\colorlet{TokenColor}{black}
}

\iftoggle{color-macro}{
\colorlet{MathSubColor}{KUPurple}
}{
\colorlet{MathSubColor}{black}
}

\iftoggle{color-macro}{
\colorlet{RedditColor}{black}
}{
\colorlet{RedditColor}{black}
}

\iftoggle{color-macro}{
\colorlet{SchwartzProbColor}{KUBlue}
}{
\colorlet{SchwartzProbColor}{black}
}

\iftoggle{color-macro}{
\colorlet{IColor}{KUGray}
}{
\colorlet{IColor}{black}
}

\usepackage[T1]{fontenc}

\usepackage{booktabs} 

\usepackage{amsmath,amsfonts,bm}

\def\eqref#1{equation~\ref{#1}}

\def\1{\bm{1}}

\DeclareMathAlphabet{\mathsfit}{\encodingdefault}{\sfdefault}{m}{sl}
\SetMathAlphabet{\mathsfit}{bold}{\encodingdefault}{\sfdefault}{bx}{n}

\def\sX{{\mathbb{X}}}

\def\sX{{\mathbb{X}}}

\definecolor{codegreen}{rgb}{0,0.6,0}
\definecolor{codegray}{rgb}{0.5,0.5,0.5}
\definecolor{codepurple}{rgb}{0.58,0,0.82}
\definecolor{backcolour}{rgb}{0.97,0.97,0.95}

\lstdefinestyle{mystyle}{
    backgroundcolor=\color{backcolour},   
    commentstyle=\color{codegreen},
    keywordstyle=\color{magenta},
    numberstyle=\tiny\color{codegray},
    stringstyle=\color{codepurple},
    basicstyle=\ttfamily\footnotesize,
    breakatwhitespace=true,         
    breaklines=true,                 
    captionpos=b,                    
    keepspaces=true,                 
    numbers=left,                    
    numbersep=5pt,                  
    showspaces=false,                
    showstringspaces=false,
    showtabs=false,                  
    tabsize=2
}

\title{\raisebox{-0.3em}{\includegraphics[height=1.1em]{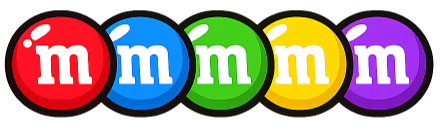}}: A Unified Taxonomy for Investigating the \textcolor{Red}{M}echanisms of \textcolor{NavyBlue}{M}ultilingual \textcolor{Green}{M}ulti\textcolor{Dandelion}{M}odal \textcolor{Purple}{M}isinformation}

\author{
Nadav Borenstein \quad
Greta Warren \quad
Desmond Elliott \quad
Isabelle Augenstein \\
University of Copenhagen \quad \\
{\tt \href{mailto:nb@di.ku.dk}{nb@di.ku.dk}} \quad
{\tt \href{mailto:grwa@di.ku.dk}{grwa@di.ku.dk}} \quad \\
{\tt \href{mailto:de@di.ku.dk}{de@di.ku.dk}}  \quad 
{\tt \href{mailto:augenstein@di.ku.dk}{augenstein@di.ku.dk}} \quad 
}

\begin{document}
\maketitle


\begin{abstract}
Multimodal misinformation on social media is highly prevalent, potent, and harmful, yet difficult to detect and counter, and still poorly understood compared to its text-only counterpart. Research on the properties and deceptive strategies of multimodal misinformation is hindered by a lack of taxonomies grounded in real-world contexts and by the limitations of current multimodal machine learning models, which prevent the automation of annotation and analysis at scale. We address these shortcomings in three steps. First, we collect a large-scale, high-quality dataset of real-world misinformation instances from Twitter/$\sX$ in seven languages. Second, we develop a novel, comprehensive taxonomy of multimodal misinformation grounded in an in-depth qualitative analysis of the data and prior theoretical work. Finally, we operationalise the taxonomy through an automated multi-step annotation pipeline using a Vision-Language Model (VLM), and perform human-validation. Our novel approach leads to previously undocumented insights about how social media users combine images with text to spread misinformation in the wild, e.g., that AI-generated content is particularly prevalent in technology and science, while vaccination misinformation disproportionately utilises images from news outlets to assert credibility.
Our method and findings provide guidance for targeted approaches for detecting multimodal misinformation, and suggest that mitigation efforts should be developed and applied strategically rather than uniformly.

\end{abstract}
\everypar{\looseness=-1}

\section{Introduction}
\label{sec:introduction}
Online misinformation poses severe real-world harms to individuals, groups and society. These risks are amplified when the misinformation is multimodal: claims accompanied by images and videos\footnote{While some elements of this study can be applied to videos, the focus of this work is images.} can be more convincing  \citep{Newman2012NonprobativeP, Newman2020TruthinessH, Hameleers03032020} and emotionally salient \cite{Li2019IsAP} than text alone. This concern is compounded by the multimodal nature of today's social media, as most popular platforms being either exclusively (TikTok, Instagram, YouTube) or predominantly (Twitter/$\sX$\footnote{We will use the name $\sX$ in this work.}, Facebook, and Reddit) multimodal.

\begin{figure}
    \centering
    \includegraphics[width=1\linewidth]{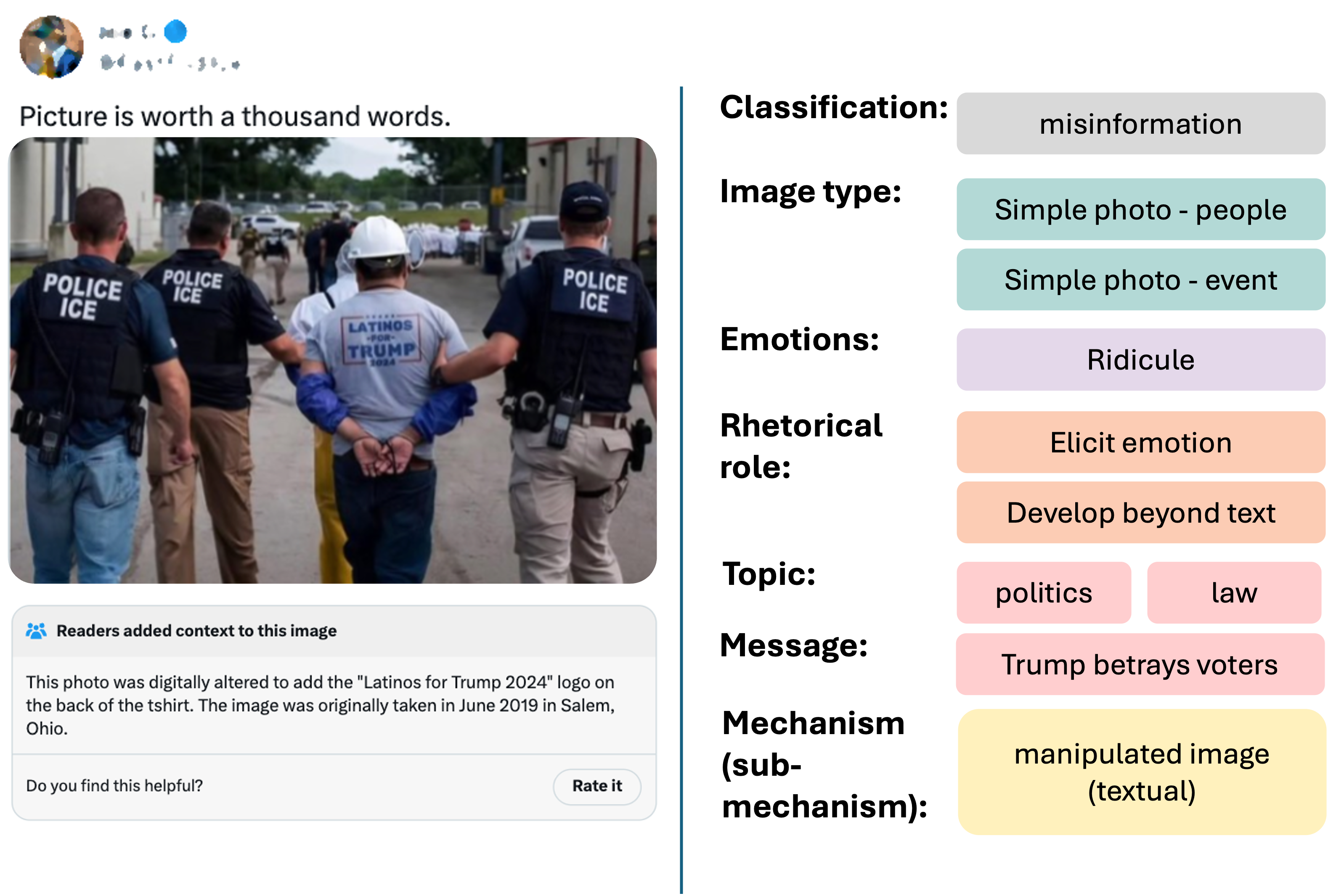}
    \caption{Example of a multimodal misinformation instance found on $\sX$, and the labels assigned to it from our taxonomy. Refer to \cref{tab:taxonomy_with_definitions} in \cref{app:full_taxonomy} for a detailed description of the taxonomy and its labels.}
    \label{fig:misinfo-example}
\end{figure}

Developing a comprehensive taxonomy of how multimodal misinformation manifests on social media is a crucial step toward designing effective mitigation strategies. In particular, categorising the \textbf{mechanisms} (i.e., how a image-text pairing produces a false or misleading impression) through which images are used to propagate false narratives (e.g., taking an image out of context or manipulating its content) can provide actionable guidance on which interventions are likely to be most impactful and cost-effective. Existing work is limited; current taxonomies of multimodal misinformation (see \cref{sec:related_work} for an overview) are disparate, contain ambiguous categories, and leave many cases unclassified.\footnote{For example, approximately 40\% of images annotated with the taxonomy defined in \citet{Dufour2024AMMeBaAL} could not be classified into any of their own pre-defined subcategories.}

Moreover, the scope of current research is limited by the difficulty of reliably analysing multimodal misinformation at scale. It is challenging for crowdsourced annotators to apply extensive taxonomies, while the complex and ambiguous nature of multimodal misinformation often requires domain-specific knowledge to accurately evaluate them. Until recently, Vision Language Models (VLMs) lacked the capabilities to reliably annotate such data \citep{qi2024sniffermultimodallargelanguage, 10.1145/3697349, liu2024mmfakebench, zhang2026imagesspeaklouderwords}. To date, this has necessitated reliance on time and resource-intensive expert annotations, limiting the scale of analysis.

This study addresses these limitations through a \textit{novel and comprehensive taxonomy of multimodal misinformation mechanisms}, including five additional orthogonal axes: Classification, Image Type, Emotions, Rhetorical Role, and Topic and Message (\cref{sec:axes}). To ensure practical relevance, the taxonomy's development was guided both by prior theoretical work and an \textit{in-depth qualitative analysis of a multilingual dataset of real-world misinformation instances} collected from $\sX$’s Community Notes dataset. As our taxonomy was developed based on a broad set of misinformation (Community Notes), it can be applied more widely than prior taxonomies based on fact-checking articles, which are necessarily narrower in terms of categories.
We then develop and evaluate an annotation pipeline based on a VLM to apply the taxonomy at scale to the Community Notes dataset (\cref{sec:experiments}).

We demonstrate the utility of our approach by examining how text and images are combined \textbf{in real-world contexts} to propagate false narratives (\cref{sec:analysis}). We show the generalisability of our taxonomy and annotation pipeline by applying it to the \textsc{AMMeBa} dataset \cite{Dufour2024AMMeBaAL}, which consists of multimodal misinformation verified by professional fact-checkers, finding that our approach shows good agreement with expert human annotators across both datasets.

Our analysis of the annotated data reveals that images that provide a slanted representation of the truth are among the most common types of multimodal information, that AI-generated content is a particularly prevalent mechanism in technology and science domains, and that vaccination-related misinformation disproportionately appropriates images from news outlets for credibility.
These findings show that misinformation mechanisms differ systematically across domains and narratives, suggesting that mitigation efforts should be developed and applied strategically rather than uniformly.

\section{Background}
\label{sec:related_work}

\subsection{Multimodal Misinformation Taxonomies}

Few works explicitly define taxonomies of multimodal misinformation mechanisms. \citet{fauxtography} propose seven categories, including \textit{Denial of image authenticity} and \textit{Manufactured media}, while \citet{NEWMAN2024101778} introduce a five-category scheme with labels such as \textit{Slanted representation}, \textit{Repurposed content}, and \textit{Decorative}. The most comprehensive effort to date is the taxonomy by \citet{Dufour2024AMMeBaAL}, resulting in the \textsc{AMMeBa} dataset. It covers several orthogonal axes, including image type (e.g., screenshot vs.\ simple photo), media type (image, video, multi-image), and mechanism. This last axis defines three top-level categories: \textit{Content manipulation} (with sub-categories such as \textit{AI-generated content} and \textit{news headline manipulation}), \textit{Context manipulation} (sub-categories: \textit{date manipulation}, \textit{identity manipulation}, etc.), and \textit{Text-based manipulation}.

A larger body of work defines taxonomies \textit{implicitly}, typically by constructing benchmark datasets. \citet{liu2024mmfakebench} introduce MM-FakeBench, covering twelve distinct mechanisms such as \textit{PS images}, \textit{scene inconsistency}, and \textit{semantic inconsistency}. \citet{nakamura-etal-2020-fakeddit} curate a six-category dataset from Reddit, and \citet{shao2023detectinggroundingmultimodalmedia} construct a benchmark by applying four manipulation types (e.g., face swaps) to benign news articles.

Finally, some broader misinformation research offers schemas that, while not exclusively multimodal, still transfer well to a multimodal setting. Most notably, \citet{wardle2017} present a broad taxonomy that spans orthogonal axes, such as the intent of the misinformation source (e.g., intentional vs accidental harm), the recipient's reaction, and the message itself. Importantly, the work also describes 10 misinformation mechanisms directly applicable to multimodal cases. See overview in \cref{tab:taxonomy_overview} in \cref{app:taxonomies}.  
However, existing taxonomies are fragmented, with various levels of scope and granularity. We develop a taxonomy which unifies prior schemes while introducing novel, empirically grounded categories derived through qualitative analysis of real-world data.

\subsection{Datasets of Multimodal Misinformation}

Multimodal misinformation datasets typically address a specific task, e.g., claim detection \citep{Nielsen2022MuMiNALA, cheema-etal-2022-mm}, out-of-context prediction \citep{luo-etal-2021-newsclippings, Aneja2021COSMOSCO}, image manipulation detection \citep{shao2023detectinggroundingmultimodalmedia, Wang2024MFCBenchBM}, and deepfake detection \citep{Wang2024MFCBenchBM, 10.1145/3746275.3762215} (see \citet{rothermel2026veritasdynamicbenchmarkmultimodal} for a broader overview).
%
Most datasets, such as \textsc{AMMeBa}, source examples from professional fact-checking organisations \citep{Dufour2024AMMeBaAL, rothermel2026veritasdynamicbenchmarkmultimodal}, whereas others rely on synthetic data generation \citep{Wang2024MFCBenchBM}.
An alternative approach is to collect instances directly from social media \citep{nakamura-etal-2022-hybridialogue, xing-etal-2026-communitynotes}, which may yield more diverse examples than those selected for professional fact-checking. A notable source of social media-based multimodal misinformation is Community Notes \citep{TwitterBirdwatch2021}, $\sX$'s community-based moderation platform. 
In this paper, we curate a dataset of misleading posts and images from $\sX$ in order to capture a broader range of multimodal misinformation than existing datasets (see \cref{sec:datasets}).

\subsection{VLMs for Multimodal Misinformation}

VLMs have increasingly been applied to multimodal misinformation tasks, including claim detection \citep{van-der-meer-etal-2025-hintsoftruth}, out-of-context identification \citep{qi2024sniffermultimodallargelanguage, 10.1007/978-3-031-57916-5_8}, image manipulation detection \citep{Wang2024MFCBenchBM, liu2024mmfakebench}, and justification generation \citep{qi2024sniffermultimodallargelanguage}. However, their limitations in multimodal reasoning and instruction following \citep{villegas2026reasoningdynamicslimitsmonitoring} have constrained their utility \citep{10.1007/978-3-031-57916-5_8, liu2024mmfakebench, zhang2026imagesspeaklouderwords}. These limitations are especially consequential for tasks requiring nuanced reasoning \citep{ai2026paradigm}, such as fine-grained analysis and categorisation of misinformation instances addressed in this study. Nevertheless, the rapid evolution of VLM capabilities has steadily expanded the range and complexity of tasks they can reliably complete, as demonstrated by our findings.

\section{Datasets}
\label{sec:datasets}

\begin{table*}[t]
  \centering
  \footnotesize
  \setlength{\tabcolsep}{4pt}
  \renewcommand{\arraystretch}{1.4}
  \begin{tabular}{
    >{\raggedright\arraybackslash\bfseries}p{1cm} 
    >{\raggedright\arraybackslash}p{3.4cm} 
    >{\raggedright\arraybackslash}p{3.4cm} 
    >{\raggedright\arraybackslash}p{3.4cm} 
    >{\raggedright\arraybackslash}p{3.4cm}
  }
    \toprule

    User name &
      @NBA\_follower &
      @Illuminati &
      @ShallowSeek &
      @QuakePredictor \\
    \addlinespace[4pt]

    Post &
      The Washington Wizards haven't won a playoff series since 1979! &
      Are you awake yet? &
      DeepSeek Now BANNED on ALL Apple + Google devices over `Safety' Concerns. &
      70 HOUR WARNING: Major earthquake is likely in Southern California - most likely in Los Angeles or Ventura County ... \\
    \addlinespace[4pt]

    Image &
      \includegraphics[width=\linewidth, keepaspectratio]{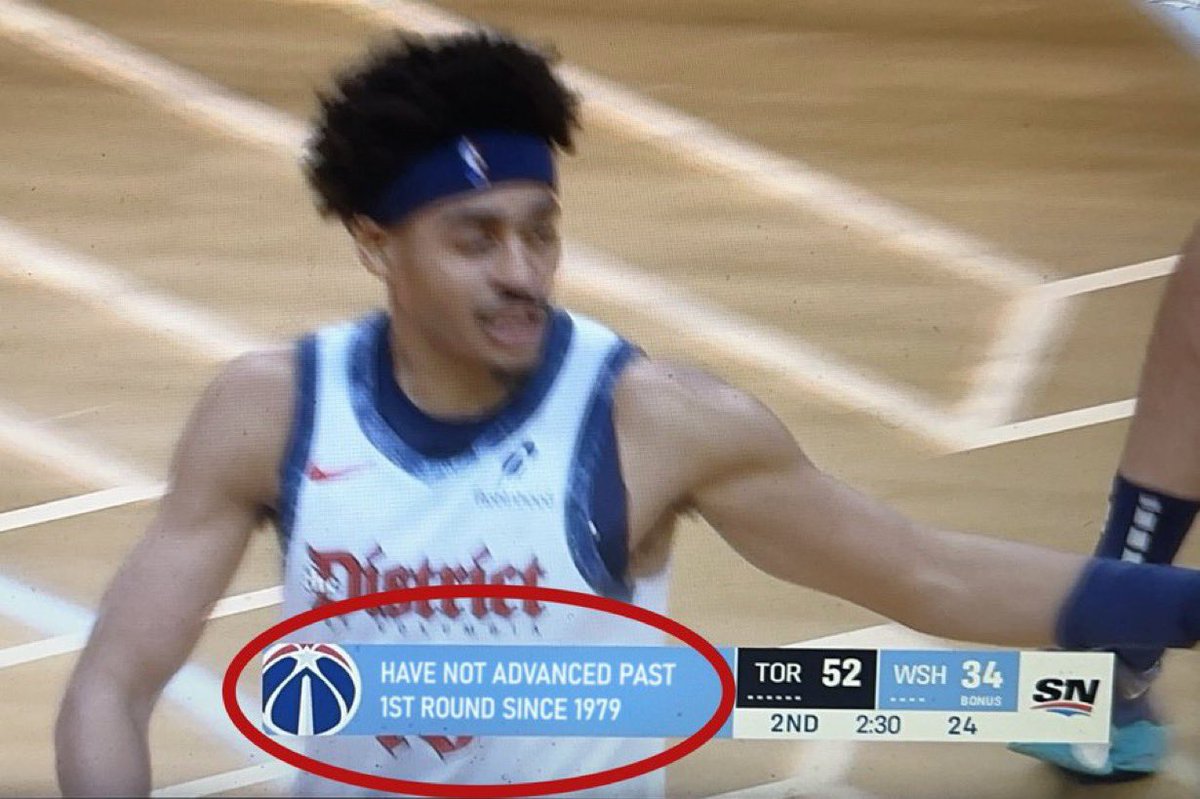} &
      \includegraphics[width=\linewidth, keepaspectratio, trim={0 3cm 0 2cm}, clip]{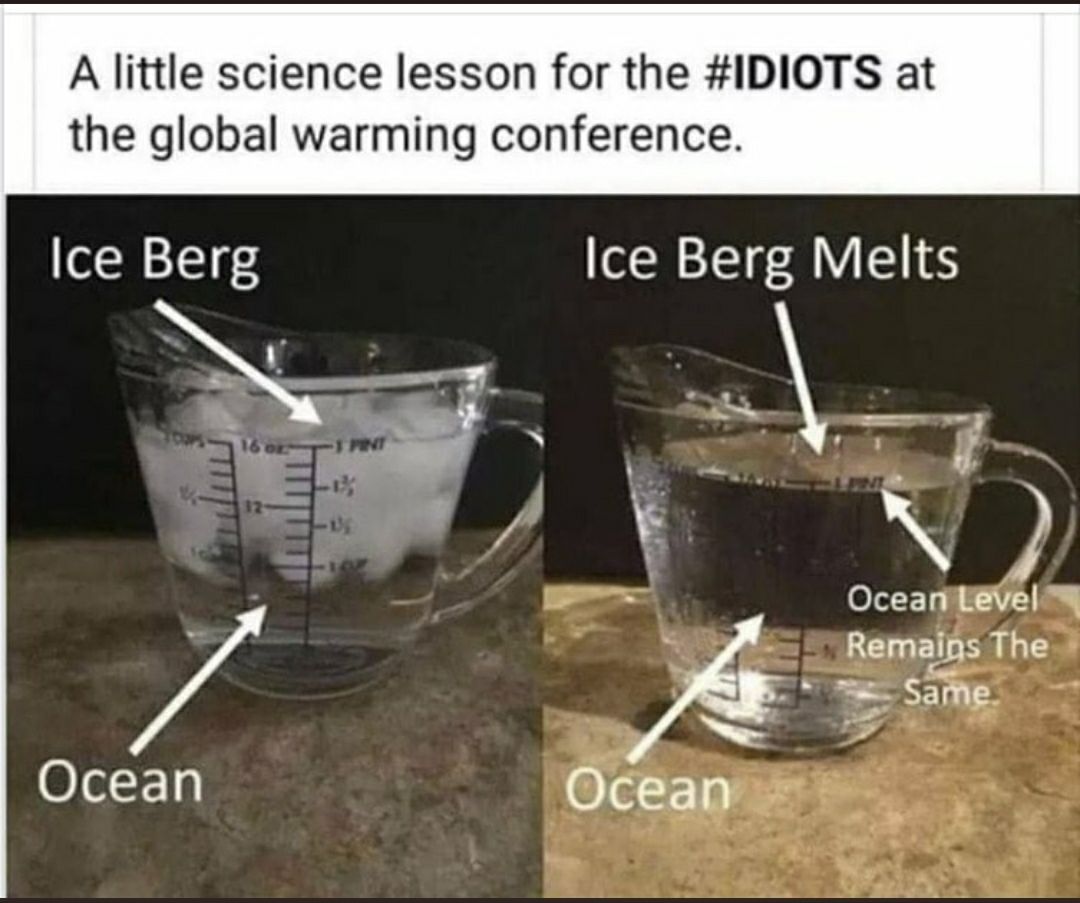} &
      \includegraphics[width=\linewidth, keepaspectratio]{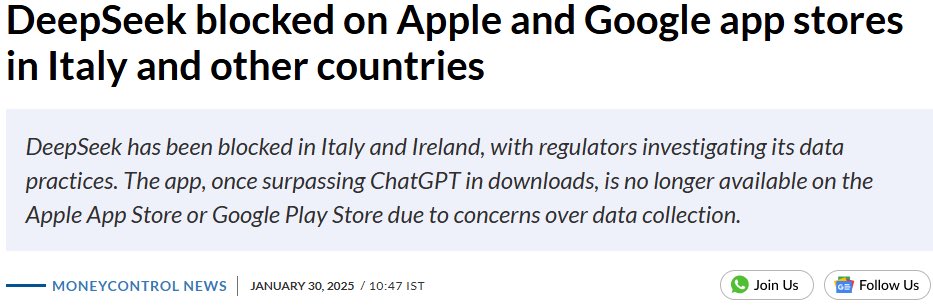} &
      \includegraphics[width=\linewidth, keepaspectratio, trim={0 10cm 0 0}, clip]{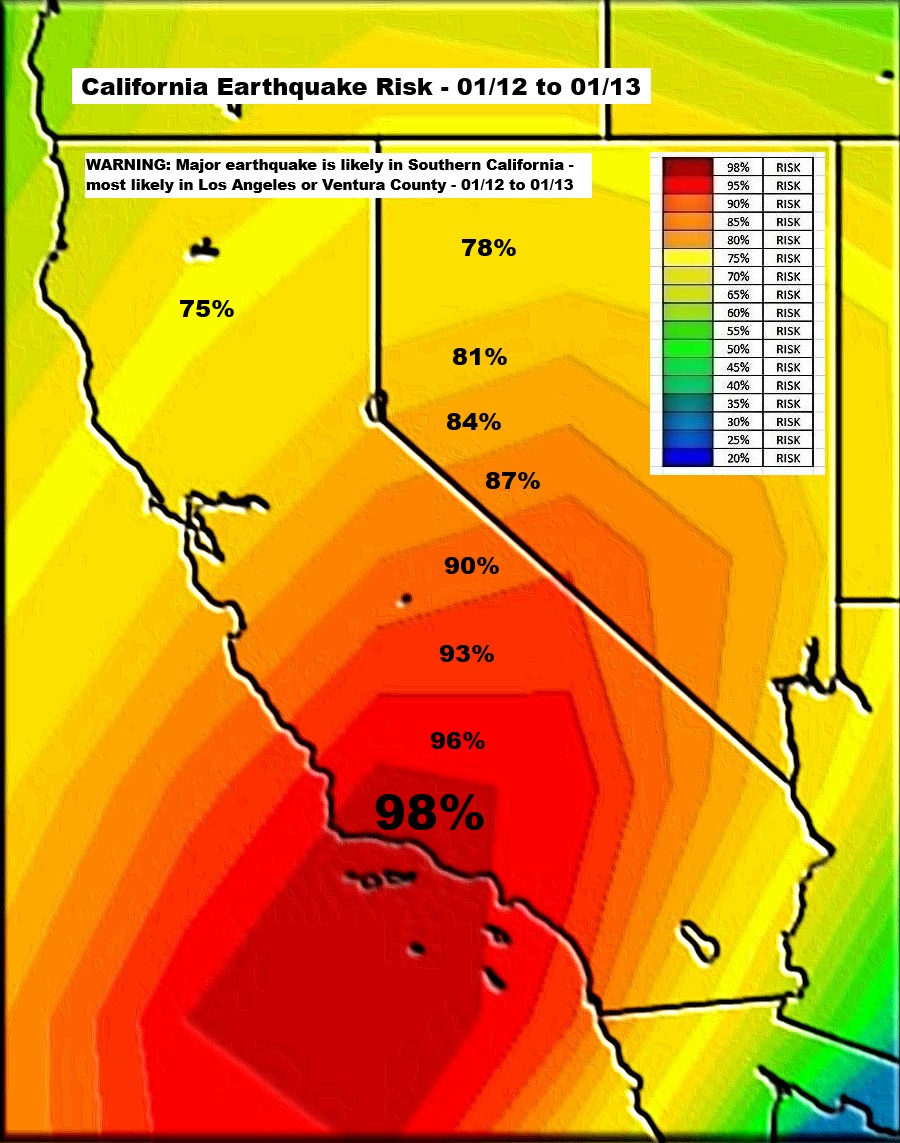} \\
    \addlinespace[4pt]

    Note &
      The wizards have made it out of the first round multiple times, including beating the Atlanta hawks in 2017. landofbasketball.com/teams/playoff\_... &
      This meme does not accurately reflect what occurs when land ice, such as glaciers, icebergs and ice sheets, melt into the ocean ... melting land ice is what contributes to the sea level rise... &
      DeepSeek is not banned on ``ALL'' devices, as specified in the image shown in the post. Only in regions of Europe... &
      There is no scientific basis for this or any earthquake prediction. Quakes are always possible in places like California, but specific forecasts perform no better than random when tested... \\

    \bottomrule
  \end{tabular}
  \caption{Community Notes samples with usernames, posts, images, and notes. Usernames are fake to retain privacy.}
  \label{tab:tweet_sample}
\end{table*}

We aimed to curate a diverse, high-quality, and contemporary dataset of multimodal misinformation samples to guide the development of our taxonomy by grounding it in real instances of online multimodal misinformation. Beyond supporting taxonomy development, such a dataset also allows us to empirically validate the taxonomy by applying it at scale, and illustrate the kinds of research questions that can now be answered (see \cref{sec:experiments} and \cref{sec:analysis} for details). $\sX$'s Community Notes is a natural candidate for the following reasons:

\noindent \textbf{Quality.} Community Notes employs strong consensus-driven self-moderation policies \citep{TwitterRatingWritingImpact, TwitterNoteRanking}, mitigating the risk of encountering low-quality notes.

\noindent \textbf{Diversity.} Notes (and corresponding posts) span a wide range of topics and languages, and capture a wide variety of misinformation 
that fact-checking agencies might overlook or deprioritise \citep{borenstein-etal-2025-community, mohammadi2026birdwatchcommunitynotestwitter,augenstein2025communitymoderationnewepistemology}.

\noindent \textbf{Relevance.} The dataset is contemporary and dynamic, with new notes contributed daily. Therefore, the taxonomy and automated annotation pipeline developed in this work can also be applied to analyse future issues and trends in misinformation.

\noindent \textbf{Format.} Each instance follows a consistent structure that reduces to a \texttt{(username, claim, image, note)} tuple, greatly simplifying their analysis (see examples in \cref{tab:tweet_sample}) by both humans and automated tools.

We base our dataset on COMMUNITYNOTES \citep{xing-etal-2026-communitynotes}, which contains over 100k \texttt{(post, note)} pairs collected from $\sX$. We adapt it by discarding pairs whose notes were rated as \textit{`unhelpful'} by the community and retaining only posts written in English, Spanish, Portuguese, Japanese, French, German, or Hebrew.\footnote{These languages were selected because of their high prevalence in the dataset.} We use web scraping to identify and retain posts containing images, and download these images. The resulting dataset contains 24,596 samples. To further expand the dataset, we collect additional examples from the official Community Notes release.\footnote{\url{https://communitynotes.x.com/guide/en/under-the-hood/download-data}} We remove deleted notes, notes rated as `\textit{unhelpful}' by the community, notes attached to posts that are labeled as `\textit{not misleading}', and notes written in languages not specified above. This public dataset, however, does not include the original posts and provides no indication of which ones contain non-textual media. We therefore use Twikit\footnote{\url{https://github.com/d60/twikit}} to recover the posts, sampling and scraping 25k instances. Finally, we retain only posts that include an image. The final merged dataset contains 26,979 \texttt{(username, claim,\footnote{Generally, the textual claim is the post's text. However, in some cases the claim is contained in the image. See \cref{sec:taxonomy}.} image, note)} tuples, spanning January 2021 to January 2026. Statistics are reported in \cref{tab:cn_statistics} and \cref{fig:date_statistics} in \cref{app:dataset_statistics}, with representative examples in \cref{tab:tweet_sample}.


To assess how well our taxonomy and pipeline generalise beyond Community Notes, we also apply them to a sample from the \textsc{AMMeBa} dataset \cite{Dufour2024AMMeBaAL}, which comprises multimodal fact-checks collected from a large set of professional fact-checking agencies. Statistics are provided in \cref{tab:amoeba_stats} (\cref{app:dataset_statistics}), and collection details in \cref{app:ammeba_collection}. Crucially, we only sampled instances in which the image plays an active role in the misinformation (\textsc{AMMeBa} includes this annotation).

\section{Taxonomy Creation}
\label{sec:taxonomy}

\begin{figure*}
    \centering
    \includegraphics[width=1\linewidth, trim={0 0 0 1.55cm},clip]{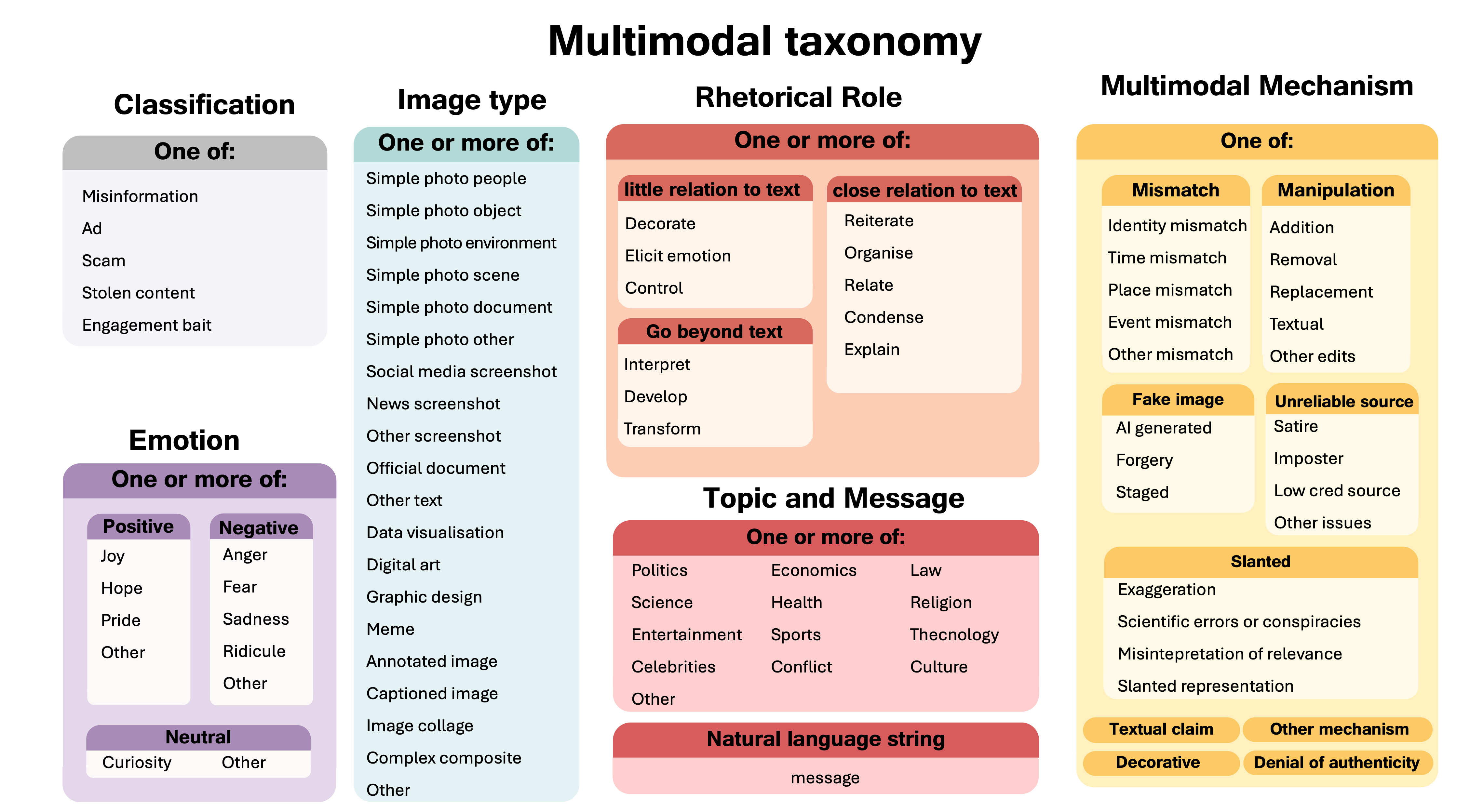}
    \caption{The six axes of our multimodal misinformation taxonomy. Definitions are given in \cref{tab:taxonomy_with_definitions} in \cref{app:full_taxonomy}.}
    \label{fig:taxonomy}
\end{figure*}

\subsection{Development Process}
\label{sec:taxonomy_development}

\cref{tab:taxonomy_overview} in \cref{app:taxonomies} summarises previous attempts at defining taxonomies for multimodal mechanisms of misinformation. These taxonomies, however, are disparate and poorly aligned: Identical mechanisms are often assigned different category names (e.g., \textit{Deepfake} vs.\ \textit{AI-generated}, or \textit{Repurposed} vs.\ \textit{Out-of-context}). Similar mechanisms can also be defined in subtle yet important ways, despite their superficial similarity (e.g., \textit{Manufactured} in \citet{fauxtography}, which can also be applied to face swaps, vs.\ \textit{Fabricated} in \citet{brennen2020types}, which cannot). Finally, the taxonomies differ in coverage, including or omitting different categories. Fourth, their structure differs -- some taxonomies are hierarchical, whereas others are flat.

We unified these taxonomies in three stages. First, we aligned exact matches: categories (or sub-categories) that share an identical mechanism, regardless of their naming. Second, we consolidated near-matches, i.e., categories with similar definitions. We ensured that each resulting (sub-)category is sufficiently distinguishable to prevent annotators from encountering ambiguous decision boundaries. Third, we organised the remaining categories into a hierarchical structure.

The unification process was not purely conceptual. We developed the taxonomy iteratively through dataset annotation, using empirical evidence to guide definitions of the taxonomy categories: Some definitions proved too vague to annotate reliably and required refining. Consider the \textit{Slanted} category \citep{NEWMAN2024101778}, defined as ``real photos that provide a biased representation of claims made in accompanying text.'' It is unclear, for example, whether the sample presented in column 2 of \cref{tab:tweet_sample}  is considered \textit{Slanted} or not. Other categories, such as \textit{Mirrored} \citep{fauxtography}, occur so rarely that we decide to exclude them. Some category pairs, though theoretically distinct, were indistinguishable in practice, prompting us to  refine or merge them. 

This iterative process also surfaced gaps in the merged taxonomy -- instances that did not neatly fit into any existing category, or that demonstrated mechanism types important enough to be included, prompting us to add several novel categories. We describe three illustrative examples:
The first is exemplified by column 4 of \cref{tab:tweet_sample}. This post claims that a major earthquake is expected in California,  supported by a visual. This claim is patently false, as earthquake prediction at this resolution is scientifically impossible, and the visual is therefore misleading. However, the \emph{mechanism} by which it misleads is hard to pin down using the existing taxonomy. The map could be fabricated or AI-generated, taken out of context, or perhaps generated using some real, pseudo-scientific methods. We therefore added a sub-category under \textit{Slanted}, named \textit{scientific errors or conspiracies}, to cover such instances. 
Second, another common pattern that lacks a dedicated category in prior taxonomies is when the textual claim exaggerates the severity, frequency, or consequences of what the image actually shows (e.g., column 3 of \cref{tab:tweet_sample}). To capture such cases, we added the sub-category \textit{exaggeration} under \textit{Slanted}. 
Column 1 of \cref{tab:tweet_sample} shows the third example, involving false textual claims embedded in the image itself, and the post text merely repeats or amplifies them. No existing category covers this setting, so we introduced the top-level category \textit{Textual claim}.

\subsection{Taxonomy Axes}
\label{sec:axes}

Inspired by work such as \citet{Dufour2024AMMeBaAL} and \citet{wardle2017}, we enrich the taxonomy with six orthogonal axes, enabling the study of phenomena beyond simple frequency analysis of mechanism types. We include:

\noindent \textbf{Classification.} Distinguishes misinformation from related but distinct content types, namely \textit{ads}, \textit{scams}, \textit{clickbait}, and \textit{stolen content}. We adopt this axis from \citet{ayton2019explainingDeceptive}, adapting it to labels already present in Community Notes.

\noindent \textbf{Image Type.} Classifies the image's visual content into categories such as \textit{social media screenshot}, \textit{document}, and \textit{collage}. We adopt this axis from \citep{Dufour2024AMMeBaAL}, greatly expanding the number of labels and allowing multilabel classification. 

\noindent \textbf{Emotion.} Describes the emotions the post author attempts to evoke through the inclusion of the image. As an image can evoke several emotions simultaneously, this axis is multilabel. We add this axis following \citet{Weikmann2022VisualDI} and the broader literature on the significance of the emotional element in misinformation and persuasiveness \citep{Hosseini2023EmotionalFI, sharma2023systematic, zollo2015emotional}.

\noindent \textbf{Rhetorical role.} The functional relationship between text-image pairs. We adopt the top-level categories of \citet{rhetorical_role}'s extensive taxonomy, omitting the finer-grained sub-categories.

\noindent \textbf{Topic and Message.} Captures \textit{what} the misinformation is about, enabling fine-grained analysis of narrative-mechanism interactions. \textit{Topic} is a high-level multilabel category, whereas \textit{Message} is a short free-text description of the specific narrative. While natural language labels are uncommon in taxonomies, this formulation is highly beneficial: misinformation narratives are rapidly evolving and cannot be captured by a fixed set of labels.  

\noindent \textbf{Multimodal Mechanism.} As described in \cref{sec:taxonomy_development}.

We deliberately omit several axes used in prior work. Some, such as \textit{Intent} \citep{ayton2019explainingDeceptive} cannot be reliably annotated from content alone, while others fall outside the scope of this study (e.g., \textit{Recipient} \citep{wardle2017}).

\subsection{Final Taxonomy}
\label{sec:final_taxonomy}

\noindent  \textbf{Final Taxonomy.} The resulting taxonomy with its six axes and 25 mechanisms is visualised in \cref{fig:taxonomy}. Detailed label definitions are given in \cref{tab:taxonomy_with_definitions} in \cref{app:full_taxonomy}.
See \cref{fig:misinfo-example} for an application example.

\subsection{Human Evaluation}
\label{sec:taxonomy_eval}

Having defined the final taxonomy, we evaluate how well it applies to the datasets introduced in \cref{sec:datasets}. We consider two properties: 1) \textit{ambiguity}, the inter-annotator agreement between annotators tasked with applying the taxonomy, and 2) \textit{coverage}, the rate at which dataset instances are assigned meaningful labels (rather than the generic label \textit{Other mechanism}). Specifically, two expert annotators (authors of this study)  labeled 100 samples from the Community Notes dataset using the full taxonomy (see \cref{fig:annotation_platform} in \cref{app:reproducibility} for the annotation interface we used).

Annotator A labeled only one instance as \textit{Other mechanism} and annotator B only three, confirming high coverage. \cref{tab:agreement-all} (top) in \cref{app:evaluation} reports inter-annotator agreement. Notably, agreement is high on \textit{Image type} (Mean Jaccard score of 0.6) and \textit{Topic} (Mean Jaccard score of 0.7), multilabel classification tasks with 20 and 13 labels, respectively. The subjective \textit{Emotion} axis scores lower, and \textit{Rhetorical role} lower still, perhaps reflecting the complexity or ambiguity of \citet{rhetorical_role}'s taxonomy. \textit{Mechanism} and \textit{Sub-mechanism} reach agreement rates of 65.2\% and 56.8\% respectively. Given the fine-grained and often ambiguous nature of multimodal misinformation, these results indicate that the taxonomy can be applied consistently in practice while still capturing meaningful distinctions between mechanisms.

\section{VLM Annotations}
\label{sec:experiments}
\subsection{Modelling}
\label{sec:modelling}

To analyse multimodal misinformation on contemporary and emerging issues, we automate this process using VLMs to apply our taxonomy to large datasets quickly and at low cost. We opted for this  over relying on experts or crowd annotators as experts are slow and resource-intensive, which is especially problematic for domains where qualified annotators are difficult to find, emerging events where speed is of the essence, and underrepresented languages and regions. Crowd annotation is challenging as labeling instances often requires fluency in a specific language, familiarity with political and cultural contexts, and domain-specific knowledge in topics such as economics or health. Moreover, several taxonomy categories require careful deliberation, difficult to sustain in a crowdsourcing setting.

We selected \texttt{Qwen3.5 27B} \citep{qwen3.5} for this task,\footnote{The model is available at \url{https://huggingface.co/Qwen/Qwen3.5-27B-FP8}.} based on its strong performance for its size on preliminary experiments, enabling efficient deployment at scale. Moreover, the model supports over 200 languages, ensuring it can effectively handle misinformation from different countries. Lastly, the VLM was released well after our dataset's cutoff date, its training data plausibly covers contemporary topics and events necessary for understanding the posts and notes.\footnote{Comparing this setup to a retrieval-augmented (RAG) approach \citep{lewis2021retrievalaugmentedgenerationknowledgeintensivenlp} with internet access is a natural extension, which we leave for future work.}

\subsection{Annotation Pipeline}
\label{sec:pipeline}

\begin{figure}[ht]
    \centering
    \includegraphics[width=0.9\linewidth]{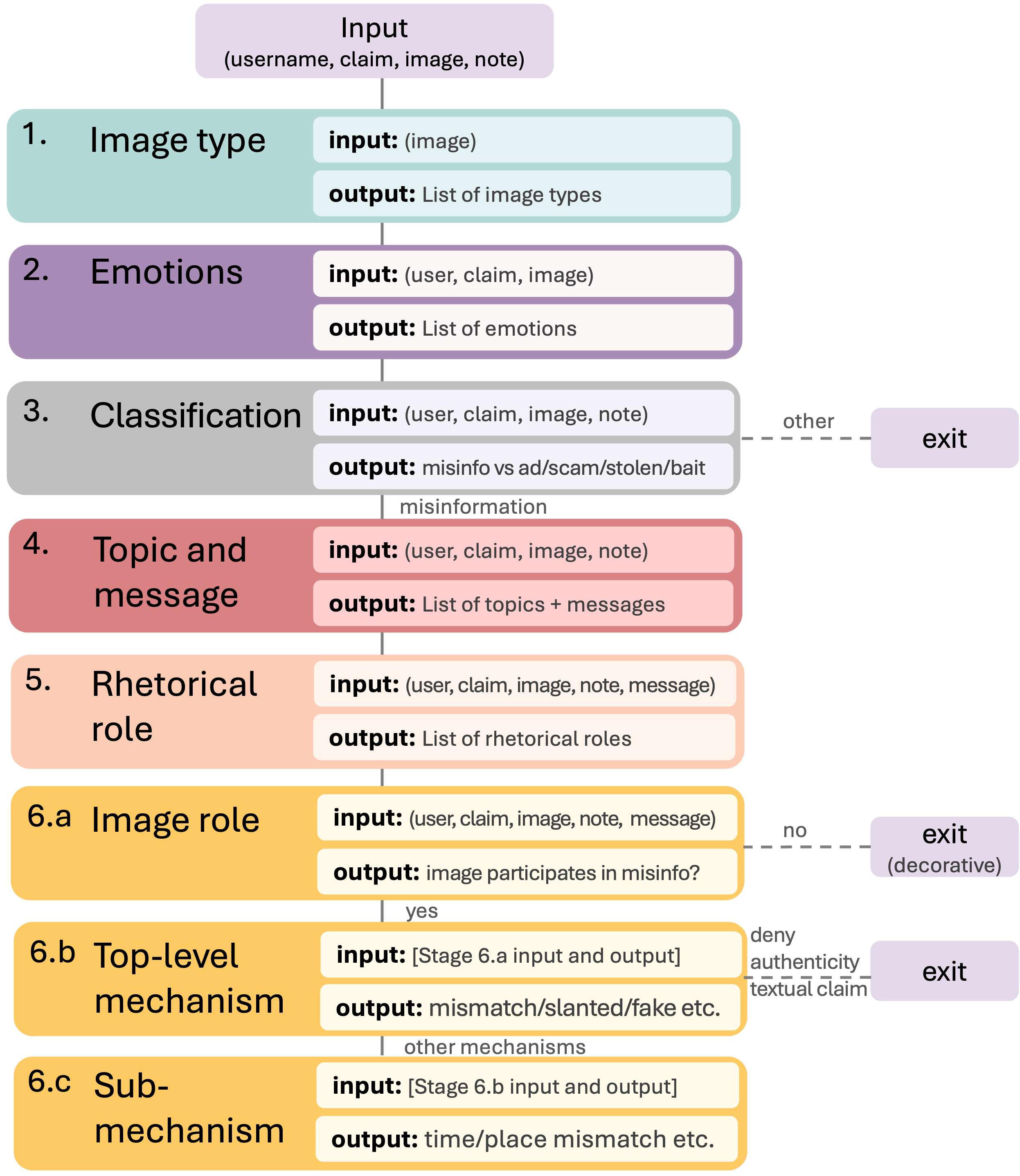}
    \caption{The annotation pipeline.}
    \label{fig:vlm_pipeline}
\end{figure}

We model the annotation process as a multi-stage pipeline, where each step addresses a specific axis of the taxonomy, as detailed below (see \cref{fig:vlm_pipeline} for an overview).

\noindent \textbf{1. Image type.} We predict the type(s) of the image, taking only the image as input. \Cref{lst:image_type_prompt} in \cref{app:prompts} details the prompt used.\footnote{As the two datasets differ in format, we use dataset-specific prompt variants for Community Notes and \textsc{AMMeBa}. The latter are provided in our GitHub: \url{https://github.com/nadavborenstein/MultimodalTaxonomy}.}

\noindent \textbf{2. Emotion.} We classify the emotion(s) the post author is attempting to evoke using the image. The input a is \texttt{(username, claim, image)} triplet. The community note is deliberately excluded as the emotions should be inferred from the post itself. See \cref{lst:emotions_prompt} for the prompt.

\noindent \textbf{3. Classification.} We determine whether the post constitutes misinformation or falls into another category (advertisement, scam, or stolen content), using \texttt{(username, claim,} \texttt{image, note)} as input. Only instances of misinformation proceed to the next steps. See \cref{lst:classification_prompt} for the prompt.

\noindent \textbf{4. Topic and message.} We extract the topic(s) and the main message of the post. The prompt (\cref{lst:topic_prompt}) also extracts abstracted versions of the message, which we further analyse in \cref{sec:analysis}.

\noindent \textbf{5. Rhetorical role.} We classify the rhetorical role of the image relative to the textual claim. Following \citet{rhetorical_role}, this is modelled as a multi-label classification task. The input is \texttt{(username, claim, image, note)} and the message extracted in step 4. See \cref{lst:rhetoric_role_prompt}.

\noindent \textbf{6. Mechanism.} Finally, we identify the multimodal misinformation mechanism category and, where applicable, its sub-category. This stage is modelled as a multi-turn conversation. First, the VLM determines whether the image actively participates in the misleading mechanism. Instances in which the image plays no deceptive role are assigned the `\textit{Decorative}' mechanism. In the second turn, the VLM predicts the top-level mechanism for all remaining instances. Finally, in the third turn, the VLM predicts a sub-mechanism for instances that were assigned the mechanism `\textit{Mismatch}', `\textit{Slanted}', `\textit{Fake image}', `\textit{Manipulated image}', or `\textit{Unreliable source}'.  Prompts are provided in \cref{lst:mechanism_stage_1_prompt}, \ref{lst:mechanism_stage_2_prompt}, and  \ref{lst:mechanism_stage_3_prompt}.

To assess prediction stability, we ran the annotation pipeline three times in total, each run taking approximately 24 hours on a single L40S GPU, using the default model and sampling parameters suggested by Qwen's creators for non-thinking mode.

\subsection{Evaluation}
\label{sec:evaluation}

We compare VLM predictions against human annotations, running the pipeline described in \cref{sec:pipeline} three times with different random seeds. \cref{tab:agreement-all} in \cref{app:evaluation} (middle) reports the mean inter-run agreement on Community Notes. The pattern matches the human IAA from \cref{sec:taxonomy_eval}: the more subjective, ambiguous axes yield lower agreement, while others are generally high. One exception is the \textit{Sub-mechanism} axis, which achieves higher agreement than the \textit{Mechanism} axis, reversing the human–human trend. \cref{tab:agreement-all} (bottom) reports VLM--human agreement, averaged over all (VLM run, human annotator) pairs, and shows the same pattern. Notably, VLM--human agreement is comparable to (albeit lower than) human--human agreement, suggesting that the model applies the taxonomy with a reliability approaching that of our expert annotators.

We further assess the model's predictions, focusing on the \textit{Mechanism} axis. We select one pipeline run at random and use stratified sampling to draw up to 20 English instances per predicted label, for a total of 630 samples. We then replace half of predicted labels with randomly drawn labels, which serve as distractors. Each data instance and label (either the model's prediction or a distractor) was annotated by two expert annotators (authors), who judged whether the label was correct.

Inter-annotator agreement is good: Cohen's $\kappa$ of 0.67 with 82.3\% raw agreement across all 630 samples. Expectedly, the agreement on non-distractor samples is lower -- 0.44 with 80.3\%, as distractor labels are easier to reject. The two annotators then resolved disagreements through discussion, after which we computed model accuracy on the non-distractor samples. 79.3\% of the predictions are judged correct, though accuracy varies considerably across labels. For example, the sub-mechanisms \textit{Manipulated image -- removal} and \textit{Unreliable source -- satire} exceeded 90\%, whereas the sub-mechanisms \textit{Slanted -- misrepresentation of relevance} and \textit{Fake image -- staged} fall below 60\%, indicating a clear avenue for future improvement. Full results are in \cref{tab:full_model_eval} in \cref{app:evaluation}.

Next, we evaluate how well our taxonomy and pipeline apply to a similar (but narrower) dataset comprised of multimodal misinformation verified by fact-checkers (\textsc{AMMeBa}). As \textsc{AMMeBa} already includes human annotations, we compare the VLM to these existing annotations, where the categories align with our taxonomy categories. \cref{tab:ammeba-agreement} (\cref{app:evaluation}) reports agreement on the seven aligned categories. F1 scores range from 0.61 to 0.83, with one exception -- \textit{Time mismatch}, at 0.26. This gap may be an artefact of our prompt design rather than a model failure: to improve consistency, we instruct the VLM to select \textit{Place mismatch} whenever both time and place mismatches apply (\cref{lst:mechanism_stage_3_prompt}). \textsc{AMMeBa} instead just selects both labels, which may account for the discrepancy. We then apply our entire taxonomy to \textsc{AMMeBa}, enabling us to observe commonalities and differences in multimodal misinformation identified by fact-checkers and Community Notes.

\section{Analysis}
\label{sec:analysis}
\subsection{General Results}

\begin{figure}[t]
    \centering
    \includegraphics[width=.9\linewidth,trim={0 0 0 1cm},clip]{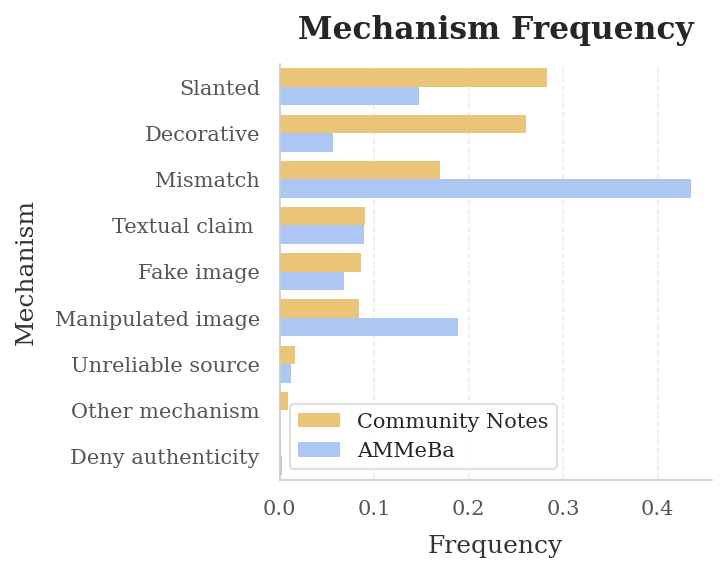}
    \caption{Frequency of the top-level mechanisms across the two datasets.}
    \label{fig:mechanism_freq}
\end{figure}

\begin{figure}[t]
    \centering
    \includegraphics[width=\linewidth,trim={0 0 0 1.65cm},clip]{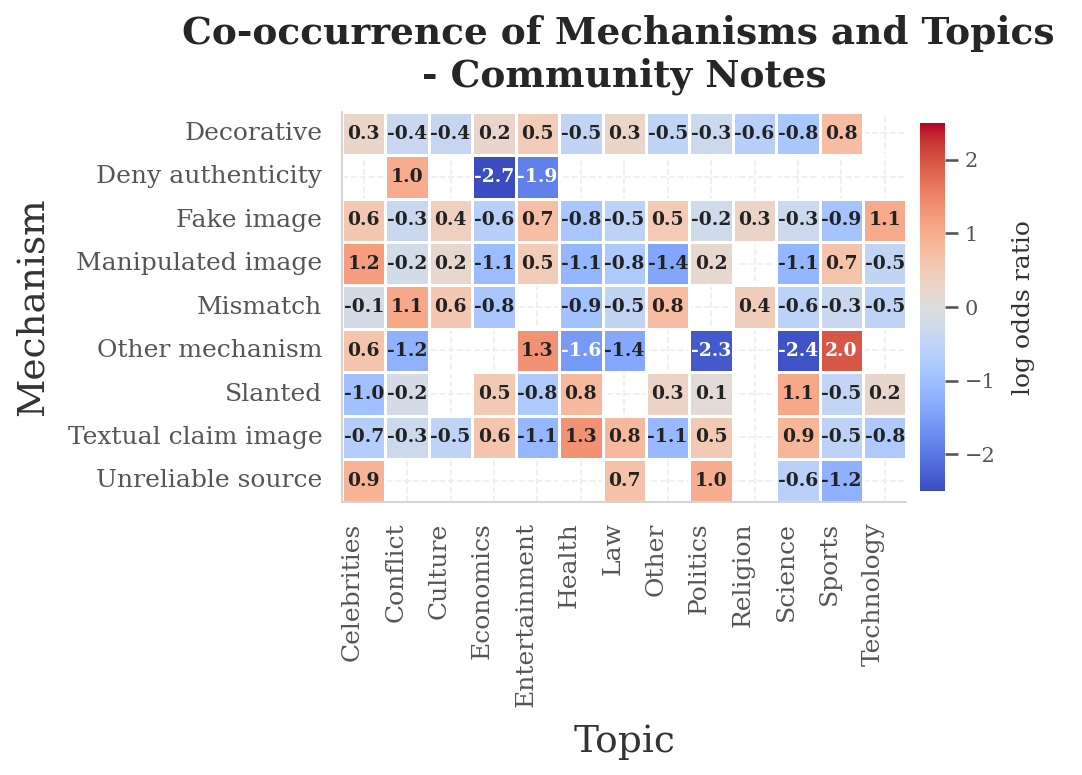}
    \caption{Co-occurrences of mechanisms vs topics on the Community Notes dataset. Values near zero indicate independence, positive values over-representation, and negative values under-representation.}
    \label{fig:co_oc_mechanism_cn}
\end{figure}

\begin{figure*}[ht]
    \centering
    \includegraphics[width=\linewidth]{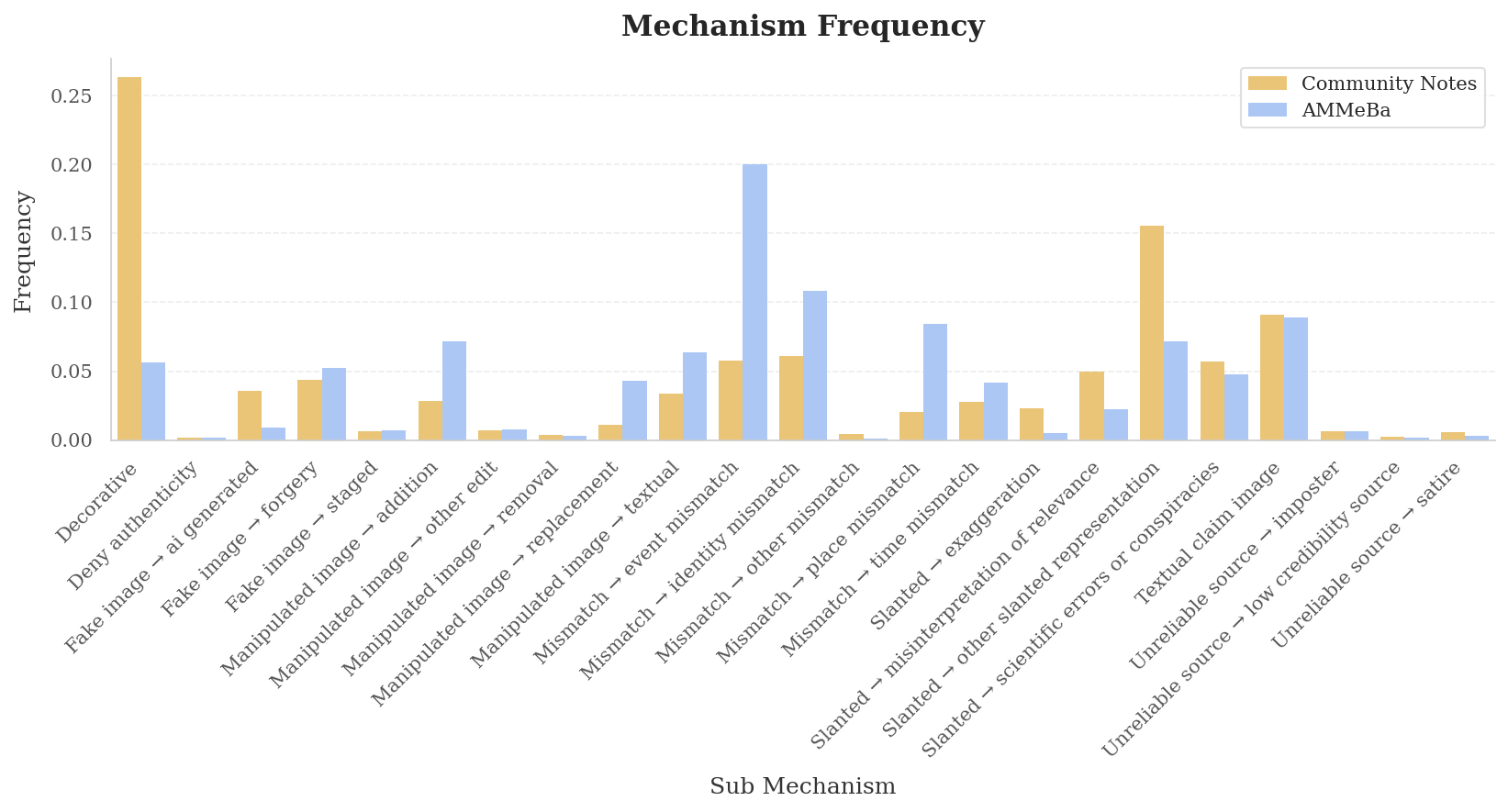}
    \caption{Frequency of the sub-mechanisms across the two datasets.}
    \label{fig:sub_mechanism_freq}
\end{figure*}

\begin{figure}[t]
    \centering
    \includegraphics[width=\linewidth]{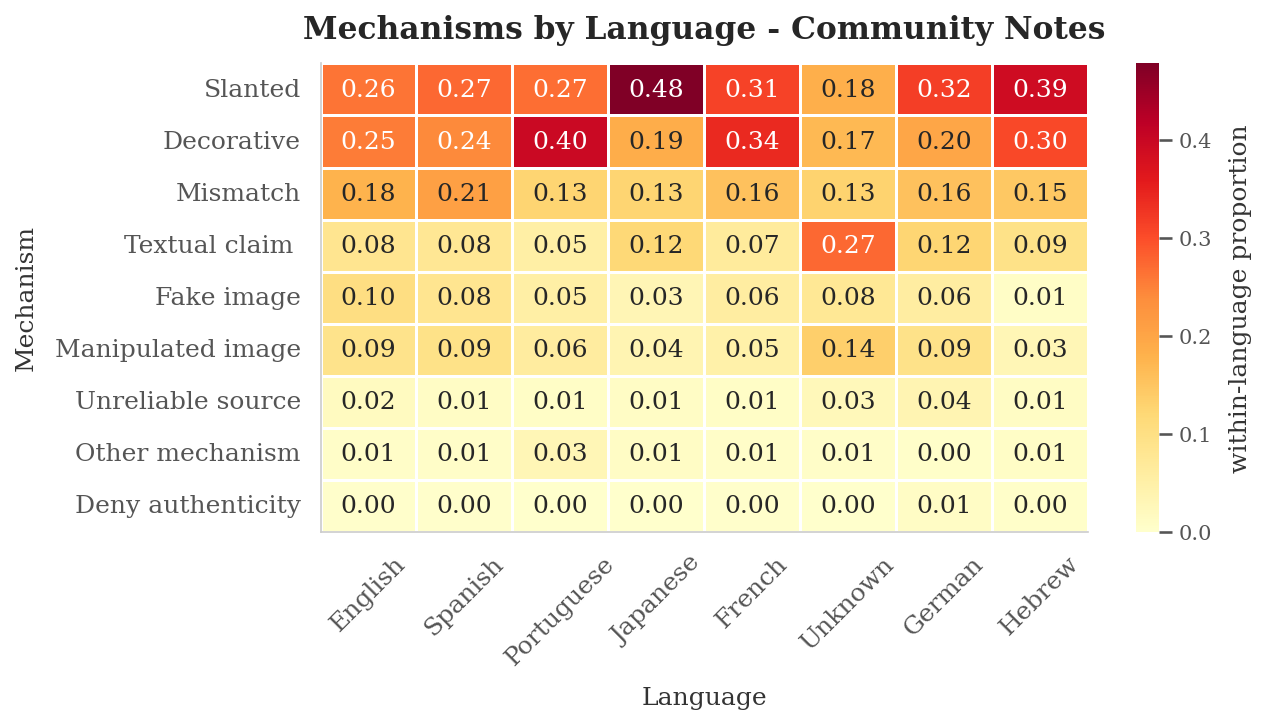}
    \caption{Co-occurrences of mechanisms vs language on the Community Notes dataset.}
    \label{fig:co_oc_mechanism_lang}
\end{figure}

\noindent \textbf{What is the distribution of deceptive mechanisms in Community Notes?} We first examine the distribution of mechanism categories predicted by the VLM on Community Notes (\cref{fig:mechanism_freq}). \textit{Slanted} is the most common category, with over 7,000 samples, followed by \textit{Decorative} and \textit{Mismatch}. Interestingly, \textit{Fake image} occurs as frequently as \textit{Manipulated image}, whereas \textit{Unreliable source}, \textit{Other mechanism}, and \textit{Deny authenticity} are rare, with less than 800 samples combined.

\cref{fig:sub_mechanism_freq} provides a finer-grained view of sub-mechanisms. Notably, \textit{AI-generated} and \textit{forged} images (e.g., fake documents) occur at comparable rates, and both exceed any form of Image manipulation with the exception of \textit{textual manipulation}. Within \textit{Context mismatch}, \textit{identity mismatch} and \textit{event mismatch} are frequent, while other mismatch types are less common. This analysis offers concrete guidance for where mitigation efforts may yield the largest practical impact.

The multilingual nature of the Community Notes dataset allows us to compare misinformation trends across languages. Two patterns are particularly notable (\cref{fig:co_oc_mechanism_lang}). First, the \textit{Decorative} mechanism is overrepresented in Portuguese. \cref{fig:topic_lang} (\cref{app:analysis_extra_figures}) provides a preliminary explanation for this: Portuguese-language misinformation tends to concentrate on \textit{Sports}, \textit{Celebrities}, and \textit{Entertainment}, which disproportionally use the \textit{Decorative} mechanism (\cref{fig:co_oc_mechanism_cn}). Second, Japanese-language misinformation disproportionately utilises the \textit{Slanted} mechanism. The sub-category heatmap (\cref{fig:co_oc_sub_mechanism_lang} in \cref{app:analysis_extra_figures}) attributes this to the \textit{scientific errors and conspiracies} mechanism. Manual inspection of samples reveals that a large portion of these posts promote pseudoscientific claims about earthquake prediction, a topic that is particularly relevant in Japan, where earthquakes are frequent.

\noindent \textbf{How does the mechanism distribution differ between Community Notes and \textsc{AMMeBa}?} The two distributions differ substantially  (\cref{fig:mechanism_freq,fig:sub_mechanism_freq}). \textit{Mismatch} is the dominant mechanism in \textsc{AMMeBa}, largely driven by the prevalence of the \textit{Event} and \textit{Identity} subcategories. In contrast, \textit{Decorative} is considerably less common in \textsc{AMMeBa}, reflecting our sampling strategy, which includes only instances where the image played an active role in the misinformation. Likewise, \textit{AI-generated} images are rarer than in Community Notes, likely due to \textsc{AMMeBa}'s earlier collection period predating the recent surge in AI image generation.

\noindent \textbf{How do deceptive mechanisms vary across topics?} Our taxonomy includes five additional axes beyond multimodal mechanisms. 
\cref{fig:co_oc_mechanism_cn} visualises topic–mechanism interactions in Community Notes. For each category pair, we apply Fisher’s exact test with Benjamini–Hochberg FDR correction and omit cells with $p \geq 0.05$.  \textit{Mismatch} (particularly \textit{Place}, \textit{Time}, and \textit{Event mismatch}, as seen in \cref{fig:co_oc_sub_mechanism_cn} in \cref{app:additional_results}) co-occurs strongly with \textit{Conflict}, consistent with the widespread use of out-of-context imagery from conflict preparations and aftermaths (\cref{tab:conflict}). \textit{Textual claim} is over-represented in \textit{Law}, \textit{Economics}, \textit{Health}, and \textit{Science}, where documents often serve as evidence.

Topic–mechanism interactions in \textsc{AMMeBa} (\cref{fig:co_oc_mechanism_ammeba} in \cref{app:analysis_extra_figures}) partially overlap with those observed in Community Notes. For example, the over-representation of \textit{Mismatch–Conflict} and the under-representation of \textit{Celebrities} and \textit{Textual claim} persist across both datasets. A main source of divergence is the \textit{Fake image} category, with \textsc{AMMeBa} reporting low correlations with almost all topics.

\subsection{Case Studies}

While valuable, Topic–Mechanism co-occurrence analyses are too coarse-grained to capture interactions related to specific misinformation narratives. The natural-language \textit{Message} annotation, however, enables us to identify misinformation linked to particular narratives of interest, enabling fine-grained analysis of Narrative-Mechanism interaction. We present two case studies demonstrating this approach. See \cref{app:case_studies_reproducability} for the procedure and \cref{app:analysis_extra_figures} for supporting figures and tables.

\noindent\textbf{Case study 1: Athletic rivalry.} As shown in \cref{fig:rivalry}, when the image plays a an active role in the misinformation, narratives about athletic rivalries disproportionately involve text-based mechanisms (\textit{Manipulated image: textual} and \textit{Textual claim in image}). Inspection of representative examples reveals that this pattern is driven by images containing fake statistics or fabricated performance metrics.

\noindent\textbf{Case study 2: Vaccination.} Unsurprisingly, misinformation related to vaccination risks is dominated by the \textit{scientific errors and conspiracies} mechanism and text-based claims (\cref{fig:vaccination}). The taxonomy, however, enables other forms of investigation. \cref{fig:vaccination_image_type} reveals that \textit{News screenshot} image types are particularly over-represented. This suggests that appropriating journalistic authority is a common deception strategy within this narrative. See examples in \cref{tab:vaccine_examples}.

\noindent\textbf{Case study 3: Anti-immigration narratives.} We examine anti-immigration misinformation, which includes narratives related to claims that immigrants are violent, a drain on financial resources, and Great-replacement conspiracies.\footnote{\url{https://en.wikipedia.org/wiki/Great_Replacement_conspiracy_theory}} \cref{fig:immigration} reports the distribution of mechanisms, showing an over-representation of \textit{Textual claim in image}, \textit{Slanted representation}, and \textit{Mismatch: time/place}. We also examine emotions associated with these narratives; \cref{fig:immigration_emotion} shows a clear reliance on negative emotions such as \textit{Anger} and \textit{Fear}.

\section{Conclusion}
\label{sec:conclusion}
We introduce a unified taxonomy of multimodal misinformation mechanisms, consolidating prior frameworks and extending them with novel categories grounded in real-world data. 
We operationalise the taxonomy through an automated VLM annotation pipeline, and validate it across two datasets: Community Notes and \textsc{AMMeBa}. 
Applying our taxonomy at scale reveals discernible variation in multimodal deceptive mechanisms, such as that AI-generated images are especially prevalent in technology and science, while vaccination-related misinformation disproportionately asserts credibility using screenshots from news organisations.
Together, our findings provide a foundation for more targeted analysis, detection, and mitigation of multimodal misinformation.

\section*{Limitations}
\label{sec:limitations}

We identify the following limitations of our work.

First, as noted in \cref{sec:taxonomy_development}, several taxonomy axes proposed in prior work were excluded from our unified taxonomy. This was done because they were either too ambiguous to annotate reliably, required access to external sources, or fell outside the scope of this work. Incorporating them is an open opportunity.

Second, although sizable, our Community Notes dataset is not large enough to support robust analysis of every narrative, and many narratives are represented by too few samples. In the same vein, the \textsc{AMMeBa} dataset we use is only a subsample. Expanding both, e.g., by improving the data collection scheme described in \cref{app:ammeba_collection}, would support fine-grained analysis of further narratives.

Third, our model selection process was limited by computational constraints and our desire to avoid dependence on proprietary systems. \texttt{Qwen3.5 27B}, while highly capable, does not match the performance of state-of-the-art VLMs. A larger model, or one fine-tuned for this task, would likely yield better predictions.

Fourth, we process multilinguality naively, relying on \texttt{Qwen3.5}'s built-in multilingual support rather than developing methods to handle it explicitly. Approaches such as translation pipelines or language-specific VLMs may perform better.

Fifth, our taxonomy treats the Multimodal Mechanism axis as a single label classification. In practice, some misinformation instances exhibit more than one mechanism type, e.g., be misattributed with respect to both time and place. Extending the taxonomy to allow multi-label assignment is a natural direction for future work.

Finally, manual inspection of model predictions revealed that some instances in both Community Notes and \textsc{AMMeBa} are of low quality. A better filtering mechanism, independent of the datasets' own quality-related annotations, would help ensure more robust results.



\section*{Acknowledgements}
$\begin{array}{l}\includegraphics[width=1cm]{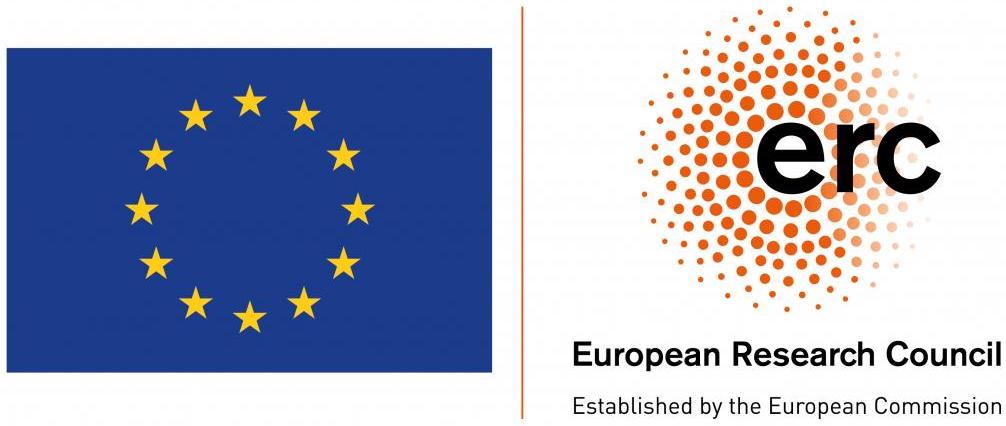} \end{array}$  This research was co-funded by the European Union (ERC, ExplainYourself, 101077481), by the European Union’s Horizon 2020 research and innovation program under grant agreement No.
101135671 (TrustLLM), and by the Pioneer Centre for AI, DNRF grant number P1.

\bibliography{anthology,custom}

\clearpage

\appendix
\section{Overview of Existing Taxonomies}
\label{app:taxonomies}

\begin{table*}[t]
\centering
\resizebox{\textwidth}{!}{%
\fontsize{8}{11}\selectfont
\begin{tabular}{@{} p{3cm}  p{2.5cm} c p{8.5cm} @{}}
\toprule
\textbf{Paper} & \textbf{Axes}  &  \textbf{\# Mechanisms} & \textbf{Mechanism Categories (sub-categories)}  \\
\midrule


\citet{rhetorical_role}  & Rhetorical Role & N/A
&  N/A \\

\citet{wardle2017}
  & Source, Recepient, Message, Intent, MM Mechanism  & 10
  & Imposter (None,  Brand, Individual), Content (satire/parody, false connection, misleading content, false context, imposter content, manipulated content, fabricated content) \\

\citet{fauxtography}
   & MM Mechanism  & 7
  &  manipulated, manufactured, recontextualised, timeshifted, extracted, mirrored, denied \\

\citet{ayton2019explainingDeceptive} & Intent, Classification, Message  & N/A
& N/A \\

\citet{nakamura-etal-2020-fakeddit}
 & MM Mechanism  & 6 
  & True, Satire/Parody, Misleading Content, Imposter Content,
    False Connection, Manipulated Content \\

\citet{Hameleers03032020}
  & MM Mechanism  & 4 
  & Decontextualisation ,
    Reframing,
    Visual doctoring,
    Multimodal doctoring \\

\citet{brennen2020types} & Metadata, Source, MM Mechanism  & 7 
& Satire/parody, False connections, Misleading content, False context, Imposter content, Fabricated content, Manipulated content \\

\citet{Weikmann2022VisualDI} & Message, Source, Emotions, MM mechanism  & 8
& Low sophistication (elimination, Cropping, decontextualisation, shallow fakes), High sophistication (PS, misleading data visualisation, deepfakes) \\


\citet{shao2023detectinggroundingmultimodalmedia}
  & MM Mechanism  & 4
  & Face Swap, Face Attribute, Text Swap, Text Attribute \\

\citet{Akhtar2023MultimodalAF}
 & MM Mechanism  & 3 
  & Manipulated content (deep fake, cheap fake), Out-of-context \\[4pt]

\citet{NEWMAN2024101778}
  & MM Mechanism  & 5
  & Probative, Slanted, Repurposed, False, Decorative \\

\citet{liu2024mmfakebench}
   & Textual, MM Mechanism & 12
  & Textual veracity distortion (real text with AI image, real text with repurposed image, artificial text with AI image, GPT-generated text with AI image, GPT-generated text with repurposed image), Visual veracity distortion (PS-edited, AI-generated), Cross-modal consistency distortion (person inconsistency, semantic inconsistency, scene inconsistency, edited text, edited image) \\

\citet{Dufour2024AMMeBaAL}  & Media Type, Image Type, MM Mechanism & 11
  & Content manipulation (text, chyron, AI, general),
    Context manipulation (date, location, identity, circumstance, self-contextualising, atypical)
    Text-based manipulation, fake official documents \\

\citet{Wang2024MFCBenchBM}  & MM Mechanism & 8
& Manipulation (face attribute edit, face swap, background change, AI generated, PS, textual entity replace, text style transfer), Out-of-context \\

\midrule

\textbf{Ours}  & Image type, Emotion, Rhetorical role, Topic, Message, MM mechanism, Classification & 25
& Decorative, Fake image (AI-generated, forgery, staged), Manipulated image (addition, removal, replacement, textual, other edit), Unreliable source (satire, imposter, low credibility source, other provenance issues), Deny authenticity, Mismatch (identity, time, place, event, other mismatch), Slanted (exaggeration, scientific errors or conspiracies, misinterpretation of relevance, slanted representation), Textual claim, Other mechanism \\

\bottomrule
\end{tabular}%
}

\smallskip

\caption{Overview of taxonomies of mechanisms of multimodal misinformation.}
\label{tab:taxonomy_overview}
\end{table*}

\noindent \textbf{\Cref{tab:taxonomy_overview}}: An overview of existing attempts and defining a taxonomy of mechanisms of multimodal misinformation. Our taxonomy is given in the bottom row.

\section{Additional Results}
\label{app:additional_results}


\subsection{Evaluation}
\label{app:evaluation}

\begin{table*}[t]
\centering
\small
\setlength{\tabcolsep}{4pt}
\resizebox{\textwidth}{!}{%
\begin{tabular}{lrrrrrrrrr}
\toprule
 & & Exact & Mean & $\geq$1 & & & Both & One & Skip \\
Step & \# & Match & Jaccard & Match & Mean$|A|$ & Mean$|B|$ & Skip & Skip & Agree \\
\midrule
\multicolumn{10}{l}{\textit{Inter-Annotator (human experts)}} \\
\midrule
Image Type    & 103 & 36.9\% & 0.6074 & 92.2\% & 1.82 & 1.43 & --   & --   & --   \\
Emotion        & 103 & 21.4\% & 0.4002 & 69.9\% & 1.45 & 1.89 & --   & --   & --   \\
Rhetoric Role  & 103 & 12.6\% & 0.3547 & 66.0\% & 1.72 & 1.76 & --   & --   & --   \\
Classification & 103 & 97.1\% & 0.9709 & ---    & 1.00 & 1.00 & --   & --   & --   \\
Topic          & 92  & 41.3\% & 0.7047 & 96.7\% & 1.75 & 2.14 & 9    & 2    & 82\% \\
Mechanism      & 92  & 65.2\% & --- & ---    & 1.00 & 1.00 & 9    & 2    & 82\% \\
Sub-mechanism  & 37  & 56.8\% & --- & ---    & 1.00 & 1.00 & 32   & 16   & 67\% \\
\midrule
\multicolumn{10}{l}{\textit{Inter-Model (3 repetitions)}} \\
\midrule
Image Type     & 5805.3 & 67.1\%  & 0.8021 & 96.5\% & -- & -- & --     & --     & --   \\
Emotion        & 5804.7 & 45.3\%  & 0.6394 & 89.8\% & -- & -- & --     & --     & --   \\
Rhetoric Role  & 5801.7 & 33.2\%  & 0.5720 & 89.1\% & -- & -- & --     & --     & --   \\
Classification & 5804.0 & 100.0\% & 1.0000 & ---    & -- & -- & --     & --     & --   \\
Topic          & 5805.3 & 62.6\%  & 0.8029 & 99.4\% & -- & -- & 44.0   & 0.7    & 99\% \\
Mechanism      & 5804.7 & 71.9\%  & --- & ---    & -- & -- & 44.0   & 1.3    & 97\% \\
Sub-mechanism  & 2852.3 & 86.1\%  & --- & ---    & -- & -- & 1540.7 & 1058.0 & 59\% \\
\midrule
\multicolumn{10}{l}{\textit{VLM--Human}} \\
\midrule
Image Type    & 106.0 & 33.3\% & 0.5653 & 86.3\% & 1.61 & 1.62 & --   & --   & --   \\
Emotion       & 106.0 & 15.6\% & 0.3939 & 69.5\% & 1.68 & 1.76 & --   & --   & --   \\
Rhetoric Role  & 106.0 & 3.3\%  & 0.1758 & 40.9\% & 1.73 & 1.81 & --   & --   & --   \\
Classification & 106.0 & 95.7\% & 0.9571 & ---    & 1.00 & 1.00 & --   & --   & --   \\
Topic          & 96.0  & 34.9\% & 0.6484 & 97.1\% & 1.95 & 2.00 & 0.0  & 10.5 & 0\%  \\
Mechanism      & 96.0  & 49.1\% & --- & ---    & 1.00 & 1.00 & 0.0  & 10.5 & 0\%  \\
Sub-mechanism  & 37.7  & 50.7\% & --- & ---    & 1.00 & 1.00 & 20.2 & 28.8 & 41\% \\
\bottomrule
\end{tabular}%
}
\caption{Agreement across all settings, by annotation step. \textit{$\geq$1 Match} is at least one match. \textit{Mean |A|} and \textit{Mean |B|} are the mean number of labels per annotator. The ``Skip'' columns correspond to the annotation steps where there is a stop condition which one or more of the annotators encounter. \textit{Classification}, \textit{Mechanism}, and \textit{Sub-mechanism} are modelled as a single label prediction task, so the mean Jaccard score and \textit{$\geq$1 Match} are not informative metrics, thus removed from the table.}
\label{tab:agreement-all}
\end{table*}

\begin{table}[t]
\centering
\small
\begin{tabular}{lrr}
\toprule
\textbf{Label} & \textbf{Acc.} & \textbf{$n$} \\
\midrule
\multicolumn{3}{l}{\textit{Mechanism}} \\
\midrule
Other mechanism        & 69\% & 13 \\
Textual claim image   & 70\% & 10 \\
Decorative                 & 84\% &  9 \\
Mismatch                & 86\% &  7 \\
Manipulated image      & 88\% &  8 \\
Slanted                 & 89\% &  9 \\
Deny authenticity      & 89\% &  9 \\
Fake image             & 92\% & 13 \\
Unreliable source      & 100\% & 11 \\
\midrule
\multicolumn{3}{l}{\textit{Sub-mechanism}} \\
\midrule
Fake image - Staged                              & 38\% &  8 \\
Slanted - Misinterpretation of relevance    & 38\% &  8 \\
Manipulated image - Addition                            & 60\% & 15 \\
Slanted - Exaggeration                        & 67\% &  9 \\
Slanted - Other slanted representation      & 70\% & 10 \\
Mismatch - Event mismatch                     & 77\% & 13 \\
Mismatch - Place mismatch                     & 78\% &  9 \\
Fake image - Forgery                             & 82\% & 11 \\
Mismatch - Time mismatch                      & 82\% & 11 \\
Manipulated image - Other edit                         & 82\% & 11 \\
Manipulated image -Textual                             & 82\% & 11 \\
Unreliable source - Imposter                            & 83\% &  6 \\
Mismatch - Identity mismatch                  & 83\% & 12 \\
Fake image - AI generated                       & 86\% &  7 \\
Slanted -  Scientific errors or conspiracies & 89\% &  9 \\
Manipulated image - Replacement                         & 90\% & 10 \\
Unreliable source - Low credibility source            & 90\% & 10 \\
Unreliable source - Other provenance issues           & 100\% &  5 \\
Manipulated image - Removal                             & 100\% &  5 \\
Unreliable source - Satire                              & 100\% &  5 \\
\bottomrule
\end{tabular}
\caption{Model prediction accuracy by mechanism and sub-mechanism label.}
\label{tab:full_model_eval}
\end{table}

\begin{table}[t]
\centering
\resizebox{\linewidth}{!}{%
\fontsize{8}{11}\selectfont
\begin{tabular}{lrrrrr}
\toprule
Category & Acc. & Prec. & Rec. & F1 & Support \\
\midrule
Mismatch                     & 0.78 & 0.98 & 0.72 & 0.83 & 1723 \\
Identity mismatch                & 0.85 & 0.77 & 0.68 & 0.72 & 348  \\
Time mismatch                    & 0.55 & 0.87 & 0.16 & 0.26 & 638  \\
Place mismatch                   & 0.80 & 0.78 & 0.50 & 0.61 & 385  \\
Manipulated image                & 0.93 & 0.86 & 0.83 & 0.84 & 590  \\
Text manipulation                    & 0.76 & 0.97 & 0.50 & 0.66 & 278  \\
AI manipulation                      & 0.99 & 0.78 & 0.86 & 0.82 & 21   \\
\bottomrule
\end{tabular}
}
\caption{Agreement with \textsc{AMMeBa}'s existing human annotation on aligned categories.}
\label{tab:ammeba-agreement}
\end{table}

\noindent \textbf{\Cref{tab:agreement-all}}: Agreement across all settings, by annotation step. \textit{$\geq$1 Match} is at least one match. \textit{Mean |A|} and \textit{Mean |B|} are the mean number of labels per annotator. The ``Skip'' columns correspond to the annotation steps where there is a stop condition which one or more of the annotators encounter. \textit{Classification}, \textit{Mechanism}, and \textit{Sub-mechanism} are modelled as a single label prediction task, so the mean Jaccard score and \textit{$\geq$1 Match} are not informative metrics, thus removed from the table. 

\noindent \textbf{\Cref{tab:full_model_eval}}: Model prediction accuracy by mechanism and sub-mechanism label.

\noindent \textbf{\Cref{tab:ammeba-agreement}}: Agreement with \textsc{AMMeBa}'s existing human annotation on aligned categories.

\subsection{Dataset Statistics}
\label{app:dataset_statistics}

\begin{table}[ht]
\centering
\resizebox{\linewidth}{!}{%
\fontsize{8}{11}\selectfont
    \begin{tabular}{lrr}
    \toprule
    \textbf{Statistic} & \textbf{Value} & \textbf{\%} \\
    \midrule
    \multicolumn{3}{l}{\textit{General}} \\
    \quad Instances & 26,979 &  \\
    \quad Avg.\ post length (words) & 19.4 &  \\
    \quad Avg.\ note length (words) & 28.2 &  \\
    \quad Date range & 2021-01 -- 2026-01 &  \\
    \midrule
    \multicolumn{3}{l}{\textit{Language}} \\
    \quad English & 15,822 & 58.6 \\
    \quad Spanish & 3,312 & 12.3 \\
    \quad Portuguese & 2,254 & 8.4 \\
    \quad Japanese & 2,018 & 7.5 \\
    \quad French & 1,581 & 5.9 \\
    \quad Unknown & 1,023 & 3.8 \\
    \quad German & 712 & 2.6 \\
    \quad Hebrew & 257 & 1.0 \\
    \bottomrule
\end{tabular}
}
\caption{Statistics for the Community Notes dataset.}
\label{tab:cn_statistics}
\end{table}

\begin{figure*}[ht]
    \centering
    \includegraphics[width=\linewidth]{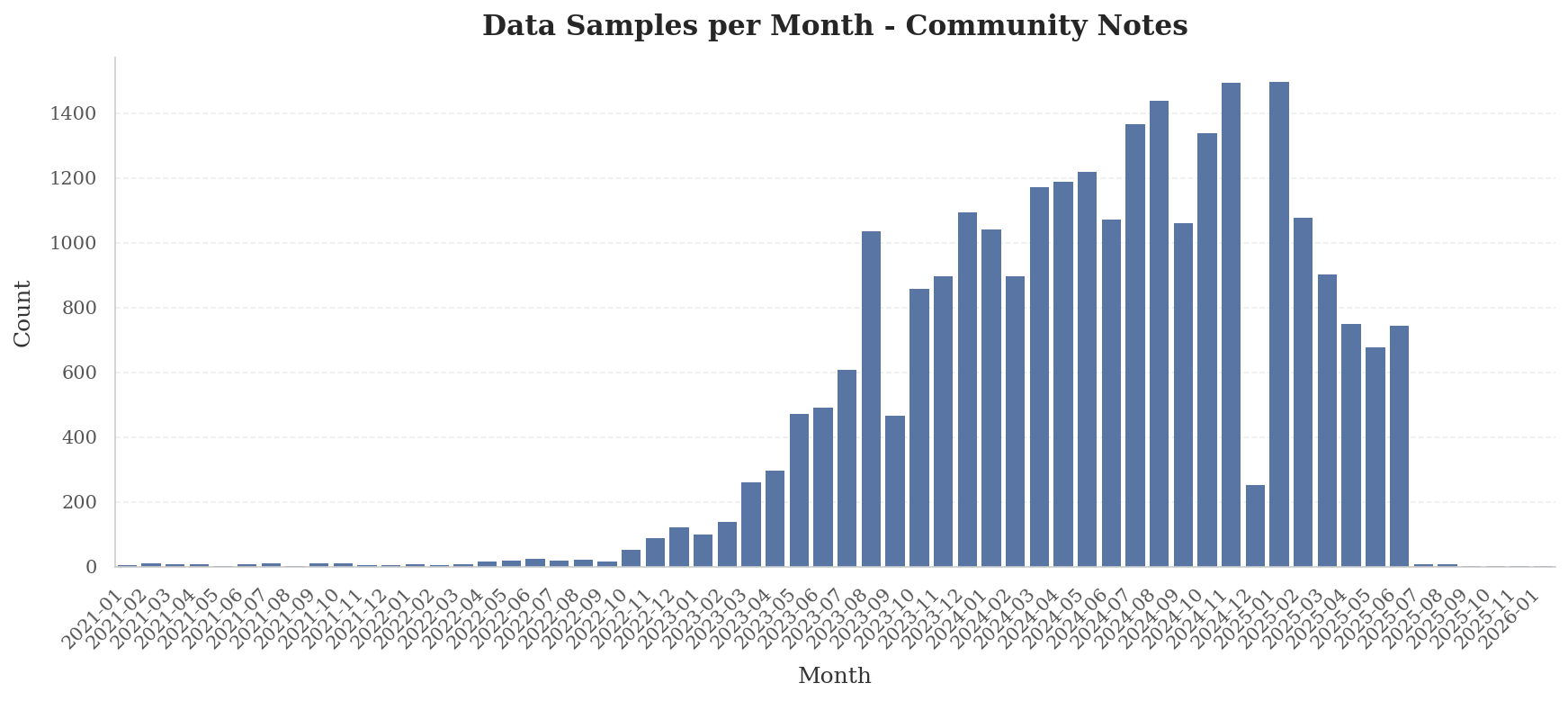}
    \caption{Distribution of dataset samples over time.}
    \label{fig:date_statistics}
\end{figure*}

\begin{table}[ht]
\centering
\centering
\resizebox{\linewidth}{!}{%
\begin{tabular}{lrr}
\toprule
\textbf{Statistic} & \textbf{Value} & \textbf{\%} \\
\midrule
\multicolumn{3}{l}{\textit{General}} \\
\quad Instances & 3,629 &  \\
\quad Avg.\ claim length & 13.2 &  \\
\quad Avg.\ fact-check title length & 11.2 &  \\
\quad Avg.\ article length (non-null) & 573 & \\
\quad Date range & 2022-03 -- 2023-10 &  \\
\midrule
\multicolumn{3}{l}{\textit{Language}} \\
\quad English & 3505 & 96.5 \\
\quad Hindi & 114 & 3.1 \\
\quad Other & 10 & 0.4 \\
\midrule
\multicolumn{3}{l}{\textit{Misinformation source platform}} \\
\quad Facebook & 1,676 & 29.7 \\
\quad Web archive & 1,767 & 31.3 \\
\quad Twitter/X & 1,038 & 18.4 \\
\quad Instagram & 193 & 3.4 \\
\quad Other & 973 & 17.2 \\
\midrule
\multicolumn{3}{l}{\textit{Fact-checking outlet (top 5)}} \\
\quad AFP FactCheck & 1,188 & 21.0 \\
\quad Factly & 737 & 13.0 \\
\quad AltNews & 576 & 10.2 \\
\quad PolitiFact & 511 & 9.0 \\
\quad BOOM Live & 492 & 8.7 \\
\bottomrule
\end{tabular}
}
\caption{Dataset statistics for \textsc{AMMeBa}.}
\label{tab:amoeba_stats}
\end{table}

\noindent \textbf{\Cref{tab:cn_statistics}}: Statistics for the Community Notes dataset.

\noindent \textbf{\Cref{fig:date_statistics}}: Distribution of dataset samples over time for the Community Notes dataset.

\noindent \textbf{\Cref{tab:amoeba_stats}}: Dataset statistics for \textsc{AMMeBa}.

\subsection{Analysis}
\label{app:analysis_extra_figures}

\begin{table*}[ht]
  \centering
  \renewcommand{\arraystretch}{1.8}
  \begin{tabular}{
    >{\raggedright\arraybackslash}p{2.5cm}
    >{\raggedright\arraybackslash}p{3.2cm}
    >{\raggedright\arraybackslash}p{3.5cm}
    >{\raggedright\arraybackslash}p{5cm}
  }
    \toprule
    \textbf{Username} & \textbf{Post} & \textbf{Image} & \textbf{Community Note} \\
    \midrule

      \begin{minipage}[t]{2cm}\vspace{0pt}\raggedright\small
        @jackson
      \end{minipage} &
      \begin{minipage}[t]{3.2cm}\vspace{0pt}\raggedright\small
        Yemen is READY!
      \end{minipage} &
      \begin{minipage}[t]{3.5cm}\vspace{0pt}
        \includegraphics[width=\linewidth, keepaspectratio]{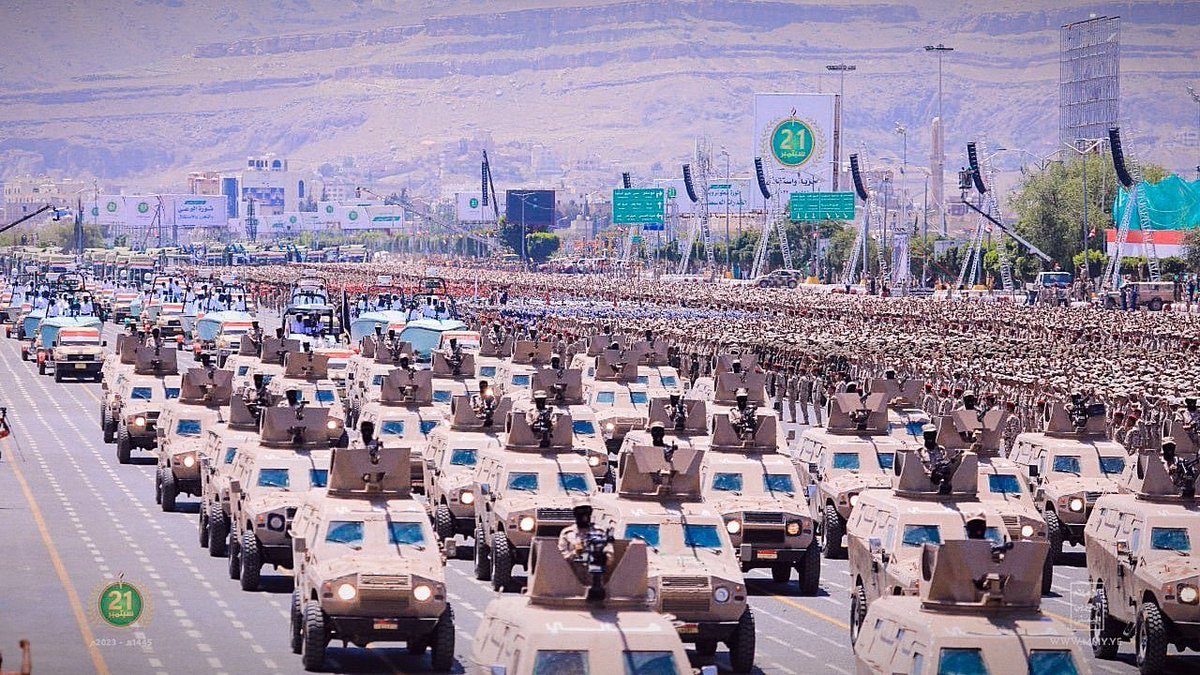}
      \end{minipage} &
      \begin{minipage}[t]{5cm}\vspace{0pt}\raggedright\small
        This image is from a Houthi military parade held in Yemen on September 2023. It is being recirculated in 2025 without context, which may mislead viewers into thinking it is a recent event...
      \end{minipage} \\

      \begin{minipage}[t]{2cm}\vspace{0pt}\raggedright\small
        @sulaimani
      \end{minipage} &
      \begin{minipage}[t]{3.2cm}\vspace{0pt}\raggedright\small
        JUST IN: Half of the stores have closed their doors in Tel Aviv due to Iranian strikes and the markets are empty
        
        Source: Israeli media Maariv 
      \end{minipage} &
      \begin{minipage}[t]{3.5cm}\vspace{0pt}
        \includegraphics[width=\linewidth, keepaspectratio]{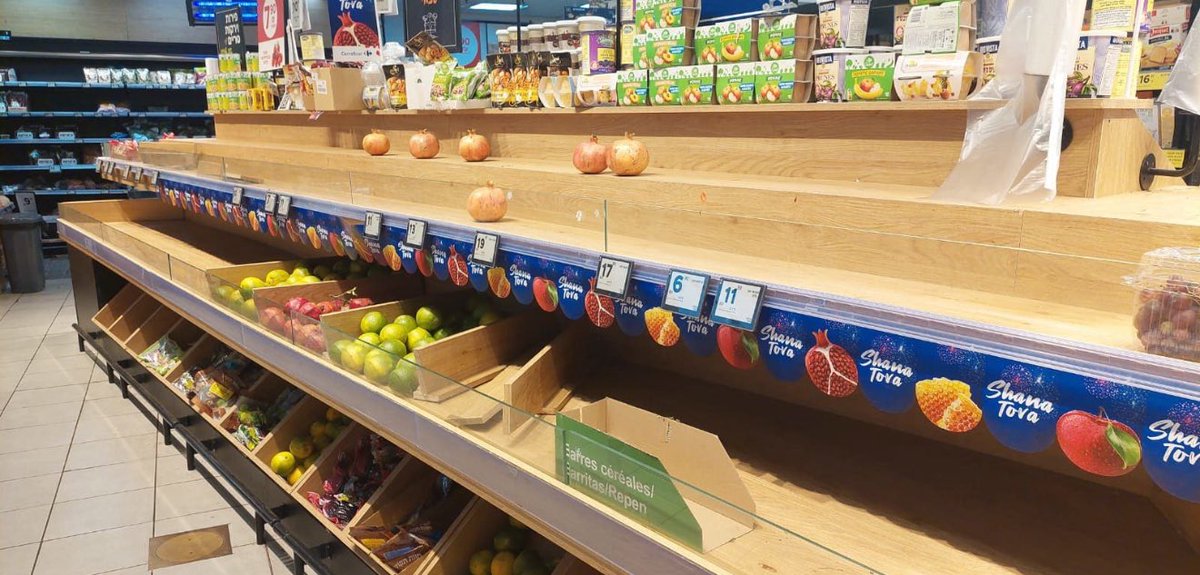}
      \end{minipage} &
      \begin{minipage}[t]{5cm}\vspace{0pt}\raggedright\small
        This image was previously posted in April 2024 and does not depict current events. archive.ph...
      \end{minipage} \\

     \begin{minipage}[t]{2cm}\vspace{0pt}\raggedright\small
        @south\_asia
      \end{minipage} &
      \begin{minipage}[t]{3.2cm}\vspace{0pt}\raggedright\small
       Big Breaking 

      Pakistan claims it has shot down an Indian Rafale jet near Bahawalpur...
      \end{minipage} &
      \begin{minipage}[t]{3.5cm}\vspace{0pt}
        \includegraphics[width=\linewidth, keepaspectratio]{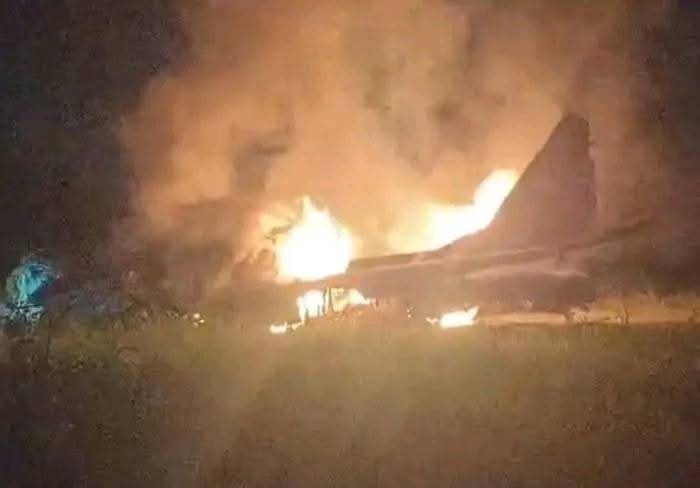}
      \end{minipage} &
      \begin{minipage}[t]{5cm}\vspace{0pt}\raggedright\small
        Misinformation! This is an old image from an earlier incident involving an Indian Air Force (IAF) MiG-29 fighter jet that crashed in Barmer, Rajasthan, in September 2024. x.com/dintentdata/st...
      \end{minipage} \\

    \bottomrule
  \end{tabular}
  \caption{Examples of misinformation instances from the Community Notes dataset classified with the topic \textit{Conflict} and the sub-mechanisms \textit{Place mismatch}, \textit{Time mismatch}, or \textit{Event mismatch}.}
  \label{tab:conflict}
\end{table*}

\begin{table*}[ht]
  \centering
  \renewcommand{\arraystretch}{1.8}
  \begin{tabular}{
    >{\raggedright\arraybackslash}p{2.5cm}
    >{\raggedright\arraybackslash}p{3.2cm}
    >{\raggedright\arraybackslash}p{3.5cm}
    >{\raggedright\arraybackslash}p{5cm}
  }
    \toprule
    \textbf{Username} & \textbf{Post} & \textbf{Image} & \textbf{Community Note} \\
    \midrule

      \begin{minipage}[t]{2cm}\vspace{0pt}\raggedright\small
        @healthbot1
      \end{minipage} &
      \begin{minipage}[t]{3.2cm}\vspace{0pt}\raggedright\small
        Wow.
      \end{minipage} &
      \begin{minipage}[t]{3.5cm}\vspace{0pt}
        \includegraphics[width=\linewidth, keepaspectratio]{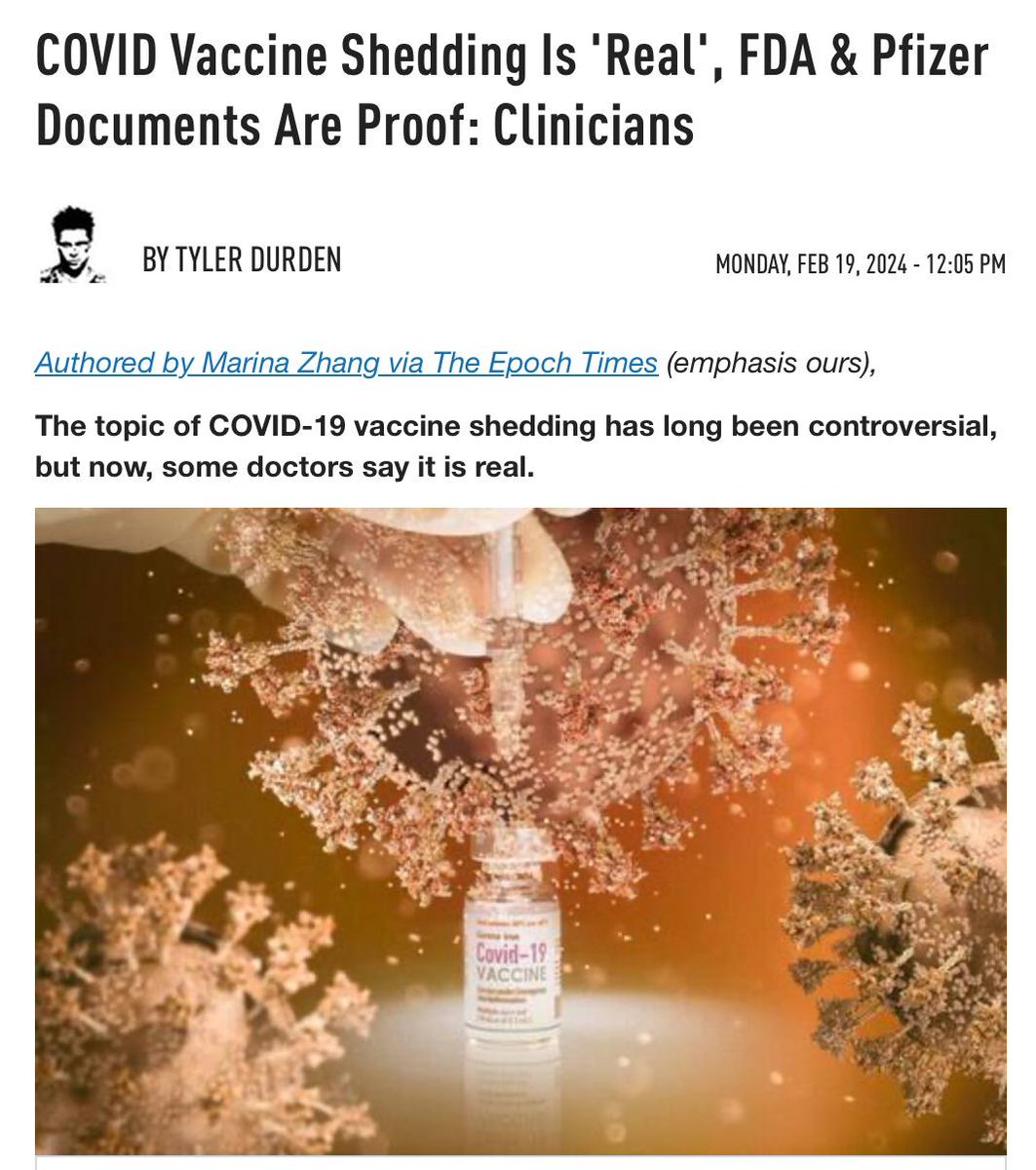}
      \end{minipage} &
      \begin{minipage}[t]{5cm}\vspace{0pt}\raggedright\small
    COVID-19 vaccines being used in the U.S. do NOT contain live virus, so they are incapable of causing shedding:
    chop.edu/parents-pack/p...
      \end{minipage} \\

      \begin{minipage}[t]{2cm}\vspace{0pt}\raggedright\small
        @truthseeker
      \end{minipage} &
      \begin{minipage}[t]{3.2cm}\vspace{0pt}\raggedright\small
        ``Sudden And Unexpected''
        The European Court of Justice has ruled that healthcare professionals who urged or coerced people to take the mRNA vaccines can be criminally prosecuted...
      \end{minipage} &
      \begin{minipage}[t]{3.5cm}\vspace{0pt}
        \includegraphics[width=\linewidth, keepaspectratio]{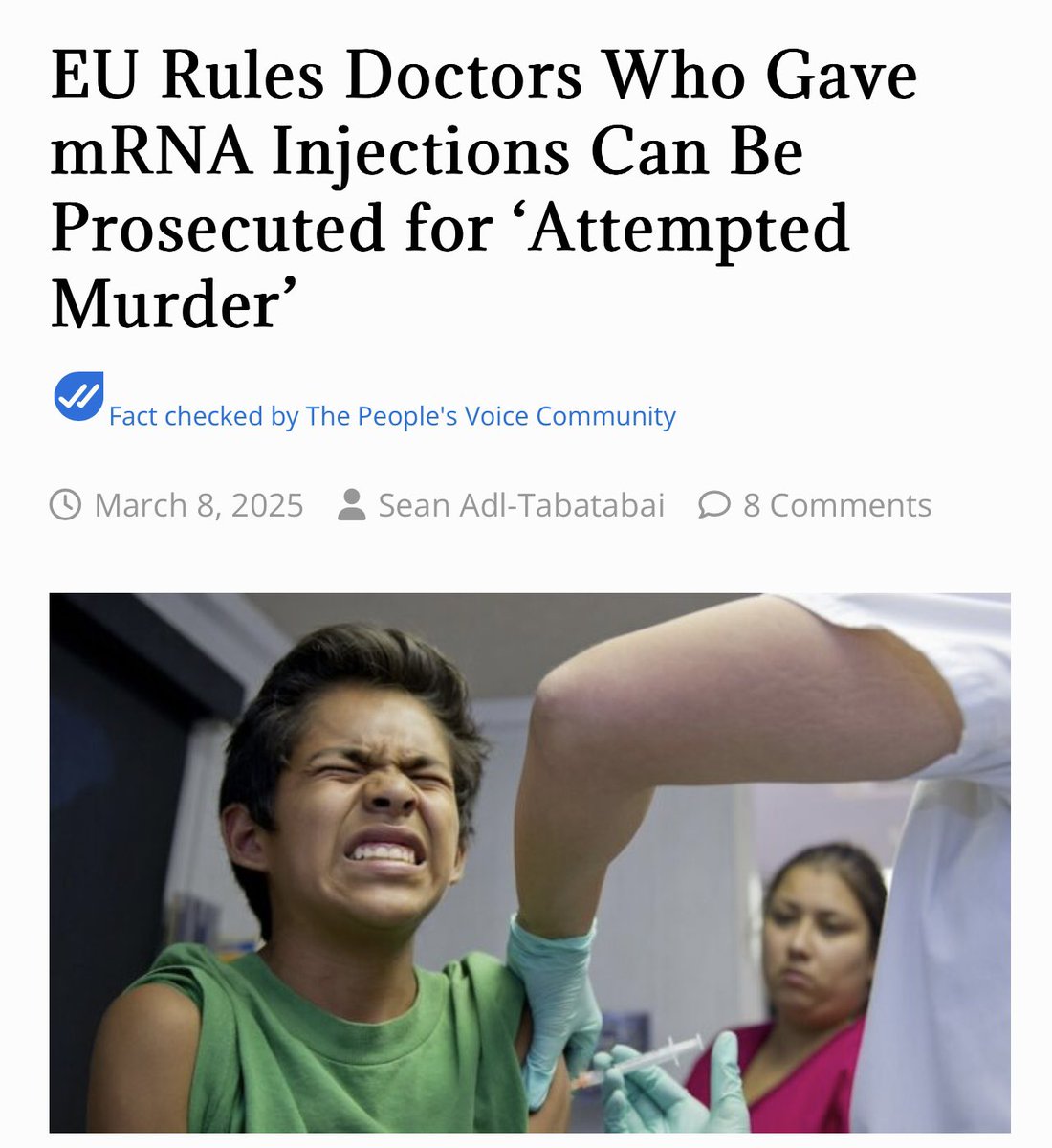}
      \end{minipage} &
      \begin{minipage}[t]{5cm}\vspace{0pt}\raggedright\small
    The judgement (curia.europa.eu/juris/document...) rejects every appeal by Frajese and specifically explains why doctors aren't liable (paragraphs 34 to 39 and 49 to 50)
      \end{minipage} \\

     \begin{minipage}[t]{2cm}\vspace{0pt}\raggedright\small
        @healthbot1
      \end{minipage} &
      \begin{minipage}[t]{3.2cm}\vspace{0pt}\raggedright\small
       ->
      \end{minipage} &
      \begin{minipage}[t]{3.5cm}\vspace{0pt}
        \includegraphics[width=\linewidth, keepaspectratio]{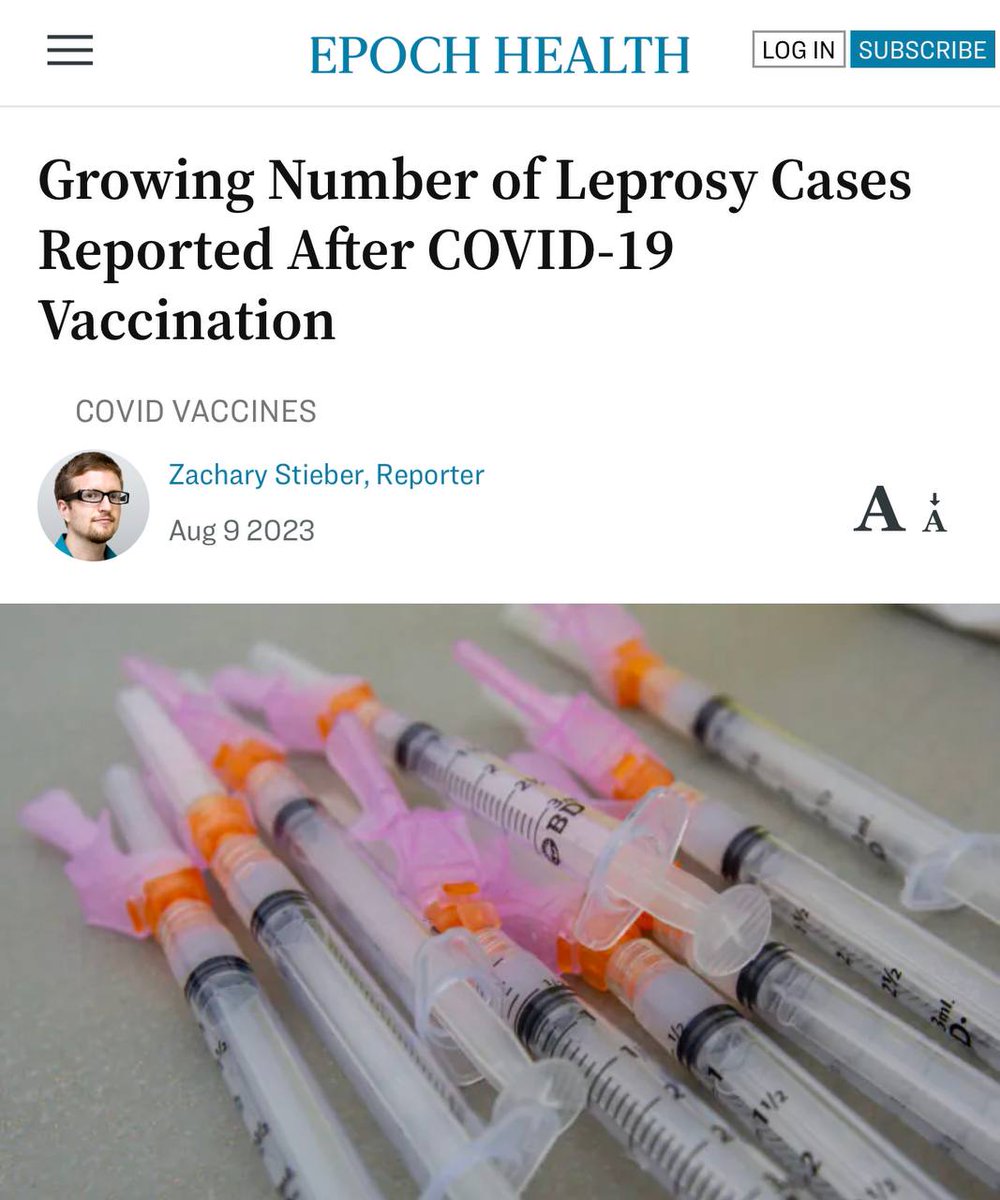}
      \end{minipage} &
      \begin{minipage}[t]{5cm}\vspace{0pt}\raggedright\small
    While there have been reports of leprosy reactions in vaccinated individuals, these are NOT new cases of leprosy caused by the vaccine. Instead, they are immune responses in individuals already carrying the bacteria. This has been debunked several times:
techarp.com/facts/pfizer-v...
      \end{minipage} \\

    \bottomrule
  \end{tabular}
  \caption{Examples of misinformation instances from the Community Notes dataset related to vaccination, with image type \textit{News Screenshot}.}
  \label{tab:vaccine_examples}
\end{table*}

\begin{figure}[ht]
    \centering
    \includegraphics[width=\linewidth]{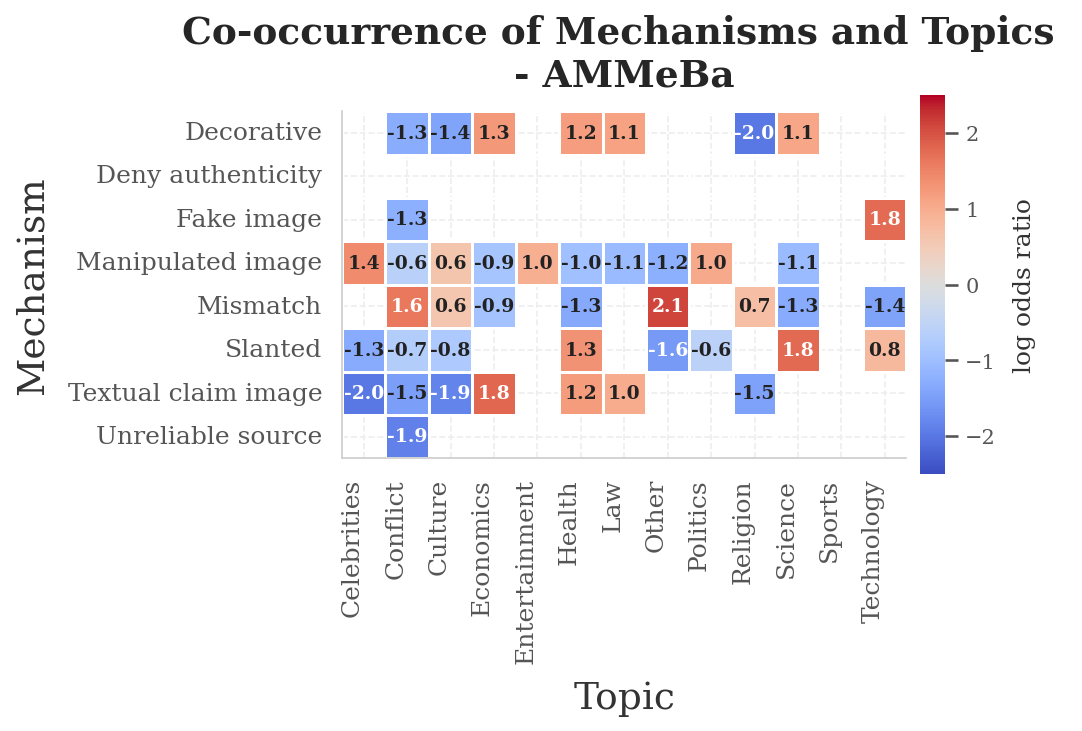}
    \caption{Co-occurrences of mechanisms vs topics on \textsc{AMMeBa}.}
    \label{fig:co_oc_mechanism_ammeba}
\end{figure}

\begin{figure}[ht]
    \centering
    \includegraphics[width=\linewidth]{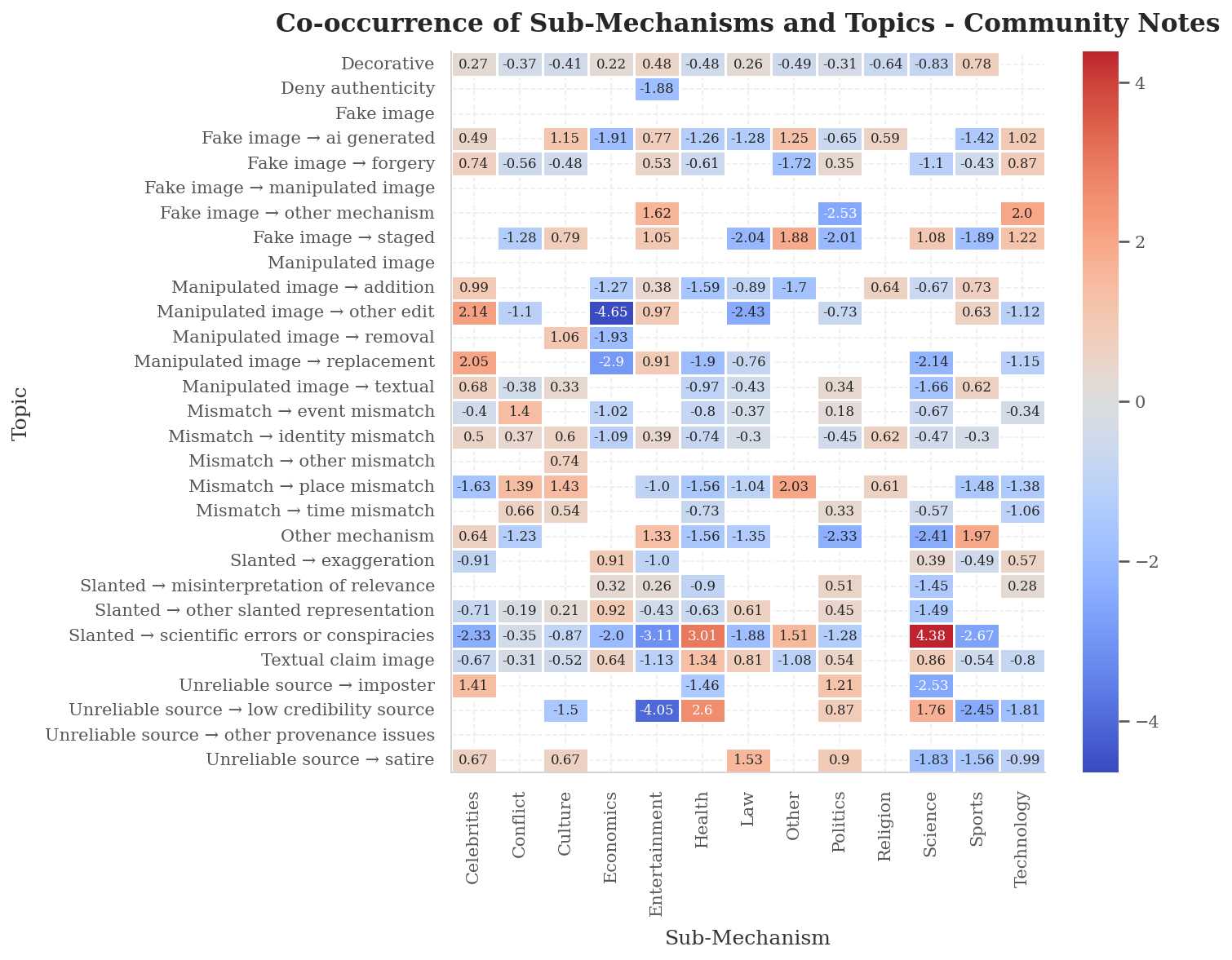}
    \caption{Co-occurrences of sub-mechanisms vs topics on Community Notes.}
    \label{fig:co_oc_sub_mechanism_cn}
\end{figure}

\begin{figure}[ht]
    \centering
    \includegraphics[width=\linewidth]{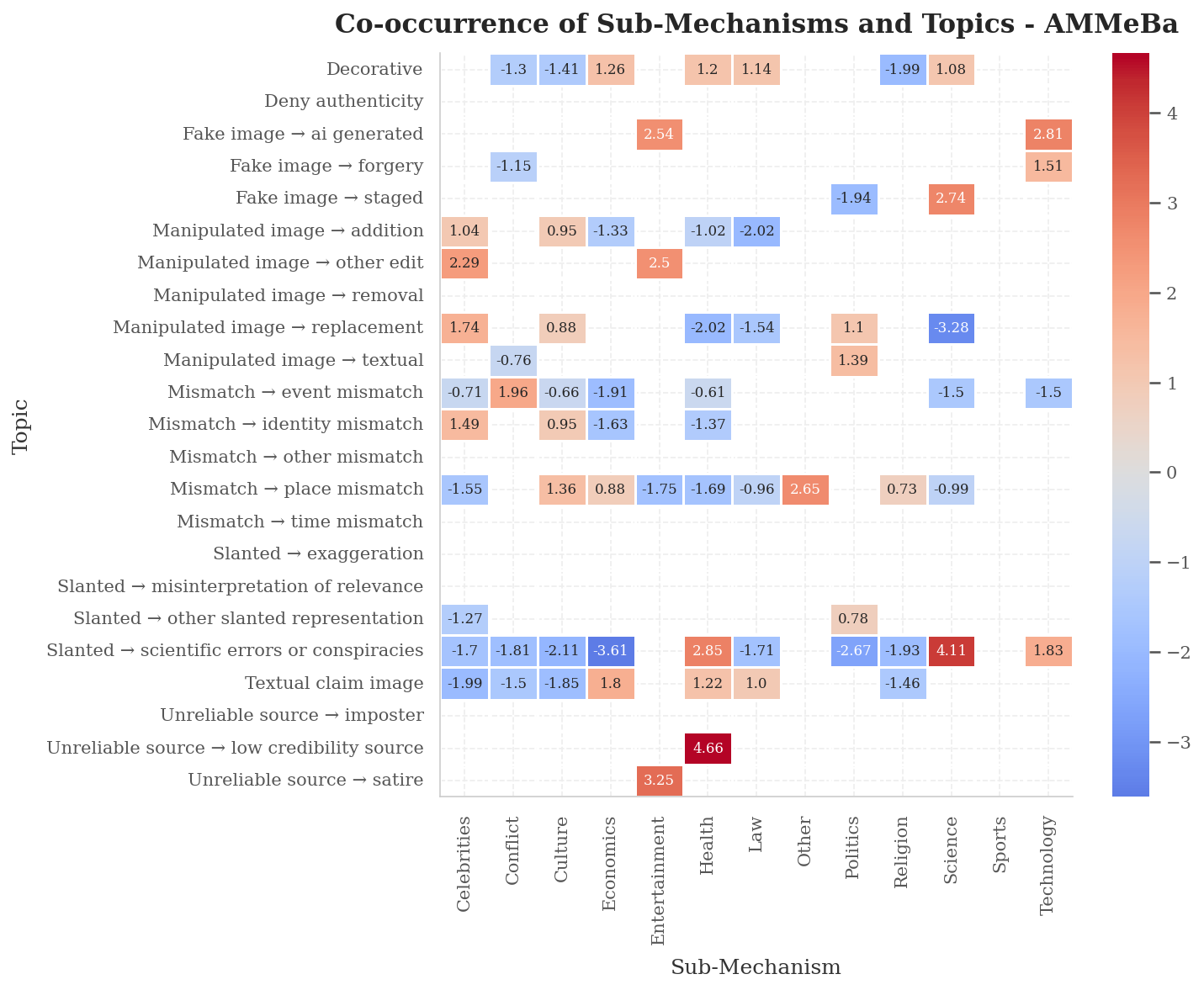}
    \caption{Co-occurrences of sub-mechanisms vs topics on \textsc{AMMeBa}.}
    \label{fig:co_oc_sub_mechanism_ammeba}
\end{figure}

\begin{figure}[ht]
    \centering
    \includegraphics[width=\linewidth]{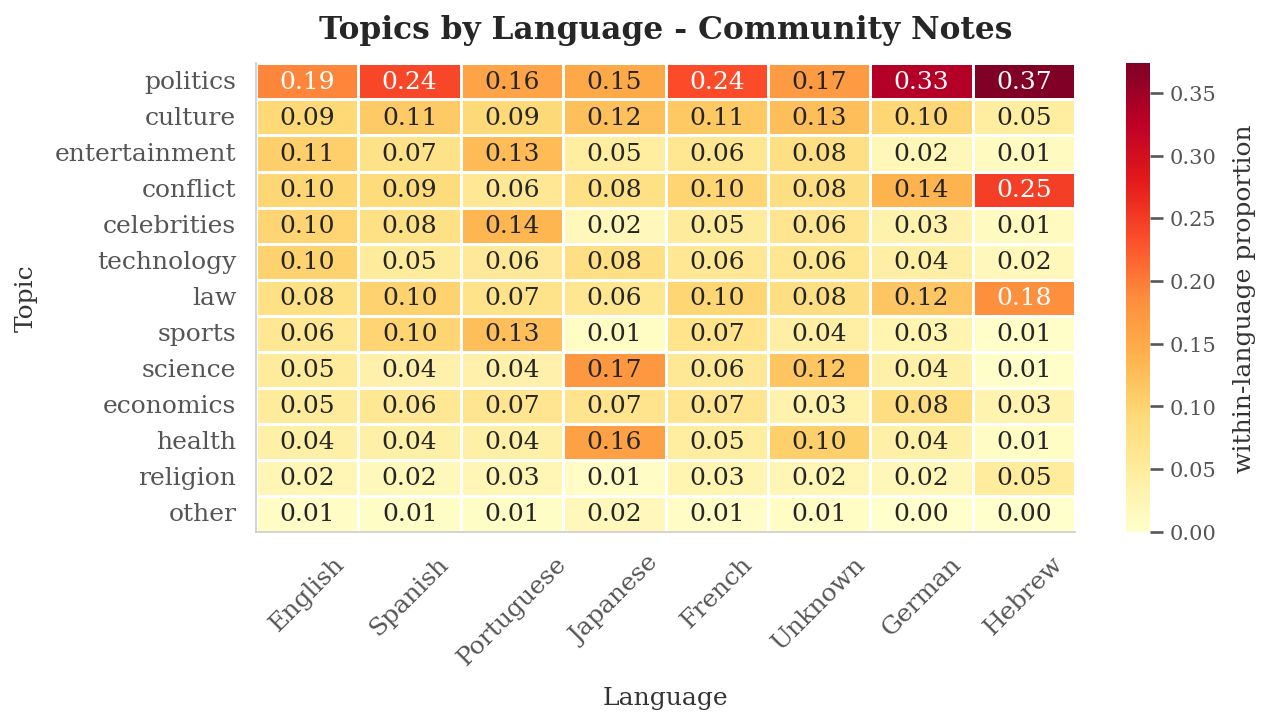}
    \caption{Distribution of topics across languages on Community Notes.}
    \label{fig:topic_lang}
\end{figure}

\begin{figure}[ht]
    \centering
    \includegraphics[width=\linewidth]{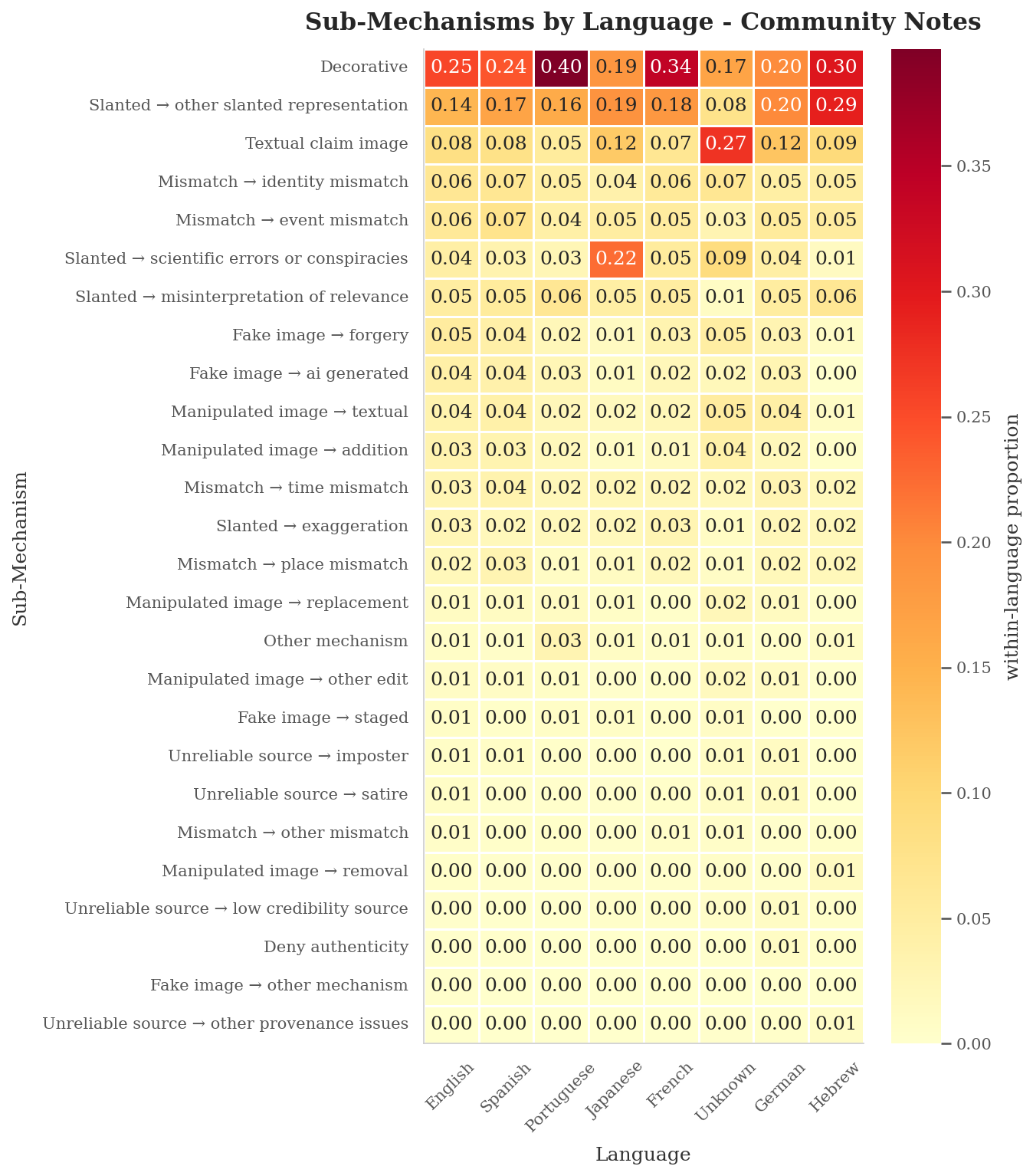}
    \caption{Co-occurrences of sub-mechanisms vs language on the Community Notes dataset.}
    \label{fig:co_oc_sub_mechanism_lang}
\end{figure}

\begin{figure*}[ht]
    \centering
    \includegraphics[width=\linewidth]{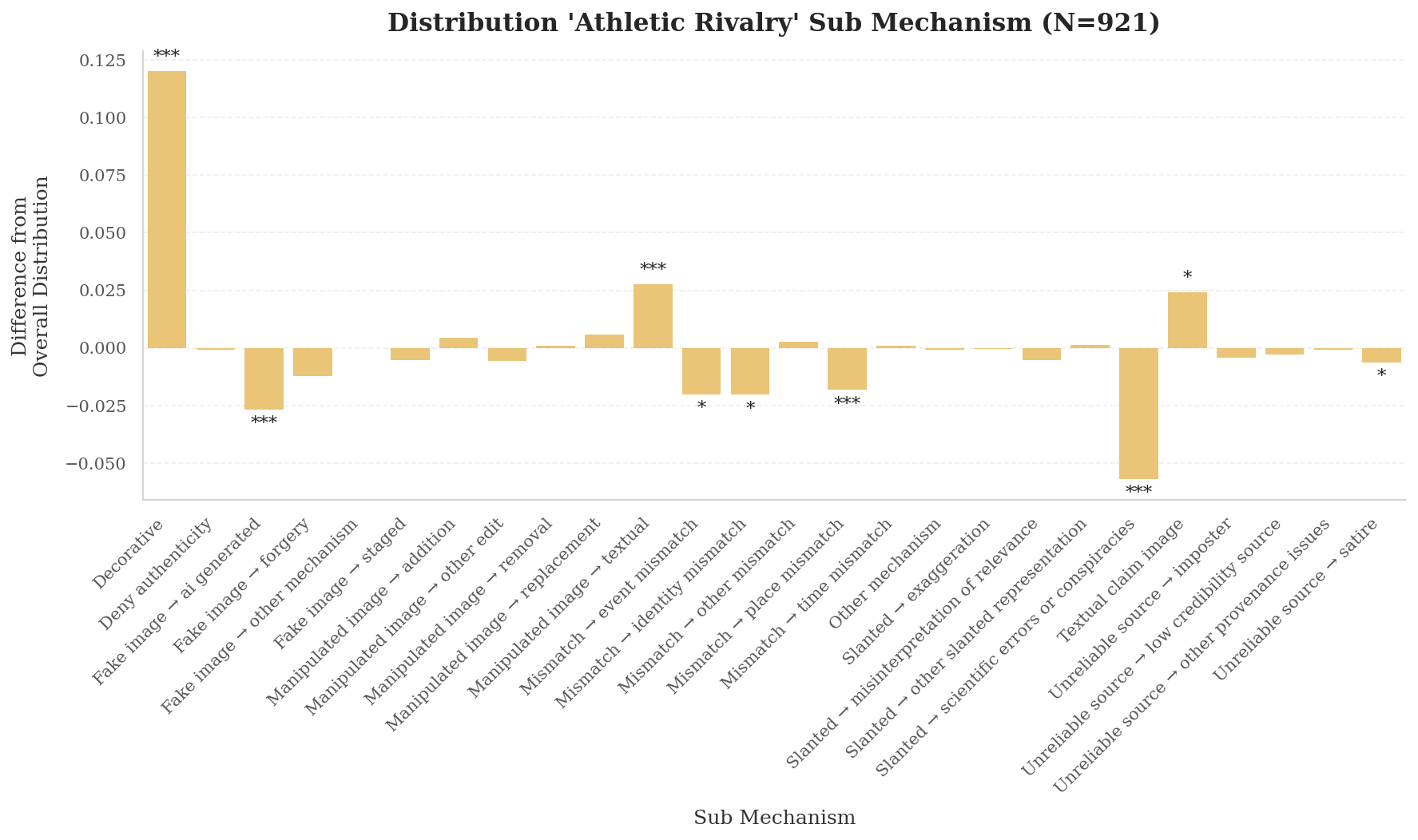}
    \caption{Sub-mechanism distribution on Community Notes for misinformation instances related to athletic rivalry. We show divergence from the overall distribution across all misinformation instances. Stars indicate statistical significance. }
    \label{fig:rivalry}
\end{figure*}

\begin{figure*}[ht]
    \centering
    \includegraphics[width=\linewidth]{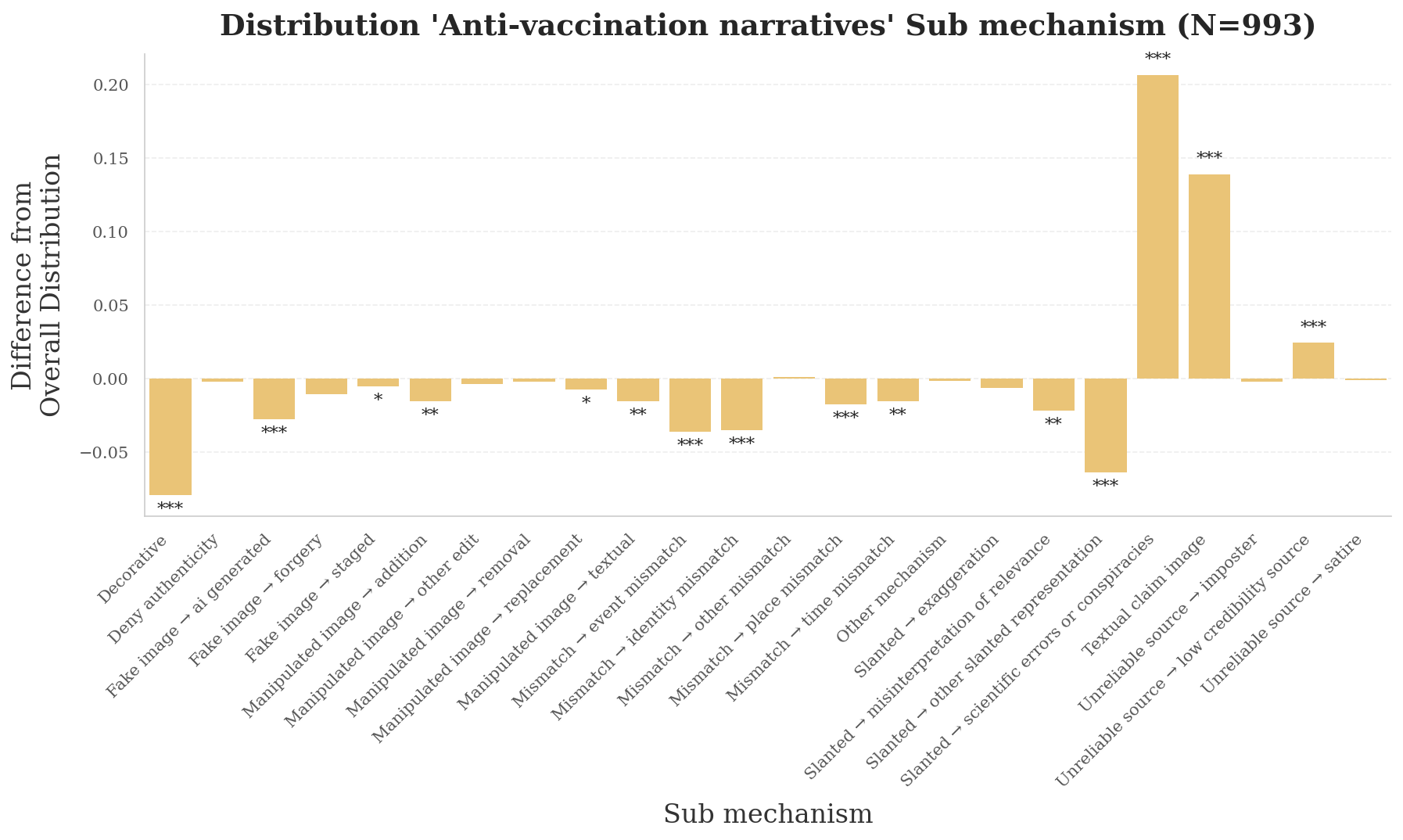}
    \caption{Sub-mechanism distribution on Community Notes for misinformation instances related to vaccination.}
    \label{fig:vaccination}
\end{figure*}

\begin{figure*}[ht]
    \centering
    \includegraphics[width=\linewidth]{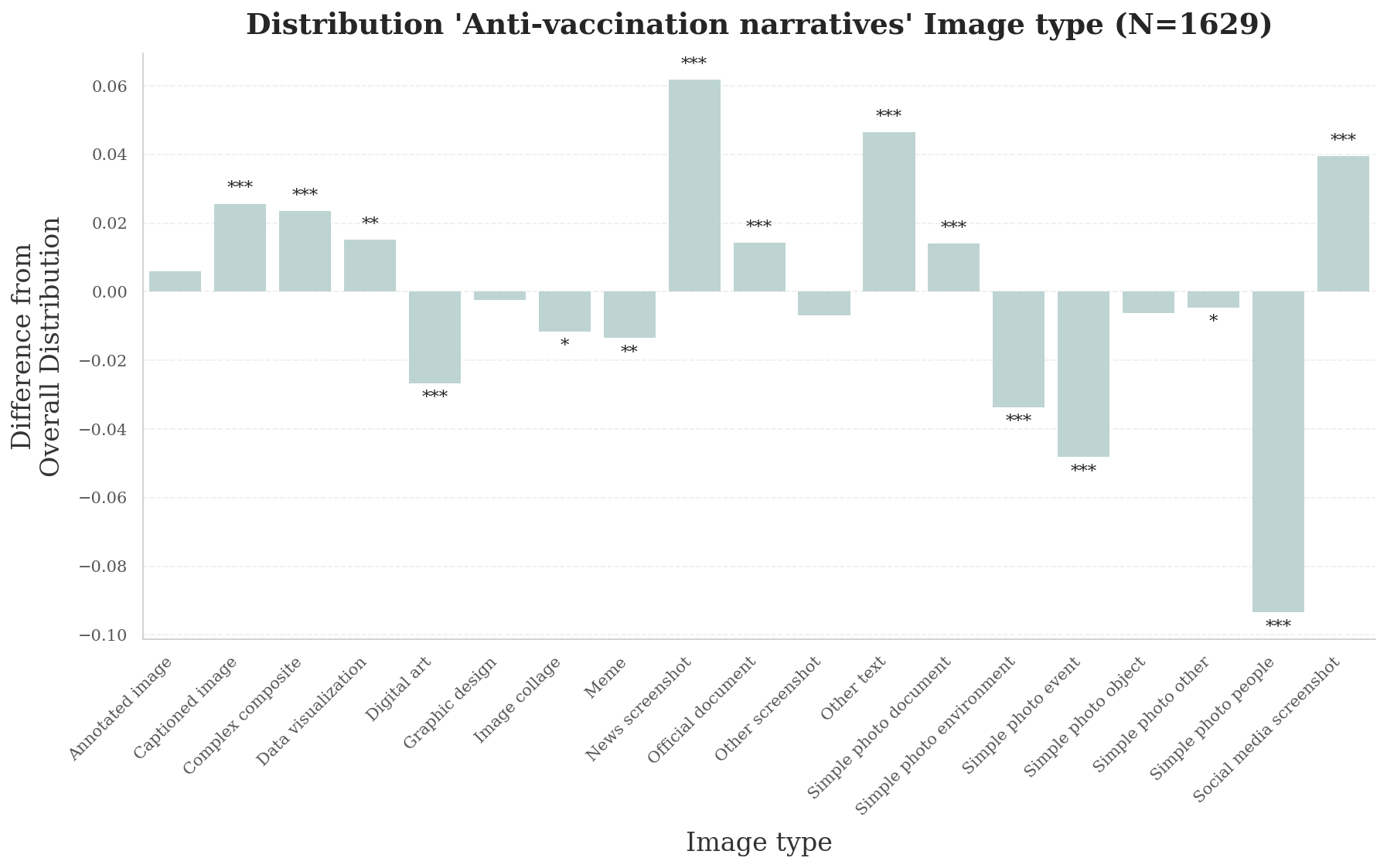}
    \caption{Image type distribution on Community Notes for misinformation instances related to vaccination. }
    \label{fig:vaccination_image_type}
\end{figure*}

\begin{figure*}[ht]
    \centering
    \includegraphics[width=\linewidth]{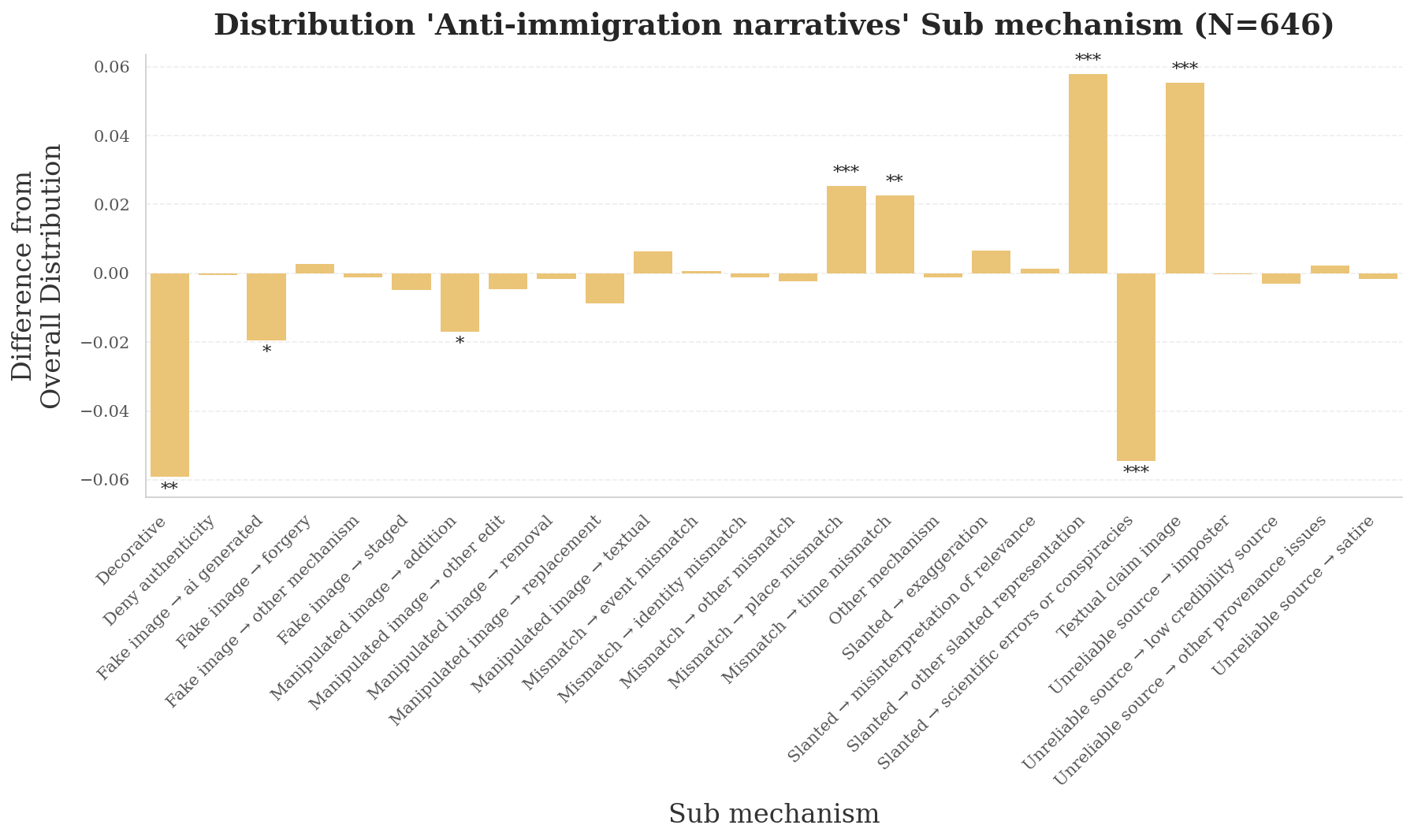}
    \caption{Sub-mechanism distribution on Community Notes for misinformation instances related to immigration. }
    \label{fig:immigration}
\end{figure*}

\begin{figure*}[ht]
    \centering
    \includegraphics[width=\linewidth]{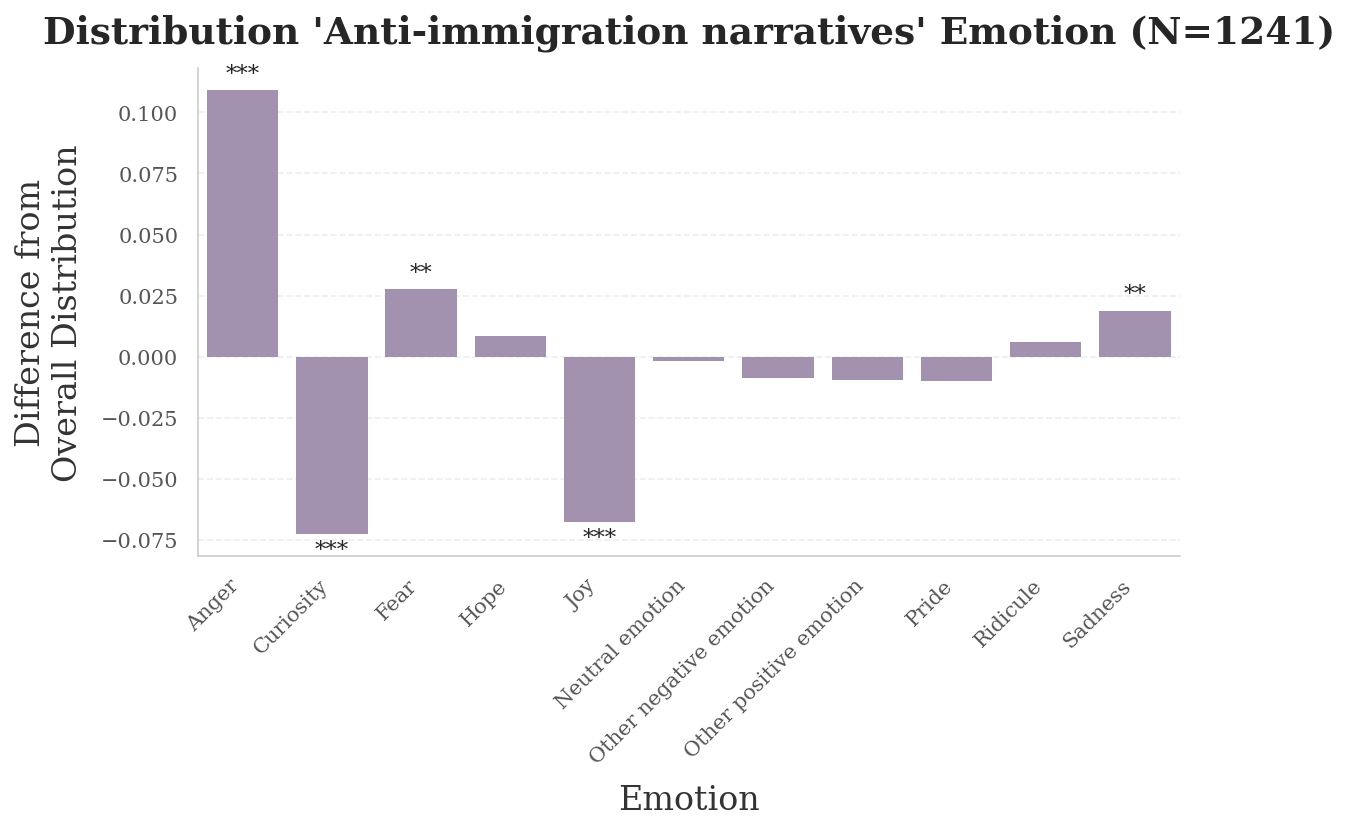}
    \caption{Emotion distribution on Community Notes for misinformation instances related to immigration.}
    \label{fig:immigration_emotion}
\end{figure*}

\noindent \textbf{\Cref{fig:co_oc_mechanism_ammeba}}: Co-occurrences of mechanisms vs topics on \textsc{AMMeBa}. Only cells with $p<0.05$ are shown. Refer to \cref{sec:analysis} for details.

\noindent \textbf{\Cref{fig:co_oc_sub_mechanism_cn}}: Co-occurrences of sub-mechanisms vs topics on Community Notes.

\noindent \textbf{\Cref{fig:co_oc_sub_mechanism_ammeba}}: Co-occurrences of sub-mechanisms vs topics on \textsc{AMMeBa}.

\noindent \textbf{\cref{fig:topic_lang}} Distribution of topics across languages on Community Notes.

\noindent \textbf{\cref{fig:co_oc_sub_mechanism_lang}} Co-occurrences of sub-mechanisms vs language on the Community Notes dataset.


\noindent \textbf{\Cref{tab:conflict}}: Examples of misinformation instances from the Community Notes dataset classified with the topic \textit{Conflict} and the sub-mechanisms \textit{Place mismatch}, \textit{Time mismatch}, or \textit{Event mismatch}.

\noindent \textbf{\Cref{fig:rivalry}}: Sub-mechanism distribution on Community Notes for misinformation instances related to athletic rivalry. We show divergence from the overall distribution across all misinformation instances.

\noindent \textbf{\Cref{fig:vaccination}}: Sub-mechanism distribution on Community Notes for misinformation instances related to vaccination. We show divergence from the overall distribution across all misinformation instances.

\noindent \textbf{\Cref{fig:vaccination_image_type}}: Image type distribution on Community Notes for misinformation instances related to vaccination. We show divergence from the overall distribution across all misinformation instances.

\noindent \textbf{\Cref{tab:vaccine_examples}}: Examples of misinformation instances from the Community Notes dataset related to vaccination, with image type \textit{News Screenshot}.

\noindent \textbf{\Cref{fig:immigration}}: Sub-mechanism distribution on Community Notes for misinformation instances related to immigration.

\noindent \textbf{\Cref{fig:immigration_emotion}}: Emotion distribution on Community Notes for misinformation instances related to immigration.

\section{Prompts}
\label{app:prompts}

\begin{figure*}[t]
\begin{lstlisting}[basicstyle=\ttfamily\tiny, label=lst:image_type_prompt, caption=The prompt used to classify the type of the image., numbers=none]
SYSTEM_PROMPT
You are an expert annotator specialising in analysing image content on social media. You answer questions with precision and conciseness.
 

USER_PROMPT
You will be given an IMAGE that was posted on a social media platform. Your task is to classify the IMAGE into one or more categories from the taxonomy below.
This is a multilabel task: an IMAGE may belong to multiple categories simultaneously. For example, a meme that is also a screenshot of a social media post should be labeled as both MEME and SOCIAL_MEDIA_SCREENSHOT.

## Taxonomy
- SIMPLE_PHOTO_PEOPLE: A photograph captured with a camera (with at most minimal editing such as cropping or color correction) that primarily depicts one or more people. Examples: a portrait, a candid shot of a crowd, a selfie
- SIMPLE_PHOTO_OBJECT: A minimally edited photograph primarily depicting a physical object or product. Examples: a photo of a weapon, a consumer product, a vehicle.
- SIMPLE_PHOTO_EVENT: A minimally edited photograph documenting a specific event or happening. Examples: a protest, a ceremony, an accident scene.
- SIMPLE_PHOTO_ENVIRONMENT: A minimally edited photograph of a place, landscape, or setting. Examples: a cityscape, a flooded area, a building exterior.
- SIMPLE_PHOTO_DOCUMENT: A photograph of a physical document, sign, or handwritten note. Examples: a photo of a printed receipt taken with a phone, a photo of a whiteboard.
- SIMPLE_PHOTO_OTHER: A minimally edited photograph that does not fit any of the above subcategories. Examples: food, animals, abstract patterns.
- SOCIAL_MEDIA_SCREENSHOT: A screen capture of content from a social media platform (e.g., X/Twitter, Facebook, Instagram, TikTok, Reddit, Telegram). The platform UI (e.g., like/share buttons, usernames, timestamps) should be at least partially visible.
- NEWS_SCREENSHOT: A screen capture from a news source: a news website, TV broadcast, news app, or newspaper.
- OTHER_SCREENSHOT: A clear screenshot that is not of a social media platform or a news source. The UI should be at least partially visible.
- OFFICIAL_DOCUMENT: A digital rendering of a formal institutional document such as a government form, court filing, or official report. Distinguished from SIMPLE_PHOTO_DOCUMENT by being a direct digital reproduction rather than a casual photograph.
- OTHER_TEXT: An image whose content is primarily text but does not fit the above text-bearing categories. Examples: a quote on a plain background, a text-only listicle, a plain text statement.
- DATA_VISUALIZATION: A chart, graph, map, or table presenting structured quantitative or geospatial data. Examples: bar charts, line graphs, choropleth maps, statistical tables.
- DIGITAL_ART: An illustration or rendering created from scratch using a software. Examples: digital paintings, vector art, 3D renders.
- GRAPHIC_DESIGN: A designed visual composition for communication or promotion, combining typography, imagery, and/or layout elements. Examples: advertisements, campaign posters, branded infographics, flyers.
- MEME: An image following a recognizable memetic template or internet humor convention, typically combining a stock or reaction image with overlaid text for humorous, satirical, or rhetorical effect.
- ANNOTATED_IMAGE: A base image overlaid with explanatory graphical elements such as arrows, circles, labels, or highlights to draw attention to or explain specific features. Examples: a satellite image with circled areas, a photo with arrows pointing to details.
- CAPTIONED_IMAGE: A simple structure in which a single image is paired with a text element (headline, caption, or label), typically placed above or below the image. The composition contains no more than one image and one text block.
- IMAGE_COLLAGE: A composition of multiple photographs arranged together (e.g., in a grid or side-by-side layout) without significant additional graphic or textual elements. Examples: before/after comparisons, photo grids.
- COMPLEX_COMPOSITE: A multi-element composition combining three or more distinct content types (e.g., photographs, text blocks, data visualizations, graphic elements, arrows) into a single structured image. Examples: conspiratorial infographics, elaborate explainer images mixing photos with maps and quoted text.
- OTHER: Any image that does not fit the categories above.

## Instructions
- Examine the IMAGE carefully and assign ALL categories that apply. Do not limit yourself to a single label.
- Apply the SIMPLE_PHOTO_* labels only to photographs that appear to be camera-captured with minimal editing. If a photo is embedded within a larger composition, label the composition type (e.g., CAPTIONED_IMAGE, COMPLEX_COMPOSITE, IMAGE_COLLAGE) but do NOT label the embedded photo separately.
- When in doubt between CAPTIONED_IMAGE and COMPLEX_COMPOSITE, use CAPTIONED_IMAGE only if the structure is limited to one image and one text element. If there are additional elements, use COMPLEX_COMPOSITE.
- When in doubt between CAPTIONED_IMAGE and MEME, apply MEME only if the image follows a recognizable memetic template or internet humor convention.
- Here are some examples of common label co-occurrences:
  - A news screenshot containing a data visualization -> NEWS_SCREENSHOT + DATA_VISUALIZATION
  - An annotated photograph of people -> ANNOTATED_IMAGE + SIMPLE_PHOTO_PEOPLE
  - A collage with added text headers -> IMAGE_COLLAGE + CAPTIONED_IMAGE
  - A screenshot of a Wikipedia article -> OTHER_SCREENSHOT + OTHER_TEXT
  - A table in an official document -> OFFICIAL_DOCUMENT + DATA_VISUALIZATION


## Output format
Return a JSON object with the following fields:
- "labels": a list of applicable category labels (use the exact uppercase identifiers above).
- "reasoning": a brief explanation (2-3 sentences) justifying why each label was assigned.
Output ONLY a JSON object with the following structure. Do not include any preamble, explanation, or markdown formatting.


Example output:
{
  "labels": ["SOCIAL_MEDIA_SCREENSHOT", "MEME"],
  "reasoning": "The image is a screenshot of a Twitter post, as indicated by the visible Twitter UI elements. The content of the post is a meme using the 'distracted boyfriend' template with overlaid text."
}


Now classify the following IMAGE:

## Input
IMAGE: {{image}}

\end{lstlisting}
\end{figure*}

\begin{figure*}[t]
\begin{lstlisting}[basicstyle=\ttfamily\footnotesize, label=lst:emotions_prompt, caption=The prompt used to classify the emotions the user is trying to evoke by including the image in their post., numbers=none]

SYSTEM_PROMPT
You are an expert annotator specialising in analysing multimodal content on social media. You answer questions with precision and conciseness.
 
 
USER_PROMPT
## Instructions
You will be given a POST, IMAGE, and USERNAME. The POST and IMAGE are the text and image posted by USERNAME on a social media platform containing some claim, which might be either correct or incorrect.
Carefully evaluate the POST and determine the relationship between the POST and the IMAGE. Specifically, determine which **emotion** USERNAME is trying to evoke by including the IMAGE.
Importantly, try to estimate which emotions **USERNAME** is trying to evoke, not which emotions a random social media user might feel. For example, if the POST contain anti-LGBT content, and the IMAGE contains the pride flag, USERNAME is probably trying to evoke *anger*.


Select any number of items from the following list:
  - joy: A feeling of pleasure and happiness.
  - hope: A feeling of expectation and desire for a certain thing to happen.
  - pride: A feeling of deep pleasure or satisfaction derived from one's own achievements, the achievements of those with whom one is closely associated, or from qualities or possessions that are widely admired.
  - curiosity: A strong desire to know or learn something.
  - other_positive_emotion: Any other positive emotion not listed above.
  - anger: A strong feeling of annoyance, displeasure, or hostility.
  - sadness: A feeling of sorrow or unhappiness.
  - fear: An unpleasant emotion caused by the threat of danger, pain, or harm.
  - ridicule: The subjection of someone or something to contemptuous and dismissive language or behavior.
  - other_negative_emotion: Any other negative emotion not listed above.
  - neutral_emotion: The IMAGE evokes a neutral emotion


## Output format
Return a JSON object with the following fields:
- "emotions": a list of applicable category labels.
- "reasoning": a brief explanation (2-3 sentences) justifying why each label was assigned.
Output ONLY a JSON object with the following structure. Do not include any preamble, explanation, or markdown formatting.

Example output:
{
  "emotions": ["anger", "fear", "sadness"],
  "reasoning": "The IMAGE is of a plane crash and the POST claims that the crush was caused by malpractice from the engineering team. The IMAGE is used to evoke anger at the engineers who failed maintaining the plane, sadness for the lost life, and fear that such accident might happen again."
}


## Input
USERNAME: {{username}}
POST: {{post}}
IMAGE: {{image}}

\end{lstlisting}
\end{figure*}

\begin{figure*}[t]
\begin{lstlisting}[basicstyle=\ttfamily\footnotesize, label=lst:classification_prompt, caption=The prompt used to determine whether the post is considered misinformation., numbers=none]

SYSTEM_PROMPT
You are an expert fact-checker specialising in analysing multimodal misinformation on social media. You answer questions with precision and conciseness.
 
 
USER_PROMPT

## Instructions
You will be given a POST, IMAGE, USERNAME, and ADDITIONAL_CONTEXT. The POST and IMAGE are the text and image posted by USERNAME on a social media platform containing some claim. ADDITIONAL_CONTEXT provides background on the POST and IMAGE and explains how they may be misleading.
Carefully evaluate ADDITIONAL_CONTEXT and determine the specific nature of POST and the IMAGE. Select *exactly one* of the following options. They are listed in priority order: if multiple could apply, select the first one that fits.

- misinformation: The POST and IMAGE spread misinformation. USERNAME is trying to promote some incorrect message or narrative.
- ad: The ADDITIONAL_CONTEXT claims that the POST and IMAGE contain an undisclosed advertisement. USERNAME is trying to sell or promote a product or a service.
- scam: The ADDITIONAL_CONTEXT claims that the POST and IMAGE contain a scam of some sort.
- stolen_content: The ADDITIONAL_CONTEXT claims that the IMAGE was posted without properly crediting its original creator. USERNAME is pretending to post an original work.
- engagement_bait: The ADDITIONAL_CONTEXT claims that the POST and IMAGE are engagement bait -- trying to provoke a strong reaction from the reader.
 
## Output Format
Output ONLY a JSON object with the following structure. Do not include any preamble, explanation, or markdown formatting.
 
{
  "classification": "misinformation|ad|scam|stolen_content|engagement_bait",
  "reasoning": "<reasoning>"
}

## Input
USERNAME: {{username}}
POST: {{post}}
IMAGE: {{image}}
ADDITIONAL_CONTEXT: {{note}}

\end{lstlisting}
\end{figure*}

\begin{figure*}[t]
\begin{lstlisting}[basicstyle=\ttfamily\tiny, label=lst:topic_prompt, caption=The prompt used to extract the topic and message of a post., numbers=none]

SYSTEM_PROMPT
You are an expert annotator specialising in analysing multimodal content on social media. You answer questions with precision and conciseness.
 
USER_PROMPT
## Instructions
You will be given a POST, IMAGE, USERNAME, and ADDITIONAL_CONTEXT. The POST and IMAGE are the text and image posted by USERNAME on a social media platform containing some claim. ADDITIONAL_CONTEXT provides background on the POST and IMAGE and explains how they may be misleading.
Your goal is to analyze the POST, IMAGE, USERNAME and provide the following information:
 
- topic: The topic of the post. Choose from: politics, economics, science, health, entertainment, sports, celebrities, conflict, law, religion, technology, culture, other. This is a multilabel task -- some posts might belong to several topics.
- message: Describe the main message USERNAME is trying to convey by posting the POST and IMAGE. Treat the input as correct and true. Do not debunk, fact-check, or invalidate the message. Do not include caveats or disclaimers.
- message_abstract_level_[1-3]: Paraphase the message to be more more anstract and generic, while still capturing the essence of the original message. There are three levels of abstractation: from 1, somewhat abstract (no named entities are mentioned) to 3, extremely abstract (only the most high-level idea is conveyed). Importantly, abstraction level 1 should have no more than 8 words, abstractation level 2 no more than 6 words, and abstractation level 3 no more than 4 words.

## Examples
Example 1:
USERNAME: PatriotVoices
POST: "Senator Smith just voted against the veterans' benefits bill for the THIRD time!"
IMAGE: A photo of Senator Smith walking out of the Senate building.
ADDITIONAL_CONTEXT: The bill was a broader spending package, not solely a veterans' benefits bill.

Output:
{
    "topic": ["politics", "law"],
    "message": "Senator Smith has repeatedly voted against legislation meant to support veterans.",
    "message_abstract_level_1": "A politician repeatedly opposes veteran support legislation",
    "message_abstract_level_2": "Politician opposed veteran support bill",
    "message_abstract_level_3": "Politician harms veterans"
}

Example 2:
USERNAME: HealthAlerts
POST: "Shocking discovery"
IMAGE: A screenshot of a headline reading "Scientists confirm: eating breakfast daily reduces heart disease risk by 40%"
ADDITIONAL_CONTEXT: The study cited in the headline had a small sample size and the 40% figure was not statistically significant.

Output:
{
    "topic": ["health", "science"],
    "message": "Scientists have found that eating breakfast every day significantly reduces the risk of heart disease.",
    "message_abstract_level_1": "Eating breakfast daily significantly reduces heart disease risk",
    "message_abstract_level_2": "Eating breakfast reduces disease risk",
    "message_abstract_level_3": "Dietary habit prevents disease"
}


## Important Notes
- Often, the POST doesn't contain any textual claim, and instead the claim appears as text in the IMAGE. In such cases, answer as if the text in the IMAGE also appears in the POST.
- Characterize the post's intended message as if it were genuine. Do NOT fact-check or debunk.


## Output format
Output ONLY a JSON object with the following structure. Do not include any preamble, explanation, or markdown formatting. Output your analysis *in English*.

{
    "topic": [topic1, topic2,...],
    "message": <message>,
    "message_abstract_level_1": <message_abstract_level_1>,
    "message_abstract_level_2": <message_abstract_level_2>,
    "message_abstract_level_3": <message_abstract_level_3>,
}


## Input
USERNAME: {{username}}
POST: {{post}}
IMAGE: {{image}}
ADDITIONAL_CONTEXT: {{note}}

\end{lstlisting}
\end{figure*}

\begin{figure*}[t]
\begin{lstlisting}[basicstyle=\ttfamily\tiny, label=lst:rhetoric_role_prompt, caption=The prompt used to classify the rhetorical role employed by the user., numbers=none]

SYSTEM_PROMPT
You are an expert annotator specializing in analyzing multimodal content on social media. You have deep knowledge of image-text relationships and visual rhetoric. You answer questions with precision and conciseness.

USER_PROMPT
## Instructions
You will be given a POST, IMAGE, and USERNAME. The POST and IMAGE are the text and image posted by USERNAME on a social media platform. Your goal is to analyze the functional relationship between the IMAGE and the POST - that is, what role does the IMAGE play relative to the textual content of the POST? Classify the IMAGE-POST pair into **one or more** of the following relationship categories, based on the taxonomy of Marsh & White (2003). The categories are organized into three high-level groups (A, B, C) according to how closely the image relates to the text.

### A - Functions expressing little relation to the text
The IMAGE has a weak conceptual connection to the POST's content. It does not directly represent or extend the claims in the text.
- **A1_decorate**: The IMAGE makes the POST more visually attractive or attention-grabbing without meaningfully affecting the reader's understanding of the text. It may also interrupt continuity or match the stylistic tone of the POST.
- **A2_elicit_emotion**: The IMAGE encourages an emotional response (e.g., shock, sympathy, outrage, humor) through its content or style, but is not closely tied to the specific factual claims in the POST. This includes cases where the IMAGE creates tension or mood through contrast in style or tone with the text.
- **A3_control**: The IMAGE directs, holds, or regulates the reader's attention, or motivates a response (e.g., engagement, sharing, clicking). Use this when the IMAGE primarily serves to engage or motivate the reader rather than to inform.

### B - Functions expressing close relation to the text
The IMAGE has a direct conceptual connection to the POST's content. It visually restates, organizes, or explains what the text says.
- **B1_reiterate**: The IMAGE visually restates or makes concrete what the text says, with minimal added interpretation. This includes showing what the text describes (concretizing), providing a visual translation of written content, showing a specific example or instance of what the text references, depicting the person or source mentioned, or presenting data as a graph.
- **B2_organize**: The IMAGE provides spatial, temporal, or conceptual structure to the text's content - e.g., locating events in time or place, establishing a setting, containing information in a structured format (flowchart, timeline, Venn diagram), or inducing perspective by encouraging the reader to see things in their relative importance.
- **B3_relate**: The IMAGE explicitly compares, contrasts, or draws a parallel between elements described in the text. The comparison stays within the scope of the text's own content.
- **B4_condense**: The IMAGE reduces the text's content to its most essential or critical elements, presenting a simplified or concentrated version of the information.
- **B5_explain**: The IMAGE makes the text's content clearer or more understandable. This includes defining concepts, or complementing the text so that image and text together convey the message more effectively than either alone.

### C - Functions that go beyond the text
The IMAGE adds meaning, interpretation, or content that is not present in the text itself.
- **C1_interpret**: The IMAGE provides force, importance, or emphasis to the text's content, or documents it with factual/visual evidence not mentioned in the text. The image illustrates complex ideas in a concrete form that goes beyond what the text states.
- **C2_develop**: The IMAGE expands on the text by introducing new details, comparisons, or contrasts using elements not present in the text. This includes analogizing (drawing a resemblance to something outside the text) or contrasting with external content for rhetorical effect.
- **C3_transform**: The IMAGE recodes, reorganizes, or reinterprets the text in a fundamentally new way. This includes modeling cognitive or physical processes not directly observable, alternating with the text to progress a narrative, or using the text as a starting point to introduce substantially new content.

## Examples
Example 1 - **A2_elicit_emotion**:
USERNAME: TruthPatriot99
POST: "Our border is under siege. When will our leaders act?"
IMAGE: A dramatic aerial photo of a large crowd of people at a border fence.

Output:
{
    "image_text_relationship": ["A2_elicit_emotion"],
    "reasoning": "The POST makes a general political claim about border policy. The IMAGE - a dramatic, large-scale aerial shot - is selected primarily to evoke alarm and urgency rather than to document a specific event referenced in the text. The emotional impact of the image is the primary function."
}

...
(6 more examples)
...

## Important Notes
- This is a **multilabel** task: a single IMAGE-POST pair can fulfill multiple functions simultaneously. Assign all that apply.
- Focus on the **functional relationship** between the IMAGE and the POST's textual content - not on the image's subject matter in isolation.
- The categories refer to the image's role relative to the text, not to the image's inherent quality or genre.
- When the IMAGE contains text (e.g., a screenshot of a headline, a meme with overlaid text), treat the text within the IMAGE as part of the visual content and assess its relationship to the POST.
- If the POST has minimal or no text, base your analysis on whatever text is present (including hashtags, captions, or emojis).
- Classify at the most specific applicable level. For example, if an image both reiterates and explains the text, assign both B1_reiterate and B5_explain.
- Prefer labels where the function is clearly present. Do not assign a label based on weak or speculative connections.

## Output Format
Output ONLY a single valid JSON object with this structure. Do not include any preamble, explanation, or markdown formatting.

{"image_text_relationship": ["label_1", "label_2", ...], "reasoning": "<reasoning>"}

## Input
USERNAME: {{username}}
POST: {{post}}
IMAGE: {{image}}


\end{lstlisting}
\end{figure*}

\begin{figure*}[t]
\begin{lstlisting}[basicstyle=\ttfamily\tiny, label=lst:mechanism_stage_1_prompt, caption=The prompt used to the multimodal misinformation mechanism. Stage 1., numbers=none]

SYSTEM_PROMPT
You are an expert fact-checker analyzing multimodal misinformation on social media. You ONLY output valid JSON - no preamble, markdown, or explanation outside the JSON object.
 
 
USER_PROMPT
## Task
Analyze a social media post (text + image) in two phases:
- Phase 1: Characterize the post's intended message as if it were genuine. Do NOT fact-check or debunk.
- Phase 2: Using ADDITIONAL_CONTEXT, diagnose whether the image contributes to the misleading nature of the post.

Complete Phase 1 reasoning fully before considering ADDITIONAL_CONTEXT for Phase 2.

## Field Definitions
- topic: Primary topic. Choose from: politics, economics, science, health, entertainment, sports, celebrities, conflict, law, religion, technology, culture, other
- message: The main message USERNAME intends to convey, in under 15 words. Treat the post as genuine. No caveats or debunking.
- image_misleads: Does the IMAGE itself contribute to making the post misleading - e.g., by being fabricated, manipulated, taken out of context? 
Answer "YES" if removing or replacing the image would change the message being conveyed by USERNAME or prevent the message from being understood. Similarly, answer "YES" if the ADDITIONAL_CONTEXT claims that the USERNAME describes the IMAGE in a misleading way, such as misrepresenting its meaning or consequences, or drawing false or exaggerated conclusions from it.
Conversely, answer "NO" if the image is merely illustrative (e.g., a stock photo of a person mentioned in the POST), and is not described in a misleading way by USERNAME. 
- reasoning: 1-3 sentences explaining your image_misleads judgment.

## Examples
Example 1:
POST: "Breaking: [Politician] arrested for fraud!"
IMAGE: An unedited official headshot of the politician.
ADDITIONAL_CONTEXT: The politician was not arrested. The claim is fabricated.

Output:
{
    "topic": "politics",
    "message": "Politician was arrested for committing fraud",
    "image_misleads": "NO",
    "reasoning": "The image is a generic headshot that merely identifies the politician. The misleading claim is entirely textual - removing the image would not reduce the post's deceptive effect."
}

Example 2:
POST: "Look at the massive crowd at the rally today!"
IMAGE: A photo of a large crowd, but it is actually from a concert 3 years ago.
ADDITIONAL_CONTEXT: The image is from a 2021 music festival, not the rally referenced.

Output:
{
    "topic": "politics",
    "message": "A massive crowd attended today's rally",
    "image_misleads": "YES",
    "reasoning": "The image is taken out of context - it depicts a different event entirely. Without this misattributed image, the post's claim of a large crowd has no visual evidence."
}

## Important Notes

- The IMAGE is provided as a visual input. If relevant text appears within the image, treat it as part of the post's claim.
- If ADDITIONAL_CONTEXT is missing or unclear, base your Phase 2 analysis on what you can observe and state your uncertainty in the reasoning field.

## Output Format
Output ONLY a single valid JSON object with this structure:

{
    "topic": "<topic>",
    "message": "<message>",
    "image_misleads": "YES|NO",
    "reasoning": "<reasoning>"
}

## Inputs
USERNAME: {{username}}
POST: {{post}}
IMAGE: {{image}}
ADDITIONAL_CONTEXT: {{note}}

\end{lstlisting}
\end{figure*}

\begin{figure*}[t]
\begin{lstlisting}[basicstyle=\ttfamily\tiny, label=lst:mechanism_stage_2_prompt, caption=The prompt used to the multimodal misinformation mechanism. Stage 2., numbers=none]

## Further instructions

You have stated that the IMAGE itself contribute to making the post misleading. Your reasoning is "{{reasoning}}".

Using the ADDITIONAL_CONTEXT and your answers to the previous questions, determine the specific mechanism that USERNAME is employing to mislead with *the IMAGE*. Choose exactly one of the following labels:
  - fake_image: The ADDITIONAL_CONTEXT claims that the IMAGE is fabricated, AI-generated, staged, or otherwise completely fake.
  - manipulated_image: The IMAGE is a real image, but the ADDITIONAL_CONTEXT claims it was digitally altered or edited in some manner.
  - unreliable_source: The IMAGE itself depicts content from an unreliable source - e.g., it is a screenshot of a satirical account, a low-credibility outlet, or an impersonator account. The issue concerns the source shown *within* the IMAGE, not USERNAME's account or the text in the POST.
  - deny_authenticity: The POST incorrectly denies the authenticity of the IMAGE, but the ADDITIONAL_CONTEXT claims that the IMAGE is real.
  - mismatch: The ADDITIONAL_CONTEXT explains that the POST makes a factually incorrect claim about the content or context of the IMAGE, such as incorrect time, place, person, or event. Use this when the IMAGE is misidentified or taken from a different context entirely.
  - slanted: The ADDITIONAL_CONTEXT claims that the POST describes the IMAGE in a misleading way, such as misrepresenting its meaning or consequences, or drawing false or exaggerated conclusions from it. Use this when the IMAGE is correctly identified but the interpretation or conclusion is misleading.
  - textual_claim_image: The IMAGE contains the claim in a textual form, and the claim is false. The IMAGE contains no important visual elements apart from the text.
  - other_mechanism: The additional context describes a different mechanism that is not listed above.
  - nothing: Select this option if you think your answer to the question "image_misleads" is incorrect.

## Examples
Example 1 - **mismatch** (image correctly depicts something real, but is misattributed to the wrong event/time/place):
POST: "Incredible footage of Russian troops retreating outside Kyiv!"
IMAGE: A video still showing military vehicles on a road.
ADDITIONAL_CONTEXT: The footage is from a 2019 military exercise in Belarus, unrelated to the referenced conflict.
Previous reasoning: "The image is presented as live evidence of a current military event, but it is archival footage from a different country and time - removing it would eliminate the only 'proof' of the claim."

Output:
{
    "mechanism": "mismatch",
    "reasoning": "The image is real footage but is misattributed to a completely different event, time, and location. This is a context mismatch: the image is not fake or edited, it simply does not depict what the POST claims."
}

Example 2 - **slanted** (image is correctly identified, but the POST draws false or exaggerated conclusions from it):
POST: "Thousands flood the streets demanding the president's resignation - the regime is crumbling!"
IMAGE: A genuine photo of a protest in the capital city.
ADDITIONAL_CONTEXT: The protest shown is real and correctly located, but attendance was estimated at a few hundred people, and it was one of many routine demonstrations.
Previous reasoning: "The image is a real, correctly identified photo of the protest, but the POST wildly exaggerates its scale and frames it as a historic turning point that the image does not support."

Output:
{
    "mechanism": "slanted",
    "reasoning": "The image is authentic and correctly identified, but the POST misrepresents its significance - inflating the crowd size and framing a routine protest as a regime-collapsing event. The misleading element is interpretive, not factual misattribution."
}

Example 3 - **unreliable_source** (the IMAGE itself shows content from a low-credibility or satirical source):
POST: "Even the CDC admits vaccines cause autism - finally the truth comes out!"
IMAGE: A screenshot of a tweet from the account @CDCofficial_news.
ADDITIONAL_CONTEXT: The account shown in the screenshot is a parody account impersonating the CDC. The real CDC account is @CDCgov and has never made such a claim.
Previous reasoning: "The image displays what appears to be an official government statement, lending false authority to the claim - but the account depicted is a known impersonator."

Output:
{
    "mechanism": "unreliable_source",
    "reasoning": "The IMAGE is a screenshot of a parody/impersonator account, not the real CDC. The misleading element is the source depicted within the image itself - it creates the false impression of an authoritative endorsement."
}

## Important Notes
- Select "nothing" *only if* you think your answer to the question "image_misleads" is incorrect.
- Use **mismatch** when the image is real but depicts the *wrong* event, person, time, or place - the factual identification is incorrect.
- Use **slanted** when the image is real and correctly identified, but the POST *interprets* it in a misleading way (exaggerated conclusions, false framing).
- Use **unreliable_source** when the IMAGE shows a screenshot or depiction of content *from* a satirical, parody, or low-credibility source - the issue is the source shown inside the image, not USERNAME's own account.

## Output Format
Output ONLY a single valid JSON object with this structure:

{
    "mechanism": "fake_image|manipulated_image|unreliable_source|deny_authenticity|mismatch|slanted|textual_claim_image|other_mechanism|nothing",
    "reasoning": "<reasoning>"
}

\end{lstlisting}
\end{figure*}

\begin{figure*}[t]
\begin{lstlisting}[basicstyle=\ttfamily\tiny, label=lst:mechanism_stage_3_prompt, caption=The prompt used to the multimodal misinformation mechanism. Stage 3. The \{\{SUB\_MECHANISM\_OPTIONS\}\} and \{\{EXAMPLES\}\} are replaced by mechanism-specific options and examples. To save space we show only the options and examples of the Mismatch category here. The other categories can be found in the GitHub at REDACTED., numbers=none]


USER_PROMPT
## Further instructions
The mechanism you selected is "{{mechanism}}". Your reasoning is "{{reasoning}}".
Your task now is to further classify the used sub mechanism. Provide the following fields:

- sub_mechanism: Using the ADDITIONAL_CONTEXT, determine the *specific* sub mechanism USERNAME is using to mislead with *the IMAGE*. Choose exactly one of the following labels:

{{SUB_MECHANISM_OPTIONS}}

- sub_mechanism_reasoning: Explain your reasoning behind your selection of sub_mechanism.

## Examples
{{EXAMPLES}}


## Output Format
Output ONLY a JSON object with the following structure. Do not include any preamble, explanation, or markdown formatting.

{
  "sub_mechanism": "<sub_mechanism>",
  "sub_mechanism_reasoning": "<sub_mechanism_reasoning>"
}



SUB_MECHANISM_OPTIONS:
---------------

<MISMATCH>
-- identity_mismatch: The POST incorrectly names an entity in the IMAGE, such as a person, a brand, or an object.
-- time_mismatch: The POST incorrectly describes the time/date of when the IMAGE was captured.
-- place_mismatch: The POST incorrectly describes the location of where the IMAGE was captured.
-- event_mismatch: There is a discrepancy regarding the event depicted in the IMAGE and the textual description in the POST. (e.g., a concert is described as a protest).
-- other_mismatch: Any other type of mismatch.
</MISMATCH>

.
.
.

EXAMPLES
---------------

<MISMATCH_EXAMPLES>
Example - identity_mismatch:
POST: "Senator Johnson leaving the courthouse after his indictment."
IMAGE: A man in a suit exiting a building.
ADDITIONAL_CONTEXT: The person pictured is Senator Williams, not Senator Johnson.
Mechanism: mismatch | Previous reasoning: "The POST names a specific politician, but the person shown is a different official."
Output:
{
  "sub_mechanism": "identity_mismatch",
  "sub_mechanism_reasoning": "The core error is a misnamed person - the event and location may be accurate, but the wrong entity is identified."
}

Example - event_mismatch (note: if place is also wrong, still prefer event_mismatch when the nature of the event is the primary distortion):
POST: "Violent riots erupting downtown as protesters clash with police."
IMAGE: A large crowd gathered outdoors.
ADDITIONAL_CONTEXT: The image is from a music festival held in the same city the previous year - no riot occurred.
Mechanism: mismatch | Previous reasoning: "The POST describes a riot, but the crowd is at a music festival."
Output:
{
  "sub_mechanism": "event_mismatch",
  "sub_mechanism_reasoning": "The event type is fundamentally mischaracterised - a peaceful festival is presented as a violent riot. Although the location may also differ, the dominant mismatch is the nature of the event, not the place."
}


## Important Notes

- If both place and event are wrong, pick whichever is the *primary* distortion (usually `event_mismatch`)
- If both time and place are wrong, pick "place_mismatch".
- `identity_mismatch` applies to misnamed persons, brands, or objects - not to mischaracterised events

</MISMATCH_EXAMPLES>


\end{lstlisting}
\end{figure*}

\noindent \textbf{\Cref{lst:image_type_prompt}}: The prompt used to classify the \textit{type} of the image.

\noindent \textbf{\Cref{lst:emotions_prompt}}: The prompt used to classify the emotions the user is trying to evoke by including the image in their post.

\noindent \textbf{\Cref{lst:classification_prompt}}: The prompt used to determine whether the post is considered misinformation.

\noindent \textbf{\Cref{lst:topic_prompt}}: The prompt used to extract the topic and message of a post. The prompt also extracts abstracted versions of the message, which are useful for high-level analysis (\cref{sec:analysis}).

\noindent \textbf{\Cref{lst:rhetoric_role_prompt}}: The prompt used to classify the rhetorical role employed by the user, following the taxonomy defined by \citet{rhetorical_role}.

\noindent \textbf{\Cref{lst:mechanism_stage_1_prompt}}: The prompt used for the multimodal misinformation mechanism, stage 1. Here, we determine whether the image participates in the misleading.

\noindent \textbf{\Cref{lst:mechanism_stage_2_prompt}}: The prompt used for the multimodal misinformation mechanism, stage 2. This prompt predicts the top-level mechanism.

\noindent \textbf{\Cref{lst:mechanism_stage_3_prompt}}:  The prompt used for the multimodal misinformation mechanism, stage 3. This prompt predicts the sub-mechanism.

\section{Reproducibility}
\label{app:reproducibility}

\subsection{\textsc{AMMeBA} Collection}
\label{app:ammeba_collection}

We construct our \textsc{AMMeBA} from the original files released in \url{https://www.kaggle.com/datasets/googleai/in-the-wild-misinformation-media}, using \texttt{stage\_2.csv}, \texttt{image\_metadata.csv}, and \texttt{fact\_checks.csv}, which we join on \texttt{image\_id} and \texttt{fact\_check\_url}. We filter ``disqualified'' entries and rows with missing columns. \num{59,358} candidate remain. For each dataset instance, we 1) download the misinformation image directly from its original source URL; 2) extract structured claim metadata (claimant, claim text, and review title) using Google's Fact Check Tools API\footnote{\url{https://developers.google.com/fact-check/tools/api}}; 3) download the full text of each fact-check article from its publisher's website using \texttt{trafilatura} \citep{barbaresi-2021-trafilatura}. We further clean the raw article texts by removing empty or corrupted entries, or those that contain fewer than 100 words. After merging all downloaded content, filtering and deduplicating, we left with \num{3,629} dataset samples. In future work we aim to improve our data collection scheme to ensure more instances from \textsc{AMMeBA} are considered.

\subsection{Case Studies}
\label{app:case_studies_reproducability}

\begin{figure*}[t]
\begin{lstlisting}[basicstyle=\ttfamily\tiny, label=lst:cluster_summ_prompt, caption=The prompt used to summarise clusters for the analysis., numbers=none]

SYSTEM_PROMPT 
You are an expert analyst of patterns in social media.
You will receive a list of short sentences describing tweets. In particular, the sentences describe the core message or narrative of a tweet, i.e., what the poster is trying to say or convey.
Identify the common theme that unites these narratives. Importantly, do not just summarize the literal topic, but rather the underlying message or narrative. For example, if the claims are about "vaccine side effects", "government control", and "pharmaceutical profits", the common theme might be "Skepticism of mainstream medicine driven by distrust".
Important! Analyse these sentences as if they are correct and true. Do not try to debunk, fact-check, or invalidate the claims. Do not include any caveats or disclaimers. Instead, focus on understanding the core narrative that would make these claims compelling to a reader who believes them.
Respond ONLY with valid JSON matching the requested format."""

USER_PROMPT:
Here are {n} tweet message descriptions from a cluster:

{messages}

Identify the common theme. Respond with this JSON structure:
{{"label": "<a short sentence capturing the common narratives>", "description": "<2-3 sentence description of the common pattern>", "domain": "<one of: politics, celebrity, sports, health, science, technology, conspiracy, advertising, other>", "keywords": ["<keyword1>", "<keyword2>", "<keyword3>", "<keyword4>", "<keyword5>"]}}"""

\end{lstlisting}
\end{figure*}

We extract and analyse narratives from misinformation samples using the following pipeline. First, each post's \textit{Message} is encoded into a 768-dimensional semantic vector using a multilingual sentence-transformer model (\texttt{all-mpnet-base-v2} \citep{reimers-2020-multilingual-sentence-bert}). Where available, abstract-level paraphrases of each message (as mentioned in \cref{sec:pipeline}) are used as the embedding input to prevent surface-level lexical features from dominating the embedding. Then, we perform a UMAP dimensionality reduction on the embedding, projecting them into 10-dimensional vectors.

Next, we cluster the embeddings into a two-level hierarchy. The top-level simply uses the pre-existing \textit{Topic} labels. As each post may belong to more than one topic (as described in \cref{sec:pipeline}), we ``explode'' our dataset, so that each (post, topic) pair is treated as an independent observation throughout all downstream analyses. We then use KMeans to further divide each top-level cluster into sub-clusters, with the number of sub-clusters proportional to topic size so that each sub-cluster contains approximately the same number of members ($\sim$50 per sub-cluster).

Finally, each sub-cluster is summarised by a large language model (Claude Sonnet 4.6, version April 20th, 2026) as follows. The 20 messages closest to the sub-cluster centroid in the original 768-dimensional embedding space are passed to the LLM. The prompt (\cref{lst:cluster_summ_prompt}) instructs the LLM to identify the \emph{underlying narrative} rather than merely the surface topic. For each cluster, we cache its summary (a short sentence). The final output is a structured JSON file that maps each dataset instance to its cluster and its summary. To investigate narratives, we manually scan the cluster summaries and identify those related to narratives of interest.

\begin{figure*}
    \centering
    \includegraphics[width=1\linewidth]{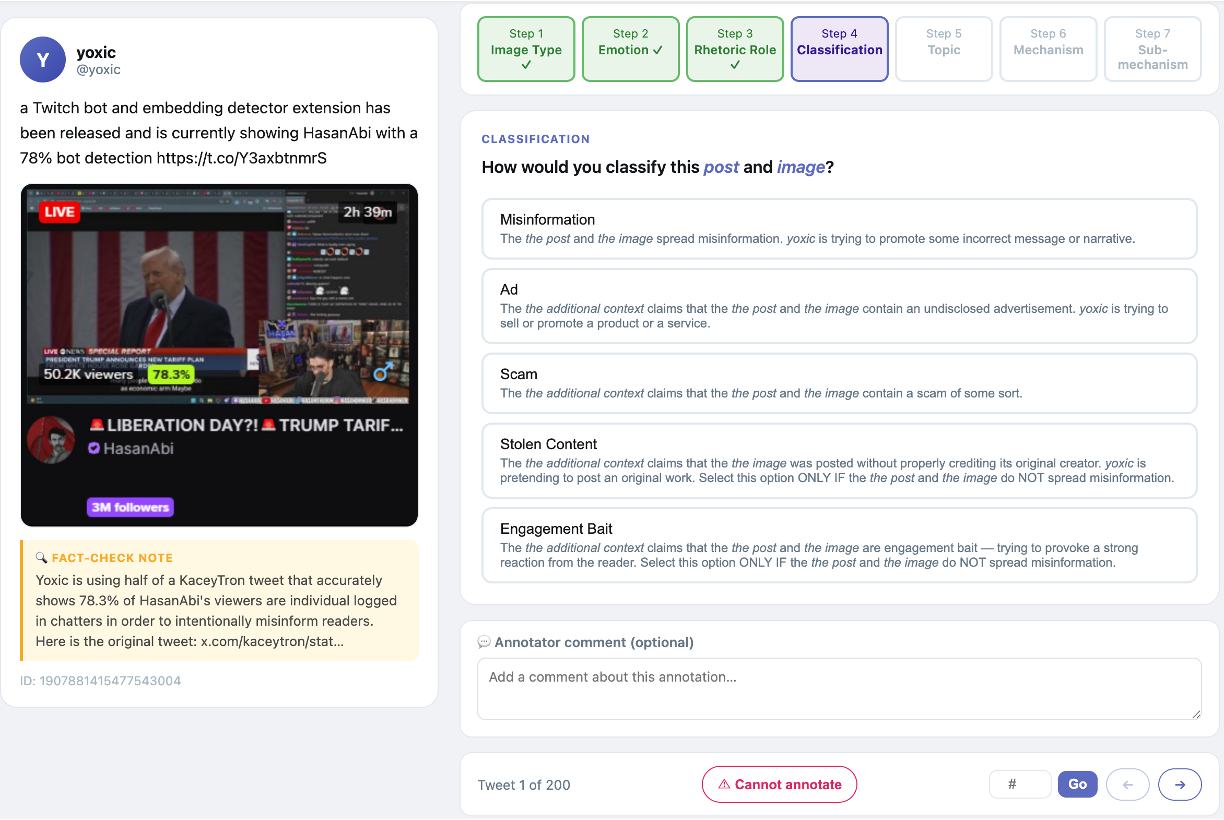}
    \caption{The annotation platform the expert annotators used to label the Community Notes and \textsc{AMeBa} with categories from our taxonomy (as described in \cref{sec:taxonomy_eval}).}
    \label{fig:annotation_platform}
\end{figure*}

\section{Full Taxonomy}
\label{app:full_taxonomy}

\onecolumn
\newcommand{\sublabel}[1]{\hspace*{0.9em}$\hookrightarrow$\,\texttt{#1}}

\newcommand{\axisrow}[1]{%
  \midrule
  \multicolumn{2}{@{}l}{\textbf{\textsc{#1}}} \\
  \midrule}
  
{\footnotesize
\renewcommand{\arraystretch}{1.15}

\begin{longtable}{@{}>{\raggedright\arraybackslash}p{4.6cm}
                    >{\raggedright\arraybackslash}p{11cm}@{}}
\caption{Our multimodal taxonomy. Each axis introduces a group of labels (sub-mechanisms are indented \,$\hookrightarrow$\, under their parent mechanism).}\label{tab:taxonomy_with_definitions}\\
\toprule
\textbf{Label} & \textbf{Definition} \\
\endfirsthead

\multicolumn{2}{@{}l}{\emph{Table~\ref{tab:taxonomy_with_definitions} (continued).}}\\
\toprule
\textbf{Label} & \textbf{Definition} \\
\endhead

\midrule
\multicolumn{2}{r}{\emph{Continued on next page}} \\
\endfoot

\bottomrule
\endlastfoot


\axisrow{Classification}
\texttt{misinformation}   & The post and image spread misinformation. The user is trying to promote some incorrect message or narrative. \\
\texttt{ad}               & The additional context claims that the post and image contain an undisclosed advertisement. The user is trying to sell or promote a product or a service. \\
\texttt{scam}             & The additional context claims that the post and image contain a scam of some sort. \\
\texttt{stolen content}  & The additional context claims that the image was posted without properly crediting its original creator. The user is pretending to post an original work. Select this option only if the post and image do not spread misinformation. \\
\texttt{engagement bait} & The additional context claims that the post and image are engagement bait\,---\,trying to provoke a strong reaction from the reader. Select this option only if the post and image do not spread misinformation. \\

\axisrow{Image type}
\texttt{simple photo people}       & A photograph captured with a camera (with at most minimal editing such as cropping or color correction) that primarily depicts one or more people. Examples: a portrait, a candid shot of a crowd, a selfie. \\
\texttt{simple photo object}       & A minimally edited photograph primarily depicting a physical object or product. Examples: a photo of a weapon, a consumer product, a vehicle. \\
\texttt{simple photo event}        & A minimally edited photograph documenting a specific event or happening. Examples: a protest, a ceremony, an accident scene. \\
\texttt{simple photo environment}  & A minimally edited photograph of a place, landscape, or setting. Examples: a cityscape, a flooded area, a building exterior. \\
\texttt{simple photo document}     & A photograph of a physical document, sign, or handwritten note. Examples: a photo of a printed receipt taken with a phone, a photo of a whiteboard. \\
\texttt{simple photo other}        & A minimally edited photograph that does not fit any of the above simple photo subcategories. Examples: food, animals, abstract patterns. \\
\texttt{social media screenshot}   & A screen capture of content from a social media platform (e.g., X/Twitter, Facebook, Instagram, TikTok, Reddit, Telegram). The platform UI (e.g., like/share buttons, usernames, timestamps) should be at least partially visible. \\
\texttt{news screenshot}            & A screen capture from a news source: a news website, TV broadcast, news app, or newspaper. \\
\texttt{other screenshot}           & A clear screenshot that is not of a social media platform or a news source. The UI should be at least partially visible. \\
\texttt{official document}          & A digital rendering of a formal institutional document such as a government form, court filing, or official report. Distinguished from \texttt{simple photo document} by being a direct digital reproduction rather than a casual photograph. \\
\texttt{other text}                 & An image whose content is primarily text but does not fit the above text-bearing categories. Examples: a quote on a plain background, a text-only listicle, a plain text statement. \\
\texttt{data visualization}         & A chart, graph, map, or table presenting structured quantitative or geospatial data. Examples: bar charts, line graphs, choropleth maps, statistical tables. \\
\texttt{digital art}                & An illustration or rendering created from scratch using software. Examples: digital paintings, vector art, 3D renders. \\
\texttt{graphic design}             & A designed visual composition for communication or promotion, combining typography, imagery, and/or layout elements. Examples: advertisements, campaign posters, branded infographics, flyers. \\
\texttt{meme}                        & An image following a recognizable memetic template or internet humor convention, typically combining a stock or reaction image with overlaid text for humorous, satirical, or rhetorical effect. \\
\texttt{annotated image}            & A base image overlaid with explanatory graphical elements such as arrows, circles, labels, or highlights to draw attention to or explain specific features. Examples: a satellite image with circled areas, a photo with arrows pointing to details. \\
\texttt{captioned image}            & A simple structure in which a single image is paired with a text element (headline, caption, or label), typically placed above or below the image. The composition contains no more than one image and one text block. \\
\texttt{image collage}              & A composition of multiple photographs arranged together (e.g., in a grid or side-by-side layout) without significant additional graphic or textual elements. Examples: before/after comparisons, photo grids. \\
\texttt{complex composite}          & A multi-element composition combining three or more distinct content types (e.g., photographs, text blocks, data visualizations, graphic elements, arrows) into a single structured image. Examples: conspiratorial infographics, elaborate explainer images mixing photos with maps and quoted text. \\
\texttt{other}                       & Any image that does not fit the categories above. \\

\axisrow{Emotion}
\texttt{joy}                        & A feeling of pleasure and happiness. \\
\texttt{hope}                       & A feeling of expectation and desire for a certain thing to happen. \\
\texttt{pride}                      & A feeling of deep pleasure or satisfaction derived from one's own achievements, the achievements of those with whom one is closely associated, or from qualities or possessions that are widely admired. \\
\texttt{curiosity}                  & A strong desire to know or learn something. \\
\texttt{other positive emotion}   & Any other positive emotion not listed above. \\
\texttt{anger}                      & A strong feeling of annoyance, displeasure, or hostility. \\
\texttt{sadness}                    & A feeling of sorrow or unhappiness. \\
\texttt{fear}                       & An unpleasant emotion caused by the threat of danger, pain, or harm. \\
\texttt{ridicule}                   & The subjection of someone or something to contemptuous and dismissive language or behavior. \\
\texttt{other negative emotion}   & Any other negative emotion not listed above. \\
\texttt{neutral emotion}           & The image evokes a neutral emotion. \\

\axisrow{Rhetorical role}
\texttt{decorate}        & The image makes the post more visually attractive or attention-grabbing without meaningfully affecting the reader's understanding of the text. \\
\texttt{elicit emotion} & The image encourages an emotional response (e.g., shock, sympathy, outrage, humor, awe) through its content or style. \\
\texttt{control}         & The image directs, holds, or regulates the reader's attention, or motivates a specific response. \\
\texttt{reiterate}       & The image visually restates what the text says, making the text's content visible with minimal added interpretation. \\
\texttt{organise}        & The image provides spatial, temporal, or conceptual structure to the text's content, helping the reader see how elements are arranged or related. \\
\texttt{relate}          & The image explicitly compares, contrasts, or draws a parallel between elements that are described \emph{within} the text itself. The relationship stays within the scope of the text's own content. \\
\texttt{condense}        & The image reduces the text's content to its most essential or critical elements (e.g., a diagram that shows only the spatial trajectory of an accident, omitting weather, motivations, and consequences described in the text). \\
\texttt{explain}         & The image makes the text's content clearer, more intelligible, or more understandable, while still following the text closely. \\
\texttt{interpret}       & The image provides illustrations of complex ideas in concrete form that go beyond what the text explicitly states. \\
\texttt{develop}         & The image expands on the text by introducing new details, perspectives, comparisons, or contrasts that use elements \emph{not present} in the text. \\
\texttt{transform}       & The image fundamentally recodes, reorganises, or reinterprets the text's content into a new form. \\

\axisrow{Topic}
\texttt{politics}      & Government, elections, political parties, policies, politicians. \\
\texttt{economics}     & Finance, trade, markets, employment, business. \\
\texttt{science}       & Research, discoveries, environment, climate, space. \\
\texttt{health}        & Medicine, public health, disease, mental health, vaccines. \\
\texttt{entertainment} & Movies, TV, music, gaming, arts. \\
\texttt{sports}        & Athletic events, teams, players, competitions. \\
\texttt{celebrities}   & Public figures, influencers, pop culture personalities. \\
\texttt{conflict}      & War, military operations, protests, violence, terrorism. \\
\texttt{law}           & Courts, crime, legislation, trials, law enforcement. \\
\texttt{religion}      & Faith, religious institutions, theology, spiritual practices. \\
\texttt{technology}    & Software, hardware, AI, internet, innovation. \\
\texttt{culture}       & Society, traditions, identity, social issues, lifestyle. \\
\texttt{other}         & The topic does not fit any of the categories above. \\

\axisrow{MM mechanism}
\texttt{decorative}                  & The image is not part of the mechanism to mislead. It provides no additional evidence for the claim and the additional context raises no issues with the image itself. \\
\texttt{fake image}                 & The additional context claims that the image is fabricated, AI-generated, staged, or otherwise completely fake. \\
\sublabel{AI generated}             & The image is AI generated. \\
\sublabel{forgery}                   & The image contains a forged document, tweet, or other text-based fake. \\
\sublabel{staged}                    & The event depicted in the image was staged. \\
\texttt{manipulated image}          & The image is a real image, but the additional context claims it was digitally altered or edited in some manner. \\
\sublabel{addition}                  & Fake details were added to the image using digital means, such as Photoshop. \\
\sublabel{removal}                   & Real details were erased or removed from the image, e.g., using cropping. \\
\sublabel{replacement}               & Entire entities or elements in the image were replaced with different, fake ones. \\
\sublabel{textual}                   & Some text in the image was edited or added. \\
\sublabel{other edit}               & The image was edited in some other way, e.g., changing the facial expression of a person or adding a filter. \\
\texttt{unreliable source}          & The additional context raises issues regarding the validity, reliability, or origin of the image. Focus only on the image, not the text in the post. \\
\sublabel{satire}                    & The source is a satirical account or outlet. \\
\sublabel{imposter}                  & The source is an imposter, impersonating an entity without clearly acknowledging it. \\
\sublabel{low credibility source}  & The image shows content from a source with a known history of publishing false or unreliable information (e.g., a screenshot of a fringe outlet or conspiracy site). \\
\sublabel{other provenance issues} & Other types of provenance issues. \\
\texttt{deny authenticity}          & The post incorrectly denies the authenticity of the image, but the additional context claims that the image is real. \\
\texttt{mismatch}                    & The additional context explains that the post makes a factually incorrect claim about the content or context of the image, such as incorrect time, place, person, or event. Use this when the image is misidentified or taken from a different context entirely. \\
\sublabel{identity mismatch}        & The post incorrectly names an entity in the image, such as a person, a brand, or an object. \\
\sublabel{time mismatch}            & The post incorrectly describes the time/date of when the image was captured. \\
\sublabel{place mismatch}           & The post incorrectly describes the location of where the image was captured. \\
\sublabel{event mismatch}           & There is a discrepancy regarding the event depicted in the image and the textual description in the post (e.g., a concert is described as a protest). \\
\sublabel{other mismatch}           & Any other type of mismatch. \\
\texttt{slanted}                     & The additional context claims that the post describes the image in a misleading way, such as misrepresenting its meaning or consequences, or drawing false or exaggerated conclusions from it. Use this when the image is correctly identified but the interpretation or conclusion is misleading. \\
\sublabel{exaggeration}              & The text in the post describes the image in an exaggerated way, such as overstating the frequency or severity of an event depicted in the image. \\
\sublabel{scientific errors or conspiracies} & The post contains a false scientific claim not grounded in scientific evidence, or promotes a known conspiracy theory. \\
\sublabel{misinterpretation of relevance}     & The post misinterprets the relevance of the image to some other event, i.e., falsely claiming that it is relevant. \\
\sublabel{slanted representation}   & The post omits other types of important context. \\
\texttt{texual claim}               & The image contains the claim in a textual form, and the claim is false. The image contains no important visual elements apart from the text. \\
\texttt{other mechanism}            & The additional context describes a different mechanism that is not listed above. \\

\end{longtable}

}

\end{document}